\documentclass[french,english]{paper}
\usepackage[T1]{fontenc}
\usepackage[latin9]{inputenc}
\usepackage{geometry}
\usepackage{babel}
\makeatletter
\addto\extrasfrench{%
   \providecommand{\fg}{\ifdim\lastskip>\z@\unskip\fi~\frqq}%
}

\makeatother
\usepackage{amsmath}
\usepackage{amssymb}
\usepackage{graphicx}
\usepackage{setspace}
\usepackage{esint}
\usepackage[unicode=true,pdfusetitle,
 bookmarks=true,bookmarksnumbered=false,bookmarksopen=false,
 breaklinks=false,pdfborder={0 0 1},backref=false,colorlinks=false]
 {hyperref}

\makeatletter

\providecommand{\tabularnewline}{\\}

\numberwithin{figure}{section}
\numberwithin{equation}{section}
\numberwithin{table}{section}
\newcommand{\lyxaddress}[1]{
\par {\raggedright #1
\vspace{1.4em}
\noindent\par}
}

\pdfoutput=1

\makeatother

\begin{document}

\title{Correlation structure of stochastic neural networks with generic
connectivity matrices}

\author{Diego Fasoli{*}, Olivier Faugeras{*}}

\maketitle

\lyxaddress{{*}NeuroMathComp Laboratory, INRIA Sophia-Antipolis, France. Email:
firstname.name@inria.fr}

\section*{Abstract}

\noindent Using a perturbative expansion for weak synaptic weights
and weak sources of randomness, we calculate the correlation structure
of neural networks with generic connectivity matrices.

\noindent In detail, the perturbative parameters are the mean and
the standard deviation of the synaptic weights, together with the
standard deviations of the background noise of the membrane potentials
and of their initial conditions.

\noindent We also show how to determine the correlation structure
of the system when the synaptic connections have a random topology.

\noindent This analysis is performed on rate neurons described by
Wilson and Cowan equations, since this allows us to find analytic
results.

\noindent Moreover, the perturbative expansion can be developed at
any order and for a generic connectivity matrix.

\noindent We finally show an example of application of this technique
for a particular case of biologically relevant topology of the synaptic
connections.

\section{\label{sec:Introduction}Introduction}

\noindent The brain is a system characterized by extremely high levels
of complexity, which is inherited from the intricate network of its
synaptic connections, known as \textit{connectome}.

\noindent Therefore it seems plausible to attribute to the connectivity
structure the incredible information processing capabilities of the
brain.

\noindent As a consequence of this point of view, an increasing effort
has been devoted to determining the connectome of different animal
species.

\noindent In particular it has been already completed for the C. elegans
\cite{White12111986}\cite{10.1371/journal.pcbi.1001066}, and partially
determined for the mouse \cite{MouseConnectomeProject}\cite{citeulike:8973484}\cite{citeulike:8973197},
the rat \cite{citeulike:9144028}\cite{23248583}, the cat \cite{ScannellBlakemoreYoung95}\cite{sca99}
and the monkey \cite{Felleman91distributedhierarchical}.

\noindent Recently the project has been started also for humans \cite{10.1371/journal.pcbi.0010042}\cite{sporns2011the}\cite{HumanConnectomeProject}.

\noindent Some topological features of these networks of connections
are already known, and the most notable are their \textit{small world}
properties and the presence of \textit{nested} structures.

\noindent The small world topology refers to the fact that even if
most nodes are not connected to the others, namely even if the network
is not fully connected, they can be reached from the other nodes traveling
along a small number of connections.

\noindent Networks with this kind of connectivity show enhanced information
processing capabilities, wiring costs, speed of propagation of the
signals and synchronizability, as discussed by many authors \cite{watts1998cds}\cite{Bassett01122006}\cite{Stam_Reijneveld_2007}.

\noindent But the brain is also characterized by a nested structure
of the synaptic connections, namely by different scales of organization.

\noindent In fact at the largest scale the brain can be seen as a
single and highly complex macroscopic system, which is able to perform
a series of differentiated tasks like learning, face recognition,
reasoning, speech, movement coordination, and so on.

\noindent Then the brain can be decomposed into many sub-regions with
specific purposes, like the cerebral cortex, the cerebellum, the hippocampus,
the brain stem, etc.

\noindent Moreover, each one of these sub-regions can be divided further
into other smaller areas with more specialized functions.

\noindent For example, wide parts of the cerebral cortex are involved
in the reception of images, smells, sounds, flavors or pain and temperature.

\noindent These areas are known respectively as visual, olfactory,
auditory, gustatory and somatosensory cortices.

\noindent If now for instance we take into account the visual cortex,
we can decompose it in even smaller parts specialized in the detection
of all the features of an image, like color, shadow, boundaries, orientations,
and so on.

\noindent However all these regions of the brain are still characterized
by a macroscopic scale.

\noindent Then we can go deeper and deeper in the subdivision process,
until we reach the mesoscopic scale, which is marked by the presence
of cortical columns \cite{Mountcastle}.

\noindent These columns are in turn formed by many interconnected
populations of neurons known as neural masses \cite{citeulike:6614677},
from which we can finally go down to the lowest level, namely the
microscopic scale of single neurons.

\noindent Moving from the macroscopic to the microscopic scale, the
density of the synaptic connections increases, therefore they actually
form a nested structure.

\noindent According to Sporns \cite{citeulike:1343837}, this topology
can be approximated by a fractal connectivity matrix, with a tunable
level of complexity.

\noindent This represents a considerable improvement in the modelization
of biologically realistic networks.

\noindent In particular, it is of extreme importance to determine
the functional and information processing capabilities that emerge
from this nested structure.

\noindent From this point of view, the first and simplest step is
to determine the relation between the pattern of the synaptic connections,
known as \textit{structural} or \textit{anatomical connectivity},
and the corresponding correlation structure of the neurons, known
as \textit{functional connectivity}.

\noindent Recently this problem has received the attention of the
scientific community \cite{Ponten_Daffertshofer_Hillebrand_Stam_2010}\cite{10.1371/journal.pcbi.1002059}\cite{citeulike:61}\cite{Honey2009}\cite{citeulike:11039744}\cite{citeulike:2332406}.

\noindent However, from a theoretical point of view, the calculation
of the correlation structure of the system is not an easy problem,
especially for highly complex connectivity matrices.

\noindent This analysis has been performed for relatively simple synaptic
topologies, like fully connected networks \cite{journals/siamads/TouboulHF12}\cite{baladron:inserm-00732288}
and connections with special kinds of invariance \cite{FasoliFaugerasStrongWeights}.

\noindent In this article we develop a perturbative approach that
allows us to determine the correlation structure of the system for
every possible topology of the synaptic connections, provided that
the synaptic weights are weak enough.

\noindent In particular, we show how to apply this technique to the
case of the fractal connectivity matrix introduced by Sporns.

\section{\label{sec:Description of the model}Description of the model}

\noindent The perturbative approach developed in this article can
be applied to any neural model, but we take into account only the
case of rate neurons described by Wilson and Cowan equations \cite{journals/siamads/TouboulHF12}\cite{FasoliFaugerasStrongWeights}\cite{Amari_1972}\cite{PhD_amari1977}\cite{Wilson72excitatoryand}\cite{WilsonCowan},
because for this kind of neural equations the perturbative method
provides analytic results. So we suppose that the neural network is
described by the following system of stochastic differential equations:

\begin{onehalfspace}
\begin{center}
{\small{
\begin{equation}
dV_{i}\left(t\right)=\left[-\frac{1}{\tau}V_{i}\left(t\right)+\sum_{j=0}^{N-1}J_{ij}\left(t\right)S\left(V_{j}\left(t\right)\right)+I_{i}\left(t\right)\right]dt+\sigma_{1}dB_{i}\left(t\right)\label{eq:exact-equation}
\end{equation}
}}
\par\end{center}{\small \par}
\end{onehalfspace}

\noindent with $i=0,1,...,N-1$, where:
\begin{itemize}
\item \noindent $N$ is the number of neurons in the network;
\item \noindent $V_{i}\left(t\right)$ is the membrane potential of the
$i$-th neuron;
\item \noindent $\tau$ is a time constant that describes the speed of convergence
to a stationary state;
\item \noindent $I_{i}\left(t\right)$ is the deterministic external input
current of the $i$-th neuron;
\item \noindent $B_{i}\left(t\right)$ is the Brownian motion that describes
the background noise of the $i$-th neuron (or equivalently the stochastic
part of the external input current);
\item \noindent $\sigma_{1}$ is the standard deviation of the Brownian
motions, which for simplicity is supposed to be the same for all the
neurons and time-independent;
\item \noindent $J_{ij}\left(t\right)$ is the random synaptic weight from
the $j$-th neuron to the $i$-th neuron;
\item \noindent $S\left(\cdot\right)$ is an activation function that converts
the membrane potential of a neuron into the rate or frequency of the
spikes it generates.
\end{itemize}
\noindent Usually in neuroscience $S\left(\cdot\right)$ is a sigmoid
function, defined as:

\begin{onehalfspace}
\begin{center}
{\small{
\begin{equation}
S\left(V\right)=\frac{T_{MAX}}{1+e^{-\lambda\left(V-V_{T}\right)}}\label{eq:sigmoid-function}
\end{equation}
}}
\par\end{center}{\small \par}
\end{onehalfspace}

\noindent where $T_{MAX}$ is the maximum amplitude of the function
(which is reached for $V\rightarrow+\infty$), $\lambda$ is the parameter
that determine its slope for $T_{MAX}$ fixed, while $V_{T}$ represents
the horizontal shift of the function along the $V$ axis.

\bigskip{}

\noindent Randomness is present in the system through three different
variables, the Brownian motions, the initial conditions and the strength
of the synaptic weights, which are treated perturbatively. Their distributions
are supposed to be normal, because this allows us to calculate analytically
the correlation structure of the network using the Isserlis' theorem
\cite{1918}. We also introduce a fourth \textit{non-perturbative}
source of randomness, namely the \textit{topology} of the synaptic
connections. This means that not only the intensities of the synaptic
weights are considered as random, but also the existence or not of
a connection between two given neurons is not certain anymore. For
the first three variables, we use the same covariance structures as
in \cite{FasoliFaugerasStrongWeights}. For the Brownian motions
it is given by the matrix $\Sigma_{1}$, whose entries are:

\begin{onehalfspace}
\begin{center}
{\small{
\begin{align}
\left[\Sigma_{1}\right]_{ij}= & Cov\left(\frac{dB_{i}\left(t\right)}{dt},\frac{dB_{j}\left(s\right)}{ds}\right)=C_{ij}^{1}\delta\left(t-s\right)\nonumber \\
\label{eq:Brownian-covariance-1}\\
C_{ij}^{1}= & \begin{cases}
1 & \begin{array}{ccc}
\mathrm{if} &  & i=j\end{array}\\
\\
C_{1} & \begin{array}{ccc}
\mathrm{if} &  & i\neq j\end{array}
\end{cases}\nonumber 
\end{align}
}}
\par\end{center}{\small \par}
\end{onehalfspace}

\noindent where $C_{1}$ is a free parameter that represents the correlation
between two Brownian motions, while $\delta\left(\cdot\right)$ is
the Dirac delta function. The matrix $\Sigma_{1}$ is a genuine covariance
matrix only if it is positive-semidefinite, namely if $\frac{1}{1-N}\leq C_{1}\leq1$.

\bigskip{}

\noindent The initial conditions are defined in terms of the following
multivariate normal process:

\begin{onehalfspace}
\begin{center}
{\small{
\begin{equation}
\overrightarrow{V}\left(0\right)\sim\mathcal{N}\left(\overrightarrow{\mu},\Sigma_{2}\right)\label{eq:initial-conditions-1}
\end{equation}
}}
\par\end{center}{\small \par}
\end{onehalfspace}

\noindent where:

\begin{onehalfspace}
\begin{center}
{\small{
\begin{equation}
\Sigma_{2}=\sigma_{2}^{2}\left[\begin{array}{cccc}
1 & C_{2} & \cdots & C_{2}\\
C_{2} & 1 & \cdots & C_{2}\\
\vdots & \vdots & \ddots & \vdots\\
C_{2} & C_{2} & \cdots & 1
\end{array}\right]\label{eq:initial-conditions-covariance-1}
\end{equation}
}}
\par\end{center}{\small \par}
\end{onehalfspace}

\noindent The parameter $\sigma_{2}$ represents the standard deviation
of the initial conditions, while $C_{2}$ is their correlation. Again
we have to choose $\frac{1}{1-N}\leq C_{2}\leq1$.

\bigskip{}

\noindent In this article we consider networks with random topologies,
which means that the fact to have or not a connection between two
given neurons is a (known) random variable: if in one realization
of the network there is a connection from the $j$-th neuron to the
$i$-th neuron, in another realization this connection could be missing.
Therefore we suppose that the synaptic weights are given by the following
formulae:

\begin{onehalfspace}
\begin{center}
{\small{
\begin{align}
J_{ij}\left(t\right)= & \begin{cases}
\frac{1}{M_{i}}\left[\sigma_{4}\overline{J}_{ij}\left(t\right)+\sigma_{3}W_{ij}\right] & \begin{array}{ccc}
\mathrm{if} &  & M_{i}\neq0\end{array}\\
\\
0 & \begin{array}{ccc}
\mathrm{if} &  & M_{i}=0\end{array}
\end{cases}\label{eq:synaptic-weights-3}\\
\nonumber \\
\overline{J}_{ij}\left(t\right)= & \widehat{\overline{J}}_{ij}\left(t\right)\circ T\label{eq:matrix-J-bar}\\
\nonumber \\
W= & \widehat{W}\circ T\label{eq:matrix-W}\\
\nonumber \\
\widehat{W}\sim & \mathcal{MN}\left(0,\Omega_{3},\Sigma_{3}\right)\label{eq:distribution-matrix-W-hat}\\
\nonumber \\
M_{i}= & \sum_{j=0}^{N-1}T_{ij}\label{eq:number-of-connections}
\end{align}
}}
\par\end{center}{\small \par}
\end{onehalfspace}

\noindent where $\sigma_{3}$ and $\sigma_{4}$ are two perturbative
parameters that represent (after the division by $M_{i}$), respectively,
the standard deviation and the mean strength of the synaptic connections.
$M_{i}$ is the number (in general random) of incoming connections
to the $i$-th neuron, and is used to prevent the explosion of the
term ${\displaystyle \sum_{j=0}^{N-1}}J_{ij}\left(t\right)S\left(V_{j}\left(t\right)\right)$
in equation \ref{eq:exact-equation} when $M_{i}$ grows arbitrarily
large. The symbol ``$\circ$'' represents the \textit{Hadamard product},
therefore $C=A\circ B$ means that $C_{ij}=A_{ij}B_{ij}$, $\forall i,j$.
$T$ is a generic binary random matrix which represents the topology
of the synaptic connections. More explicitly, we have $T_{ij}=0$
if there is no connection from the $j$-th to the $i$-th neuron (namely
if $J_{ij}\left(t\right)=0$ $\forall t$), while $T_{ij}=1$ if this
connection is present. Below we show an example of connectivity matrix
and its corresponding topology:

\begin{onehalfspace}
\begin{center}
{\small{
\[
\begin{array}{ccc}
\overline{J}\left(t\right)=\left[\begin{array}{cccc}
0 & 0 & 2cos\left(t\right) & 3.6\\
sin\left(5t\right) & 0 & 10 & 0\\
1 & \pi & 0 & arctan\left(7t\right)\\
0 & \left(1+t\right)^{-5} & e^{-3t} & 0
\end{array}\right], &  & T=\left[\begin{array}{cccc}
0 & 0 & 1 & 1\\
1 & 0 & 1 & 0\\
1 & 1 & 0 & 1\\
0 & 1 & 1 & 0
\end{array}\right]\end{array}
\]
}}
\par\end{center}{\small \par}
\end{onehalfspace}

\noindent The matrix $\hat{\overline{J}}\left(t\right)$ is completely
deterministic, while the matrix $\widehat{W}$ is random only in the
amplitudes of the synaptic weights (which follow a matrix normal distribution
$\mathcal{MN}\left(0,\Omega_{3},\Sigma_{3}\right)$ \cite{MatrixNormalDistribution}),
but not in the topology. The covariance matrices $\Omega_{3}$ and
$\Sigma_{3}$ of $\widehat{W}$ are chosen in order to have:

\begin{onehalfspace}
\begin{center}
{\small{
\begin{equation}
Cov\left(\widehat{W}_{ij},\widehat{W}_{kl}\right)=\begin{cases}
1 & \begin{array}{ccc}
\mathrm{if} &  & \left(i=k\right)\wedge\left(j=l\right)\end{array}\\
\\
C_{3} & \mathrm{otherwise}
\end{cases}\label{eq:covariance-matrix-W-hat}
\end{equation}
}}
\par\end{center}{\small \par}
\end{onehalfspace}

\noindent The free parameter $C_{3}$ represents the correlation between
two different and non-zero synaptic weights, and the range of its
plausible values depends on the topology of the connections, which
is supposed to be completely generic. Moreover we assume that $\widehat{W}$
and $T$ are independent.

\bigskip{}

\noindent To finish, we suppose that also the Brownian motions, the
initial conditions, the amplitudes of the synaptic weights and the
topology are independent from each other, therefore their reciprocal
covariances are equal to zero:

\begin{onehalfspace}
\begin{center}
{\small{
\begin{align}
 & Cov\left(B_{i}\left(t\right),V_{j}\left(0\right)\right)=Cov\left(B_{i}\left(t\right),\widehat{W}_{jk}\right)=Cov\left(B_{i}\left(t\right),T_{jk}\right)\nonumber \\
\nonumber \\
 & =Cov\left(V_{i}\left(0\right),\widehat{W}_{jk}\right)=Cov\left(V_{i}\left(0\right),T_{jk}\right)=0,\begin{array}{cc}
 & \forall i,j,k\end{array}\label{eq:mutual-covariance-2}
\end{align}
}}
\par\end{center}{\small \par}
\end{onehalfspace}

\noindent In principle, the inner and mutual covariance structure
of $B_{i}\left(t\right)$, $V_{i}\left(0\right)$ and $\widehat{W}_{ij}$
can be arbitrarily chosen. However here we use only the simple structure
defined by formulae \ref{eq:Brownian-covariance-1}, \ref{eq:initial-conditions-covariance-1},
\ref{eq:covariance-matrix-W-hat} and \ref{eq:mutual-covariance-2},
because this will generate simple analytic results for the correlation
structure of the membrane potentials.

\noindent \begin{flushleft}
\bigskip{}
We now are ready to introduce a perturbative expansion of $V_{i}\left(t\right)$
in terms of the parameters $\sigma$:
\par\end{flushleft}

\begin{onehalfspace}
\begin{center}
{\small{
\begin{equation}
V_{i}\left(t\right)\approx Y_{0}^{i}\left(t\right)+\sum_{m=1}^{4}\sigma_{m}Y_{m}^{i}\left(t\right)+{\displaystyle \sum\limits _{\substack{m,n=1\\
m\leq n
}
}^{4}}\sigma_{m}\sigma_{n}Y_{m,n}^{i}\left(t\right)\label{eq:membrane-potential-perturbative-expansion}
\end{equation}
}}
\par\end{center}{\small \par}
\end{onehalfspace}

\noindent where the functions $Y_{m}^{i}\left(t\right)$ and $Y_{m,n}^{i}\left(t\right)$
are to be determined through equation \ref{eq:exact-equation}. In
principle this expansion can be extended to any perturbative order,
but in this article we truncate it at the second because the complexity
of the results becomes quickly intractable.

\subsection{\label{sub:The system of equations}The system of equations}

\noindent In order to evaluate the functions $Y_{m}^{i}\left(t\right)$
and $Y_{m,n}^{i}\left(t\right)$, we have to replace the expansion
\ref{eq:membrane-potential-perturbative-expansion} inside the equation
\ref{eq:exact-equation}, and to identify the coefficients of the
same monomials in $\sigma$. Before doing this, we need the expansion
of the sigmoid function in terms of $\sigma$. Therefore, defining:

\begin{onehalfspace}
\begin{center}
\textit{\small{
\[
\zeta_{j}=\sum_{m=1}^{4}\sigma_{m}Y_{m}^{j}\left(t\right)+{\displaystyle \sum\limits _{\substack{m,n=1\\
m\leq n
}
}^{4}}\sigma_{m}\sigma_{n}Y_{m,n}^{j}\left(t\right)
\]
}}
\par\end{center}{\small \par}
\end{onehalfspace}

\noindent the Taylor expansion of the sigmoid function is:

\begin{onehalfspace}
\begin{center}
{\small{
\begin{align*}
S\left(\mu+\zeta_{j}\right)\approx & S\left(\mu\right)+S'\left(\mu\right)\zeta_{j}+\frac{1}{2}S''\left(\mu\right)\zeta_{j}^{2}\\
\\
\approx & S\left(\mu\right)+S'\left(\mu\right)\sum_{m=1}^{4}\sigma_{m}Y_{m}^{j}\left(t\right)\\
\\
 & +{\displaystyle \sum\limits _{\substack{m,n=1\\
m<n
}
}^{4}}\sigma_{m}\sigma_{n}\left[S'\left(\mu\right)Y_{m,n}^{j}\left(t\right)+S''\left(\mu\right)Y_{m}^{j}\left(t\right)Y_{n}^{j}\left(t\right)\right]\\
\\
 & +\sum_{m=1}^{4}\sigma_{m}^{2}\left[S'\left(\mu\right)Y_{m,m}^{j}\left(t\right)+\frac{1}{2}S''\left(\mu\right)\left(Y_{m}^{j}\left(t\right)\right)^{2}\right]
\end{align*}
}}
\par\end{center}{\small \par}
\end{onehalfspace}

\noindent having neglected the terms with order higher than $2$.
This expansion can be used provided that its radius of convergence
is large enough. The rigorous analysis can be found in \cite{FasoliFaugerasStrongWeights}
and shows that the expansion is convergent if the sigmoid function
is not too steep at $V=V_{T}$, namely if the parameter $\lambda$
is not too large. Now, if we replace this expansion and \ref{eq:membrane-potential-perturbative-expansion}
inside the equation \ref{eq:exact-equation}, comparing the coefficients
of the same monomials in $\sigma$ we obtain the following equations:

\begin{onehalfspace}
\begin{center}
{\small{
\begin{align}
dY_{0}^{i}\left(t\right)= & \left[-\frac{1}{\tau}Y_{0}^{i}\left(t\right)+I_{i}\left(t\right)\right]dt\label{eq:perturbative-equation-0}\\
\nonumber \\
dY_{1}^{i}\left(t\right)= & -\frac{1}{\tau}Y_{1}^{i}\left(t\right)dt+dB_{i}\left(t\right)\label{eq:perturbative-equation-1}\\
\nonumber \\
dY_{2}^{i}\left(t\right)= & -\frac{1}{\tau}Y_{2}^{i}\left(t\right)dt\label{eq:perturbative-equation-2}\\
\nonumber \\
dY_{3}^{i}\left(t\right)= & \left[-\frac{1}{\tau}Y_{3}^{i}\left(t\right)+\frac{1}{M_{i}}\sum_{j=0}^{N-1}W_{ij}S\left(Y_{0}^{j}\left(t\right)\right)\right]dt\label{eq:perturbative-equation-3}\\
\nonumber \\
dY_{4}^{i}\left(t\right)= & \left[-\frac{1}{\tau}Y_{4}^{i}\left(t\right)+\frac{1}{M_{i}}\sum_{j=0}^{N-1}\overline{J}_{ij}\left(t\right)S\left(Y_{0}^{j}\left(t\right)\right)\right]dt\label{eq:perturbative-equation-4}\\
\vdots\nonumber \\
dY_{1,4}^{i}\left(t\right)= & \left[-\frac{1}{\tau}Y_{1,4}^{i}\left(t\right)+\frac{1}{M_{i}}\sum_{j=0}^{N-1}\overline{J}_{ij}\left(t\right)S'\left(Y_{0}^{j}\left(t\right)\right)Y_{1}^{j}\left(t\right)\right]dt\label{eq:perturbative-equation-5}\\
\nonumber \\
dY_{2,4}^{i}\left(t\right)= & \left[-\frac{1}{\tau}Y_{2,4}^{i}\left(t\right)+\frac{1}{M_{i}}\sum_{j=0}^{N-1}\overline{J}_{ij}\left(t\right)S'\left(Y_{0}^{j}\left(t\right)\right)Y_{2}^{j}\left(t\right)\right]dt\label{eq:perturbative-equation-6}\\
\nonumber \\
dY_{3,4}^{i}\left(t\right)= & \left[-\frac{1}{\tau}Y_{3,4}^{i}\left(t\right)+\frac{1}{M_{i}}\sum_{j=0}^{N-1}\overline{J}_{ij}\left(t\right)S'\left(Y_{0}^{j}\left(t\right)\right)Y_{3}^{j}\left(t\right)+\frac{1}{M_{i}}\sum_{j=0}^{N-1}W_{ij}S'\left(Y_{0}^{j}\left(t\right)\right)Y_{4}^{j}\left(t\right)\right]dt\label{eq:perturbative-equation-7}\\
\nonumber \\
dY_{4,4}^{i}\left(t\right)= & \left[-\frac{1}{\tau}Y_{4,4}^{i}\left(t\right)+\frac{1}{M_{i}}\sum_{j=0}^{N-1}\overline{J}_{ij}\left(t\right)S'\left(Y_{0}^{j}\left(t\right)\right)Y_{4}^{j}\left(t\right)\right]dt\label{eq:perturbative-equation-8}\\
\vdots\nonumber 
\end{align}
}}
\par\end{center}{\small \par}
\end{onehalfspace}

\noindent We have only written the equations that will be used in
Section \ref{sec:Correlation structure of the network}. The others
do not influence the perturbative expansions of the variance and covariance
truncated at the $3$rd perturbative order, therefore they are not
shown here.

\subsection{\label{sub:The initial conditions}The initial conditions}

\begin{flushleft}
The perturbative expansion \ref{eq:membrane-potential-perturbative-expansion}
at $t=0$ gives:
\par\end{flushleft}

\begin{onehalfspace}
\begin{center}
{\small{
\[
V_{i}\left(0\right)\approx Y_{0}^{i}\left(0\right)+\sum_{m=1}^{4}\sigma_{m}Y_{m}^{i}\left(0\right)+{\displaystyle \sum\limits _{\substack{m,n=1\\
m\leq n
}
}^{4}}\sigma_{m}\sigma_{n}Y_{m,n}^{i}\left(0\right)
\]
}}
\par\end{center}{\small \par}
\end{onehalfspace}

\noindent From \ref{eq:initial-conditions-1} we have $V_{i}\left(0\right)\sim\mathcal{N}\left(\mu_{i},\sigma_{2}^{2}\right)=\mu_{i}+\sigma_{2}\mathcal{N}\left(0,1\right)$,
so comparing the two expressions we obtain:

\begin{onehalfspace}
\begin{center}
{\small{
\begin{align}
 & Y_{0}^{i}\left(0\right)=\mu_{i}\label{eq:initial-conditions-decomposed-0}\\
\nonumber \\
 & Y_{2}^{i}\left(0\right)\sim\mathcal{N}\left(0,1\right)\label{eq:initial-conditions-decomposed-1}\\
\nonumber \\
 & \begin{array}{ccc}
Y_{m}^{i}\left(0\right)=0, &  & m=1,3,4\end{array}\label{eq:initial-conditions-decomposed-2}\\
\nonumber \\
 & \begin{array}{ccc}
Y_{m,n}^{i}\left(0\right)=0, &  & \forall\left(m,n\right):\, m\leq n\end{array}\label{eq:initial-conditions-decomposed-3}
\end{align}
}}
\par\end{center}{\small \par}
\end{onehalfspace}

\noindent Therefore we can write the initial conditions as $V_{i}\left(0\right)=\mu_{i}+\sigma_{2}Y_{2}^{i}\left(0\right)$,
from which we obtain:

\begin{onehalfspace}
\begin{center}
{\small{
\[
Cov\left(V_{i}\left(0\right),V_{j}\left(0\right)\right)=\sigma_{2}^{2}Cov\left(Y_{2}^{i}\left(0\right),Y_{2}^{j}\left(0\right)\right)
\]
}}
\par\end{center}{\small \par}
\end{onehalfspace}

\noindent Since from \ref{eq:initial-conditions-covariance-1} we
also know that:

\begin{onehalfspace}
\begin{center}
{\small{
\[
Cov\left(V_{i}\left(0\right),V_{j}\left(0\right)\right)=\begin{cases}
\sigma_{2}^{2} & \begin{array}{ccc}
\mathrm{if} &  & i=j\end{array}\\
\\
\sigma_{2}^{2}C_{2} & \begin{array}{ccc}
\mathrm{if} &  & i\neq j\end{array}
\end{cases}
\]
}}
\par\end{center}{\small \par}
\end{onehalfspace}

\noindent from the comparison of these two expressions of the covariance
matrix of $V_{i}\left(0\right)$ we obtain:

\begin{onehalfspace}
\begin{center}
{\small{
\begin{equation}
Cov\left(Y_{2}^{i}\left(0\right),Y_{2}^{j}\left(0\right)\right)=\begin{cases}
1 & \begin{array}{ccc}
\mathrm{if} &  & i=j\end{array}\\
\\
C_{2} & \begin{array}{ccc}
\mathrm{if} &  & i\neq j\end{array}
\end{cases}\label{eq:initial-conditions-covariance-decomposed}
\end{equation}
}}
\par\end{center}{\small \par}
\end{onehalfspace}

\subsection{\label{sub:Solutions of the equations}Solutions of the equations}

\noindent Since equations \ref{eq:perturbative-equation-0} - \ref{eq:perturbative-equation-8}
are linear, they can be solved analytically, giving the following
solutions:

\begin{onehalfspace}
\begin{center}
{\small{
\begin{align}
Y_{0}^{i}\left(t\right)= & e^{-\frac{t}{\tau}}\left[\mu_{i}+\int_{0}^{t}e^{\frac{s}{\tau}}I_{i}\left(s\right)ds\right]\label{eq:solution-perturbative-equation-0}\\
\nonumber \\
Y_{1}^{i}\left(t\right)= & e^{-\frac{t}{\tau}}\int_{0}^{t}e^{\frac{s}{\tau}}dB_{i}\left(s\right)\label{eq:solution-perturbative-equation-1}\\
\nonumber \\
Y_{2}^{i}\left(t\right)= & e^{-\frac{t}{\tau}}Y_{2}^{i}\left(0\right)\label{eq:solution-perturbative-equation-2}\\
\nonumber \\
Y_{3}^{i}\left(t\right)= & \frac{e^{-\frac{t}{\tau}}}{M_{i}}\sum_{j=0}^{N-1}W_{ij}\int_{0}^{t}e^{\frac{s}{\tau}}S\left(Y_{0}^{j}\left(s\right)\right)ds\label{eq:solution-perturbative-equation-3}\\
\nonumber \\
Y_{4}^{i}\left(t\right)= & \frac{e^{-\frac{t}{\tau}}}{M_{i}}\sum_{j=0}^{N-1}\int_{0}^{t}e^{\frac{s}{\tau}}\overline{J}_{ij}\left(s\right)S\left(Y_{0}^{j}\left(s\right)\right)ds\hphantom{\,\,\,\,\,\,\,\,\,\,\,\,\,\,\,\,\,\,\,\,\,\,\,\,\,\,\,\,\,\,\,\,\,\,\,\,\,\,\,\,\,\,\,\,\,\,\,\,\,\,\,\,\,\,\,\,\,\,\,\,\,\,\,\,\,\,\,\,\,\,\,\,\,\,\,\,\,\,\,\,\,\,\,}\label{eq:solution-perturbative-equation-4}\\
\vdots\nonumber \\
Y_{1,4}^{i}\left(t\right)= & \frac{e^{-\frac{t}{\tau}}}{M_{i}}\sum_{j=0}^{N-1}\int_{0}^{t}\overline{J}_{ij}\left(s\right)S'\left(Y_{0}^{j}\left(s\right)\right)\left[\int_{0}^{s}e^{\frac{u}{\tau}}dB_{j}\left(u\right)\right]ds\label{eq:solution-perturbative-equation-5}\\
\nonumber \\
Y_{2,4}^{i}\left(t\right)= & \frac{e^{-\frac{t}{\tau}}}{M_{i}}\sum_{j=0}^{N-1}Y_{2}^{j}\left(0\right)\int_{0}^{t}\overline{J}_{ij}\left(s\right)S'\left(Y_{0}^{j}\left(s\right)\right)ds\label{eq:solution-perturbative-equation-6}
\end{align}
}}
\par\end{center}{\small \par}

\begin{center}
{\small{
\begin{align}
Y_{3,4}^{i}\left(t\right)= & \frac{e^{-\frac{t}{\tau}}}{M_{i}}\left\{ \sum_{j,k=0}^{N-1}\frac{W_{jk}}{M_{j}}\int_{0}^{t}\overline{J}_{ij}\left(s\right)S'\left(Y_{0}^{j}\left(s\right)\right)\left[\int_{0}^{s}e^{\frac{u}{\tau}}S\left(Y_{0}^{k}\left(u\right)\right)du\right]ds\right.\nonumber \\
 & \left.+\sum_{j,k=0}^{N-1}\frac{W_{ij}}{M_{j}}\int_{0}^{t}S'\left(Y_{0}^{j}\left(s\right)\right)\left[\int_{0}^{s}e^{\frac{u}{\tau}}\overline{J}_{jk}\left(u\right)S\left(Y_{0}^{k}\left(u\right)\right)du\right]ds\right\} \label{eq:solution-perturbative-equation-7}\\
\nonumber \\
Y_{4,4}^{i}\left(t\right)= & \frac{e^{-\frac{t}{\tau}}}{M_{i}}\sum_{j,k=0}^{N-1}\frac{1}{M_{j}}\int_{0}^{t}\overline{J}_{ij}\left(s\right)S'\left(Y_{0}^{j}\left(s\right)\right)\left[\int_{0}^{s}e^{\frac{u}{\tau}}\overline{J}_{jk}\left(u\right)S\left(Y_{0}^{k}\left(u\right)\right)du\right]ds\label{eq:solution-perturbative-equation-8}\\
\vdots\nonumber 
\end{align}
}}
\par\end{center}{\small \par}
\end{onehalfspace}

\noindent Now we can use these results to calculate the correlation
structure of the membrane potentials.

\section{\label{sec:Correlation structure of the network}Correlation structure
of the network}

\noindent In this section we analyze the general case of random topologies,
and we consider the networks with deterministic connections as a special
case. From the perturbative expansion \ref{eq:membrane-potential-perturbative-expansion}
with all the functions $Y_{m}^{i}\left(t\right)$ and $Y_{m,n}^{i}\left(t\right)$
evaluated as shown in Section \ref{sub:Solutions of the equations},
in order to calculate the covariance matrix of the membrane potentials
we need to determine all the pair covariances between all the possible
combinations of these functions. This is a consequence of the bilinearity
property of the covariance operator. However, using the Isserlis'
theorem and the relations \ref{eq:mutual-covariance-2}, it is easy
to see that many of these terms are equal to zero. Moreover we have
also to remove the $4$th order terms in the expression of the covariance,
like $\sigma_{1}^{2}\sigma_{3}^{2}Cov\left(Y_{1,3}^{i}\left(t\right),Y_{1,3}^{j}\left(t\right)\right)$,
since they are not complete. This is due to the fact that there are
also $4$th order terms like $\sigma_{1}^{2}\sigma_{3}^{2}Cov\left(Y_{1}^{i}\left(t\right),Y_{1,1,3}^{j}\left(t\right)\right)$.
These terms are due to $3$rd order functions, like $Y_{1,1,3}^{j}\left(t\right)$
in this case, in the perturbative expansion \ref{eq:membrane-potential-perturbative-expansion},
which have not been taken into account since we have truncated the
expansion of the membrane potential at the $2$nd order. Therefore
the expansion of the covariance must be truncated at the $3$rd order.
So, to conclude, we obtain the following result:

\textit{\small{
\begin{align*}
 & Cov\left(V_{i}\left(t\right),V_{j}\left(t\right)\right)\\
\\
 & =\sigma_{1}^{2}Cov\left(Y_{1}^{i}\left(t\right),Y_{1}^{j}\left(t\right)\right)+\sigma_{2}^{2}Cov\left(Y_{2}^{i}\left(t\right),Y_{2}^{j}\left(t\right)\right)\\
\\
 & +\sigma_{3}^{2}Cov\left(Y_{3}^{i}\left(t\right),Y_{3}^{j}\left(t\right)\right)+\sigma_{4}^{2}Cov\left(Y_{4}^{i}\left(t\right),Y_{4}^{j}\left(t\right)\right)\\
\\
 & +\sigma_{4}\left\{ \sigma_{1}^{2}\left[Cov\left(Y_{1}^{i}\left(t\right),Y_{1,4}^{j}\left(t\right)\right)+Cov\left(Y_{1,4}^{i}\left(t\right),Y_{1}^{j}\left(t\right)\right)\right]+\sigma_{2}^{2}\left[Cov\left(Y_{2}^{i}\left(t\right),Y_{2,4}^{j}\left(t\right)\right)+Cov\left(Y_{2,4}^{i}\left(t\right),Y_{2}^{j}\left(t\right)\right)\right]\right.
\end{align*}
}}{\small \par}

\begin{onehalfspace}
\begin{center}
\textit{\small{
\begin{align}
 & \left.+\sigma_{3}^{2}\left[Cov\left(Y_{3}^{i}\left(t\right),Y_{3,4}^{j}\left(t\right)\right)+Cov\left(Y_{3,4}^{i}\left(t\right),Y_{3}^{j}\left(t\right)\right)\right]+\sigma_{4}^{2}\left[Cov\left(Y_{4}^{i}\left(t\right),Y_{4,4}^{j}\left(t\right)\right)+Cov\left(Y_{4,4}^{i}\left(t\right),Y_{4}^{j}\left(t\right)\right)\right]\right\} \hphantom{\,\,\,\,\,\,\,\,\,\,\,\,\,}\label{eq:covariance}
\end{align}
}}
\par\end{center}{\small \par}
\end{onehalfspace}

\noindent where, due to formulae \ref{eq:solution-perturbative-equation-1},
\ref{eq:solution-perturbative-equation-2} and \ref{eq:solution-perturbative-equation-3},
for $i\neq j$ we obtain:

\begin{onehalfspace}
\begin{center}
{\small{
\begin{align}
Cov\left(Y_{1}^{i}\left(t\right),Y_{1}^{j}\left(t\right)\right)= & \frac{\tau C_{1}}{2}\left(1-e^{-\frac{2t}{\tau}}\right)\label{eq:covariance-part-1}\\
\nonumber \\
Cov\left(Y_{2}^{i}\left(t\right),Y_{2}^{j}\left(t\right)\right)= & C_{2}e^{-\frac{2t}{\tau}}\label{eq:covariance-part-2}\\
\nonumber \\
Cov\left(Y_{3}^{i}\left(t\right),Y_{3}^{j}\left(t\right)\right)= & C_{3}e^{-\frac{2t}{\tau}}{\displaystyle \sum\limits _{k,l=0}^{N-1}}\left[\int_{0}^{t}e^{\frac{s}{\tau}}S\left(Y_{0}^{k}\left(s\right)\right)ds\right]\left[\int_{0}^{t}e^{\frac{s}{\tau}}S\left(Y_{0}^{l}\left(s\right)\right)ds\right]\mathbb{E}\left[\frac{T_{ik}T_{jl}}{M_{i}M_{j}}\right]\label{eq:covariance-part-3}
\end{align}
}}
\par\end{center}{\small \par}
\end{onehalfspace}

\noindent and for $i=j$:

\begin{onehalfspace}
\begin{center}
{\small{
\begin{align}
Var\left(Y_{1}^{i}\left(t\right)\right)= & \frac{\tau}{2}\left(1-e^{-\frac{2t}{\tau}}\right)\label{eq:variance-part-1}\\
\nonumber \\
Var\left(Y_{2}^{i}\left(t\right)\right)= & e^{-\frac{2t}{\tau}}\label{eq:variance-part-2}\\
\nonumber \\
Var\left(Y_{3}^{i}\left(t\right)\right)= & e^{-\frac{2t}{\tau}}\left\{ {\displaystyle \sum\limits _{k=0}^{N-1}}\left[\int_{0}^{t}e^{\frac{s}{\tau}}S\left(Y_{0}^{k}\left(s\right)\right)ds\right]^{2}\mathbb{E}\left[\left(\frac{T_{ik}}{M_{i}}\right)^{2}\right]\vphantom{{\displaystyle \sum\limits _{\substack{k,l\\
k\neq l
}
}}}\right.\nonumber \\
 & \left.+C_{3}{\displaystyle \sum\limits _{\substack{k,l\\
k\neq l
}
}}\left[\int_{0}^{t}e^{\frac{s}{\tau}}S\left(Y_{0}^{k}\left(s\right)\right)ds\right]\left[\int_{0}^{t}e^{\frac{s}{\tau}}S\left(Y_{0}^{l}\left(s\right)\right)ds\right]\mathbb{E}\left[\frac{T_{ik}T_{il}}{M_{i}^{2}}\right]\right\} \hphantom{\,\,\,\,\,\,\,\,\,\,\,\,\,\,\,\,\,\,}\label{eq:variance-part-3}
\end{align}
}}
\par\end{center}{\small \par}
\end{onehalfspace}

\noindent Because of formulae \ref{eq:solution-perturbative-equation-1}
- \ref{eq:solution-perturbative-equation-8}, for all $i,j$ we obtain:

\begin{onehalfspace}
\begin{center}
{\scriptsize{
\begin{align}
 & Cov\left(Y_{4}^{i}\left(t\right),Y_{4}^{j}\left(t\right)\right)=e^{-\frac{2t}{\tau}}{\displaystyle \sum\limits _{k,l=0}^{N-1}}\left[\int_{0}^{t}e^{\frac{s}{\tau}}\hat{\overline{J}}_{ik}\left(s\right)S\left(Y_{0}^{k}\left(s\right)\right)ds\right]\left[\int_{0}^{t}e^{\frac{s}{\tau}}\hat{\overline{J}}_{jl}\left(s\right)S\left(Y_{0}^{l}\left(s\right)\right)ds\right]Cov\left(\frac{T_{ik}}{M_{i}},\frac{T_{jl}}{M_{j}}\right)\label{eq:covariance-part-4}\\
\nonumber \\
 & Cov\left(Y_{1}^{i}\left(t\right),Y_{1,4}^{j}\left(t\right)\right)=\nonumber \\
 & =\frac{\tau}{2}e^{-\frac{2t}{\tau}}\left\{ \mathbb{E}\left[\frac{T_{ji}}{M_{j}}\right]\left[\int_{0}^{t}\widehat{\overline{J}}_{ji}\left(s\right)S'\left(Y_{0}^{i}\left(s\right)\right)\left(e^{\frac{2s}{\tau}}-1\right)ds\right]+C_{1}{\displaystyle \sum\limits _{\substack{k=0\\
k\neq i
}
}^{N-1}}\mathbb{E}\left[\frac{T_{jk}}{M_{j}}\right]\left[\int_{0}^{t}\widehat{\overline{J}}_{jk}\left(s\right)S'\left(Y_{0}^{k}\left(s\right)\right)\left(e^{\frac{2s}{\tau}}-1\right)ds\right]\right\} \label{eq:covariance-part-5}\\
\nonumber \\
 & Cov\left(Y_{2}^{i}\left(t\right),Y_{2,4}^{j}\left(t\right)\right)=e^{-\frac{2t}{\tau}}\left\{ \mathbb{E}\left[\frac{T_{ji}}{M_{j}}\right]\left[\int_{0}^{t}\widehat{\overline{J}}_{ji}\left(s\right)S\left(Y_{0}^{i}\left(s\right)\right)ds\right]+C_{2}{\displaystyle \sum\limits _{\substack{k=0\\
k\neq i
}
}^{N-1}}\mathbb{E}\left[\frac{T_{jk}}{M_{j}}\right]\left[\int_{0}^{t}\widehat{\overline{J}}_{jk}\left(s\right)S\left(Y_{0}^{k}\left(s\right)\right)ds\right]\right\} \label{eq:covariance-part-6}\\
\nonumber \\
 & Cov\left(Y_{3}^{i}\left(t\right),Y_{3,4}^{j}\left(t\right)\right)\nonumber \\
 & =e^{-\frac{2t}{\tau}}\left\{ {\displaystyle \sum\limits _{k=0}^{N-1}}\mathbb{E}\left[\left(\frac{T_{ik}}{M_{i}}\right)^{2}\frac{T_{ji}}{M_{j}}\right]\left[\int_{0}^{t}e^{\frac{s}{\tau}}S\left(Y_{0}^{k}\left(s\right)\right)ds\right]\int_{0}^{t}\widehat{\overline{J}}_{ji}\left(s\right)S'\left(Y_{0}^{i}\left(s\right)\right)\left[\int_{0}^{s}e^{\frac{u}{\tau}}S\left(Y_{0}^{k}\left(u\right)\right)du\right]ds\right.\nonumber \\
 & +C_{3}{\displaystyle \sum\limits _{k,l=0}^{N-1}}\mathbb{E}\left[\frac{T_{ik}T_{il}T_{ji}}{M_{i}^{2}M_{j}}\right]\left[\int_{0}^{t}e^{\frac{s}{\tau}}S\left(Y_{0}^{k}\left(s\right)\right)ds\right]\int_{0}^{t}\widehat{\overline{J}}_{ji}\left(s\right)S'\left(Y_{0}^{i}\left(s\right)\right)\left[\int_{0}^{s}e^{\frac{u}{\tau}}S\left(Y_{0}^{l}\left(u\right)\right)du\right]ds\nonumber \\
 & +C_{3}{\displaystyle \sum\limits _{k,l=0}^{N-1}}\mathbb{E}\left[\frac{T_{ik}T_{lk}T_{jl}}{M_{i}M_{j}M_{l}}\right]\left[\int_{0}^{t}e^{\frac{s}{\tau}}S\left(Y_{0}^{k}\left(s\right)\right)ds\right]\int_{0}^{t}\widehat{\overline{J}}_{jl}\left(s\right)S'\left(Y_{0}^{l}\left(s\right)\right)\left[\int_{0}^{s}e^{\frac{u}{\tau}}S\left(Y_{0}^{k}\left(u\right)\right)du\right]ds\nonumber \\
 & \left.+C_{3}{\displaystyle \sum\limits _{k,l,m=0}^{N-1}}\mathbb{E}\left[\frac{T_{ik}T_{lm}T_{jl}}{M_{i}M_{j}M_{l}}\right]\left[\int_{0}^{t}e^{\frac{s}{\tau}}S\left(Y_{0}^{k}\left(s\right)\right)ds\right]\int_{0}^{t}\widehat{\overline{J}}_{jl}\left(s\right)S'\left(Y_{0}^{l}\left(s\right)\right)\left[\int_{0}^{s}e^{\frac{u}{\tau}}S\left(Y_{0}^{m}\left(u\right)\right)du\right]ds\right\} \label{eq:covariance-part-7}\\
\nonumber \\
 & Cov\left(Y_{4}^{i}\left(t\right),Y_{4,4}^{j}\left(t\right)\right)\nonumber \\
 & =e^{-\frac{2t}{\tau}}{\displaystyle \sum\limits _{k,l,m=0}^{N-1}}Cov\left(\frac{T_{ik}}{M_{i}},\frac{T_{jl}T_{lm}}{M_{j}M_{l}}\right)\left[\int_{0}^{t}e^{\frac{s}{\tau}}\widehat{\overline{J}}_{ik}\left(s\right)S\left(Y_{0}^{k}\left(s\right)\right)ds\right]\left[\int_{0}^{t}\widehat{\overline{J}}_{jl}\left(s\right)S'\left(Y_{0}^{l}\left(s\right)\right)\left[\int_{0}^{s}e^{\frac{u}{\tau}}\widehat{\overline{J}}_{lm}\left(s\right)S\left(Y_{0}^{m}\left(u\right)\right)du\right]ds\right]\label{eq:covariance-part-8}
\end{align}
}}
\par\end{center}{\scriptsize \par}
\end{onehalfspace}

\noindent Formula \ref{eq:covariance-part-5} is obtained using the
following identity (which is a consequence of the mutual independence
of the random variables):

\begin{onehalfspace}
\begin{center}
{\small{
\begin{align*}
Cov\left(B_{i}\left(t\right),B_{j}\left(t\right)\frac{\overline{J}_{kl}\left(t\right)}{M_{k}}\right)= & \mathbb{E}\left[B_{i}\left(t\right)B_{j}\left(t\right)\widehat{\overline{J}}_{kl}\left(t\right)\frac{T_{kl}}{M_{k}}\right]-\mathbb{E}\left[B_{i}\left(t\right)\right]\mathbb{E}\left[B_{j}\left(t\right)\widehat{\overline{J}}_{kl}\left(t\right)\frac{T_{kl}}{M_{k}}\right]\\
\\
= & \widehat{\overline{J}}_{kl}\left(t\right)\left(\mathbb{E}\left[B_{i}\left(t\right)B_{j}\left(t\right)\right]\mathbb{E}\left[\frac{T_{kl}}{M_{k}}\right]-\mathbb{E}\left[B_{i}\left(t\right)\right]\mathbb{E}\left[B_{j}\left(t\right)\right]\mathbb{E}\left[\frac{T_{kl}}{M_{k}}\right]\right)
\end{align*}
}}
\par\end{center}{\small \par}
\end{onehalfspace}

\textit{\small{
\begin{align*}
= & \widehat{\overline{J}}_{kl}\left(t\right)\mathbb{E}\left[\frac{T_{kl}}{M_{k}}\right]Cov\left(B_{i}\left(t\right),B_{j}\left(t\right)\right)
\end{align*}
}}{\small \par}

\noindent A similar relation can be found for the initial conditions
$\overrightarrow{V}\left(0\right)$ and the topology $T$:

\begin{onehalfspace}
\begin{center}
{\small{
\[
Cov\left(V_{i}\left(0\right),V_{j}\left(0\right)\frac{\overline{J}_{kl}\left(t\right)}{M_{k}}\right)=\widehat{\overline{J}}_{kl}\left(t\right)\mathbb{E}\left[\frac{T_{kl}}{M_{k}}\right]Cov\left(V_{i}\left(0\right),V_{j}\left(0\right)\right)
\]
}}
\par\end{center}{\small \par}
\end{onehalfspace}

\noindent from which we have obtained formula \ref{eq:covariance-part-6}.
Instead, in order to obtain formula \ref{eq:covariance-part-7}, we
have used the following result:

\begin{onehalfspace}
\begin{center}
{\small{
\begin{align*}
Cov\left(\frac{W_{ij}}{M_{i}},\frac{W_{kl}}{M_{k}}\frac{\overline{J}_{mn}\left(t\right)}{M_{m}}\right)= & Cov\left(\widehat{W}_{ij}\frac{T_{ij}}{M_{i}},\widehat{W}_{kl}\frac{T_{kl}}{M_{k}}\widehat{\overline{J}}_{mn}\left(t\right)\frac{T_{mn}}{M_{m}}\right)\\
\\
= & \widehat{\overline{J}}_{mn}\left(t\right)\left(\mathbb{E}\left[\widehat{W}_{ij}\frac{T_{ij}}{M_{i}}\widehat{W}_{kl}\frac{T_{kl}}{M_{k}}\frac{T_{mn}}{M_{m}}\right]-\mathbb{E}\left[\widehat{W}_{ij}\frac{T_{ij}}{M_{i}}\right]\mathbb{E}\left[\widehat{W}_{kl}\frac{T_{kl}}{M_{k}}\frac{T_{mn}}{M_{m}}\right]\right)\\
\\
= & \widehat{\overline{J}}_{mn}\left(t\right)\left(\mathbb{E}\left[\widehat{W}_{ij}\widehat{W}_{kl}\right]\mathbb{E}\left[\frac{T_{ij}T_{kl}T_{mn}}{M_{i}M_{k}M_{m}}\right]-\mathbb{E}\left[\widehat{W}_{ij}\right]\mathbb{E}\left[\frac{T_{ij}}{M_{i}}\right]\mathbb{E}\left[\widehat{W}_{kl}\right]\mathbb{E}\left[\frac{T_{kl}T_{mn}}{M_{k}M_{m}}\right]\right)\\
\\
= & \widehat{\overline{J}}_{mn}\left(t\right)Cov\left(\widehat{W}_{ij},\widehat{W}_{kl}\right)\mathbb{E}\left[\frac{T_{ij}T_{kl}T_{mn}}{M_{i}M_{k}M_{m}}\right]
\end{align*}
}}
\par\end{center}{\small \par}
\end{onehalfspace}

\noindent which is a consequence of the independence between $\widehat{W}$
and $T$. In the same way it is possible to prove that:

\begin{onehalfspace}
\begin{center}
{\small{
\[
Cov\left(\frac{W_{ij}}{M_{i}},\frac{\overline{J}_{kl}\left(t\right)}{M_{k}}\frac{\overline{J}_{mn}\left(t\right)}{M_{m}}\right)=\widehat{\overline{J}}_{kl}\left(t\right)\widehat{\overline{J}}_{mn}\left(t\right)\mathbb{E}\left[\widehat{W}_{ij}\right]\left(\mathbb{E}\left[\frac{T_{ij}T_{kl}T_{mn}}{M_{i}M_{k}M_{m}}\right]-\mathbb{E}\left[\frac{T_{ij}}{M_{i}}\right]\mathbb{E}\left[\frac{T_{kl}T_{mn}}{M_{k}M_{m}}\right]\right)=0
\]
}}
\par\end{center}{\small \par}
\end{onehalfspace}

\noindent so for this reason the term $Cov\left(Y_{3}^{i}\left(t\right),Y_{4,4}^{i}\left(t\right)\right)$
does not appear in formula \ref{eq:covariance}.

\noindent Once the covariance matrix of the membrane potentials has
been determined, we can evaluate their correlation structure using
the Pearson's correlation coefficient, defined as follows:

\begin{onehalfspace}
\begin{center}
{\small{
\begin{equation}
Corr\left(V_{i}\left(t\right),V_{j}\left(t\right)\right)=\frac{Cov\left(V_{i}\left(t\right),V_{j}\left(t\right)\right)}{\sqrt{Var\left(V_{i}\left(t\right)\right)Var\left(V_{j}\left(t\right)\right)}}\label{eq:correlation}
\end{equation}
}}
\par\end{center}{\small \par}
\end{onehalfspace}

\noindent where:

\begin{onehalfspace}
\begin{center}
\begin{equation}
Var\left(V_{i}\left(t\right)\right)=Cov\left(V_{i}\left(t\right),V_{i}\left(t\right)\right)\label{eq:variance}
\end{equation}

\par\end{center}
\end{onehalfspace}

\noindent is the variance of the stochastic process $V_{i}\left(t\right)$.

\noindent The only quantities that remain unspecified are $\mathbb{E}\left[\frac{T_{ij}}{M_{i}}\right]$,
$\mathbb{E}\left[\frac{T_{ik}T_{jl}}{M_{i}M_{j}}\right]$ and $\mathbb{E}\left[\frac{T_{ik}T_{lm}T_{jl}}{M_{i}M_{j}M_{l}}\right]$,
that depend on the distribution of the matrix $T$. This can be accomplished
by a multidimensional Taylor expansion. For example, for $\mathbb{E}\left[\frac{T_{ij}}{M_{i}}\right]$
we Taylor-expand the function:

\begin{onehalfspace}
\begin{center}
\begin{equation}
\begin{array}{cccc}
f: & \left(T_{i0},...,T_{i,N-1}\right) & \rightarrow & \frac{T_{ij}}{M_{i}}\end{array}\label{eq:function-f-random-topology}
\end{equation}

\par\end{center}
\end{onehalfspace}

\noindent at the point $\left(\mathbb{E}\left[T_{i0}\right],...,\mathbb{E}\left[T_{i,N-1}\right]\right)$
to obtain:

\begin{onehalfspace}
\begin{center}
{\scriptsize{
\begin{align}
\mathbb{E}\left[\frac{T_{ij}}{M_{i}}\right]= & \mathbb{E}\left[\frac{T_{ij}}{{\displaystyle \sum_{k=0}^{N-1}}T_{ik}}\right]\nonumber \\
\nonumber \\
= & {\displaystyle {\displaystyle \sum_{n_{0}=0}^{\infty}}}{\displaystyle {\displaystyle \sum_{n_{1}=0}^{\infty}}}\cdots{\displaystyle {\displaystyle \sum_{n_{N-1}=0}^{\infty}}}\frac{\mathbb{E}\left[\left(T_{i0}-\mathbb{E}\left[T_{i0}\right]\right)^{n_{0}}\cdots\left(T_{i,N-1}-\mathbb{E}\left[T_{i,N-1}\right]\right)^{n_{N-1}}\right]}{n_{0}!\cdots n_{N-1}!}\left(\frac{\partial^{n_{0}+...+n_{N-1}}f}{\partial T_{i0}^{n_{0}}\cdots\partial T_{i,N-1}^{n_{N-1}}}\right)\left(\mathbb{E}\left[T_{i0}\right],\cdots,\mathbb{E}\left[T_{i,N-1}\right]\right)\label{eq:Taylor-expansion-of-the-topology-1}
\end{align}
}}
\par\end{center}{\scriptsize \par}
\end{onehalfspace}

\noindent In detail, we have up to the third order:

\begin{onehalfspace}
\begin{center}
{\small{
\begin{equation}
\mathbb{E}\left[\frac{T_{ij}}{M_{i}}\right]\approx\frac{\mathbb{E}\left[T_{ij}\right]}{{\displaystyle \sum_{k=0}^{N-1}}\mathbb{E}\left[T_{ik}\right]}+\frac{1}{2}\sum_{k,l=0}^{N-1}Cov\left(T_{ik},T_{il}\right)\left(\frac{\partial^{2}f}{\partial T_{ik}\partial T_{il}}\right)\left(\mathbb{E}\left[T_{i0}\right],\cdots,\mathbb{E}\left[T_{i,N-1}\right]\right)\label{eq:Taylor-expansion-of-the-topology-2}
\end{equation}
}}
\par\end{center}{\small \par}
\end{onehalfspace}

\noindent where:

\begin{onehalfspace}
\begin{center}
{\small{
\begin{align*}
 & \left(\frac{\partial^{2}f}{\partial T_{ik}\partial T_{il}}\right)\left(\mathbb{E}\left[T_{i0}\right],\cdots,\mathbb{E}\left[T_{i,N-1}\right]\right)\\
\\
 & =\begin{cases}
\frac{2\mathbb{E}\left[T_{ij}\right]}{\left({\displaystyle \sum_{m=0}^{N-1}}\mathbb{E}\left[T_{im}\right]\right)^{3}} & \begin{array}{ccc}
\mathrm{if} &  & k,l\neq j\end{array}\\
\\
\frac{2\mathbb{E}\left[T_{ij}\right]-{\displaystyle \sum_{m=0}^{N-1}}\mathbb{E}\left[T_{im}\right]}{\left({\displaystyle \sum_{m=0}^{N-1}}\mathbb{E}\left[T_{im}\right]\right)^{3}} & \begin{array}{ccc}
\mathrm{if} &  & \left(\left(k\neq j\right)\wedge\left(l=j\right)\right)\vee\left(\left(k=j\right)\wedge\left(l\neq j\right)\right)\end{array}\\
\\
-\frac{2{\displaystyle \sum\limits _{\substack{m=0\\
m\neq j
}
}^{N-1}}\mathbb{E}\left[T_{im}\right]}{\left({\displaystyle \sum_{m=0}^{N-1}}\mathbb{E}\left[T_{im}\right]\right)^{3}} & \begin{array}{ccc}
\mathrm{if} &  & k,l=j\end{array}
\end{cases}
\end{align*}
}}
\par\end{center}{\small \par}
\end{onehalfspace}

\noindent The function \ref{eq:function-f-random-topology} is analytic
everywhere, but when $M_{i}=0$. However, we remind that for $M_{i}=0$
formula \ref{eq:synaptic-weights-3} simply gives $J_{ij}\left(t\right)=0$.
For this reason the multidimensional Taylor series of $f\left(T_{i0},...,T_{i,N-1}\right)$
has a finite radius of convergence and it does converge to {\footnotesize{$\frac{T_{ij}}{{\displaystyle \sum_{k=0}^{N-1}}T_{ik}}$
}}everywhere.

\noindent After this analysis, the conclusion is that we can calculate
$\mathbb{E}\left[\frac{T_{ij}}{M_{i}}\right]$ once we know the quantities
$\mathbb{E}\left[T_{ik}\right]$, $\mathbb{E}\left[T_{ik}T_{il}\right]$
etc. The same reasoning can be applied to $\mathbb{E}\left[\frac{T_{ik}T_{jl}}{M_{i}M_{j}}\right]$
and $\mathbb{E}\left[\frac{T_{ik}T_{lm}T_{jl}}{M_{i}M_{j}M_{l}}\right]$.
In Section \ref{sec:Fractal connectivity matrix} we show how to determine
these quantities for the fractal connectivity matrix introduced by
Sporns in \cite{citeulike:1343837}. These results can also be used
for networks with deterministic topologies, but we have to set $\mathbb{E}\left[\prod\frac{T}{M}\right]=\prod\frac{T}{M}$
in formulae \ref{eq:covariance-part-3}, \ref{eq:variance-part-3},
\ref{eq:covariance-part-5}, \ref{eq:covariance-part-6}, \ref{eq:covariance-part-7},
and we have to set to zero the covariance functions of $\frac{T}{M}$
in formulae \ref{eq:covariance-part-4} and \ref{eq:covariance-part-8}
(so that $Cov\left(Y_{4}^{i}\left(t\right),Y_{4}^{j}\left(t\right)\right)=Cov\left(Y_{4}^{i}\left(t\right),Y_{4,4}^{j}\left(t\right)\right)=0$).

\section{\label{sec:A problem with the initial conditions}A problem with
the initial conditions}

\begin{figure}
\noindent \begin{centering}
\includegraphics[scale=0.4]{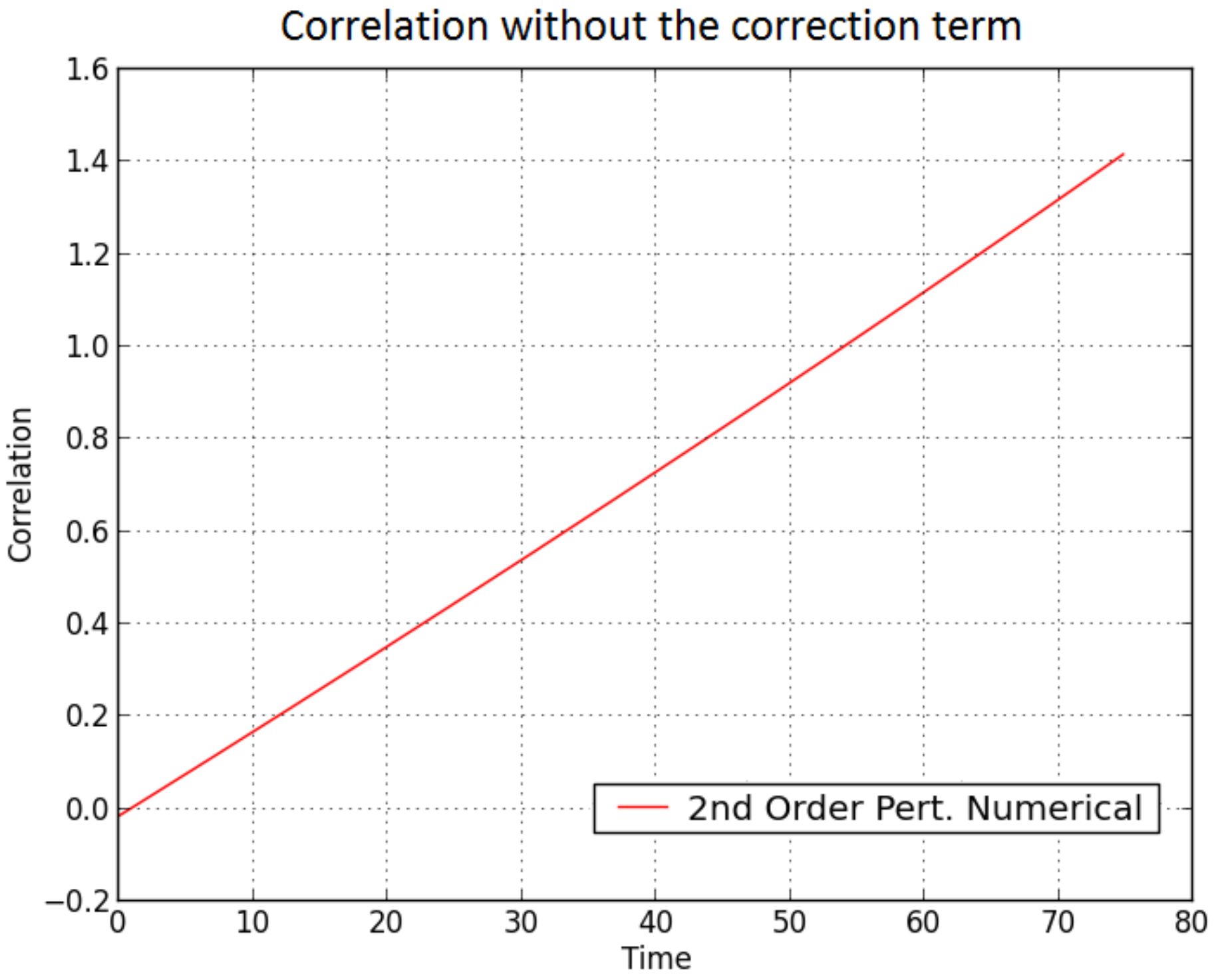}\includegraphics[scale=0.4]{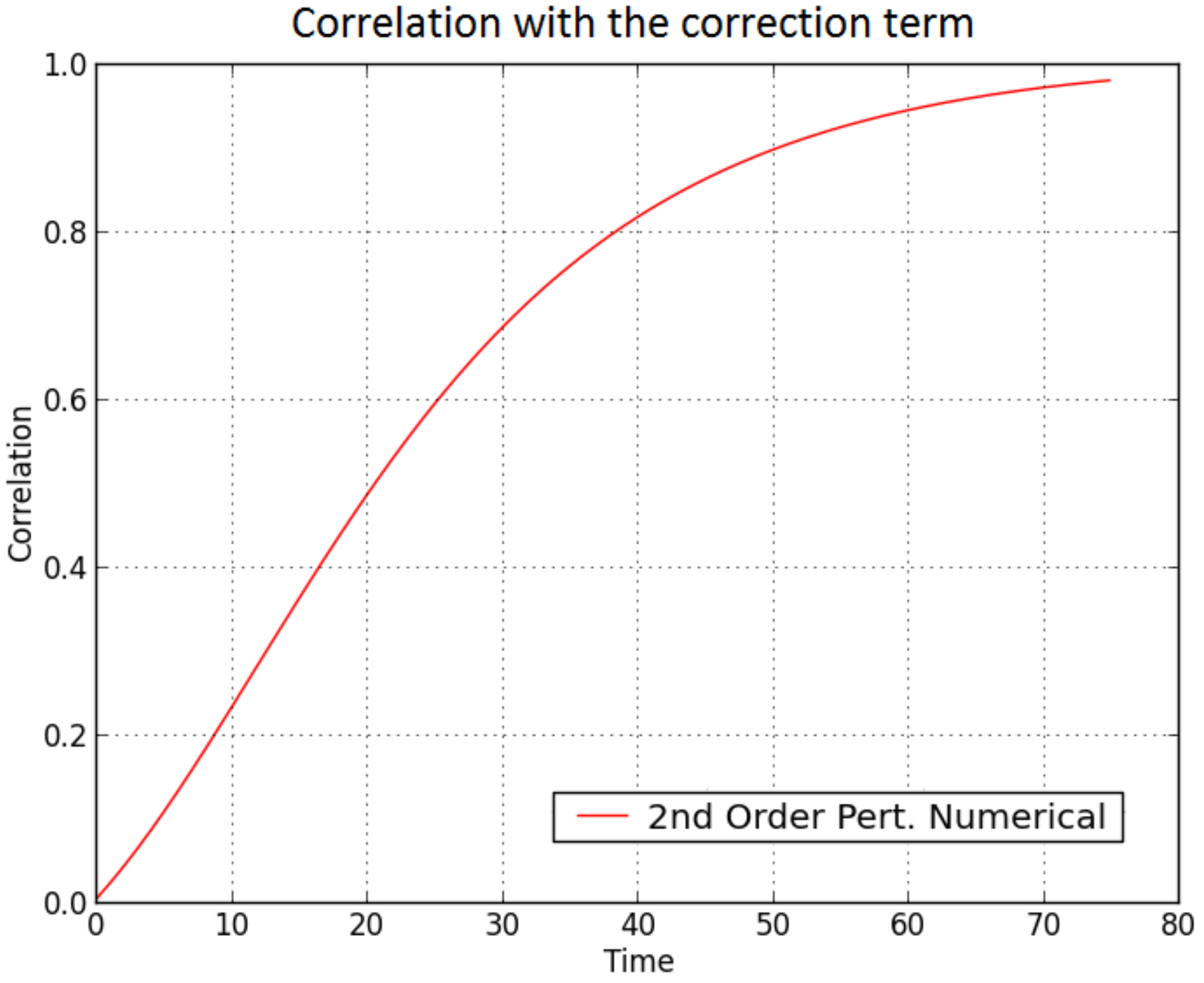}
\par\end{centering}

\caption[{\footnotesize{A problem with the initial conditions}}]{{\small{\label{fig:Correlation-problem}Correlation obtained from
formula }}\ref{eq:covariance} {\small{using the numerical simulation
of formulae \ref{eq:perturbative-equation-0} - \ref{eq:perturbative-equation-8}
(left-hand side), and the same function obtained from formula \ref{eq:covariance-fixed}
(right-hand side). The values of the parameters are shown in Table
\ref{tab:Problem-with-the-initial-conditions}, while the topology
of the network is $K_{10}$ (see Figure \ref{fig:Deterministic-topologies}).
In the first figure the correlation does not stay in the range $\left[-1,1\right]$
for all time, and the problem is corrected in the second figure, see
text.}}}
\end{figure}

\noindent Before we start to analyze a concrete example of connectivity
matrix, we have to show a problem with the initial conditions. In
fact, if we choose $\sigma_{2},\sigma_{4}\neq0$, $\sigma_{1},\sigma_{3}=0$
and $C_{2}=0$, at least in the case of a deterministic topology the
correlation function that we have calculated perturbatively is not
necessarily in the range $\left[-1,1\right]$ as required. This can
be seen from formulae \ref{eq:covariance} - \ref{eq:covariance-part-8},
which for these values of the parameters and a deterministic $T$,
give:

\begin{onehalfspace}
\begin{center}
{\small{
\begin{align}
Corr\left(V_{i}\left(t\right),V_{j}\left(t\right)\right)= & \frac{\sigma_{2}^{2}Cov\left(Y_{2}^{i}\left(t\right),Y_{2}^{j}\left(t\right)\right)+\sigma_{4}\sigma_{2}^{2}\left[Cov\left(Y_{2}^{i}\left(t\right),Y_{2,4}^{j}\left(t\right)\right)+Cov\left(Y_{2,4}^{i}\left(t\right),Y_{2}^{j}\left(t\right)\right)\right]}{\sigma_{2}^{2}Var\left(Y_{2}^{i}\left(t\right)\right)+\sigma_{4}\sigma_{2}^{2}\left[Cov\left(Y_{2}^{i}\left(t\right),Y_{2,4}^{i}\left(t\right)\right)+Cov\left(Y_{2,4}^{i}\left(t\right),Y_{2}^{i}\left(t\right)\right)\right]}\nonumber \\
\nonumber \\
= & \sigma_{4}\left\{ \frac{T_{ji}}{M_{j}}\left[\int_{0}^{t}\widehat{\overline{J}}_{ji}\left(s\right)S\left(Y_{0}^{i}\left(s\right)\right)ds\right]+\frac{T_{ij}}{M_{i}}\left[\int_{0}^{t}\widehat{\overline{J}}_{ij}\left(s\right)S\left(Y_{0}^{j}\left(s\right)\right)ds\right]\right\} \label{eq:Correlation-without-correction}
\end{align}
}}
\par\end{center}{\small \par}
\end{onehalfspace}

\noindent where for simplicity we have also supposed that all the
neurons behave in the same way, so that $Var\left(V_{i}\left(t\right)\right)=Var\left(V_{j}\left(t\right)\right)$.
Therefore, if $\widehat{\overline{J}}_{ij}\left(t\right)$, $\widehat{\overline{J}}_{ji}\left(t\right)$,
$S\left(Y_{0}^{i}\left(t\right)\right)$ and $S\left(Y_{0}^{j}\left(t\right)\right)$
are for example constant in time, from formula \ref{eq:Correlation-without-correction}
we obtain that $Corr\left(V_{i}\left(t\right),V_{j}\left(t\right)\right)$
increases linearly with time, therefore at some point it will be outside
the range $\left[-1,1\right]$. This can be seen also from Figure
\ref{fig:Correlation-problem} (left-hand side), which has been obtained
from the numerical simulation of the equations \ref{eq:perturbative-equation-0}
- \ref{eq:perturbative-equation-8} (the details of the numerical
scheme will be provided in Section \ref{sec:Numerical experiments})
for the values of the parameters reported in Table \ref{tab:Problem-with-the-initial-conditions}.

\begin{table}
\begin{centering}
\begin{tabular}{|c|c|c|c|}
\hline 
Neuron & Input & Synaptic Weights & Sigmoid Function\tabularnewline
\hline 
\hline 
$\tau=1$ & $I_{i}=0$ & $\widehat{\overline{J}}_{ij}=3$ & {\small{$T_{MAX}=1$}}\tabularnewline
\hline 
$\sigma_{2}=0.1$ & $\sigma_{1}=0$ & $\sigma_{3}=0$ & {\small{$\lambda=1$}}\tabularnewline
\hline 
 &  & $\sigma_{4}=0.1$ & {\small{$V_{T}=0$}}\tabularnewline
\hline 
$C_{2}=0$ & $C_{1}=0$ & $C_{3}=0$ & \tabularnewline
\hline 
$\mu=0$ &  &  & \tabularnewline
\hline 
\end{tabular}
\par\end{centering}

\caption[Parameters for the simulations of the problem with the initial conditions]{{\small{\label{tab:Problem-with-the-initial-conditions}Values of
the parameters used to generate Figure \ref{fig:Correlation-problem}.}}}
\end{table}

\noindent This problem does not happen when $\sigma_{1},\sigma_{4}\neq0$
and $\sigma_{2},\sigma_{3}=0$, or when $\sigma_{3},\sigma_{4}\neq0$
and $\sigma_{1},\sigma_{2}=0$, or when $\sigma_{4}\neq0$ and $\sigma_{1},\sigma_{2},\sigma_{3}=0$,
therefore it is only related to the initial conditions. It is of course
due to our approximation. In fact, if we want to calculate the variance
and covariance between two perturbative expansions of the form $F_{i}\left(t\right)=F_{0}^{i}\left(t\right)+\epsilon F_{1}^{i}\left(t\right)+\epsilon^{2}F_{2}^{i}\left(t\right)$,
where $F_{0}^{i}\left(t\right)$ is deterministic, we obtain:

\begin{onehalfspace}
\begin{center}
{\small{
\begin{align*}
Var\left(F_{i}\left(t\right)\right)= & \epsilon^{2}Var\left(F_{1}^{i}\left(t\right)\right)+2\epsilon^{3}Cov\left(F_{1}^{i}\left(t\right),F_{2}^{i}\left(t\right)\right)+\epsilon^{4}Var\left(F_{2}^{i}\left(t\right)\right)\\
\\
Cov\left(F_{i}\left(t\right),F_{j}\left(t\right)\right)= & \epsilon^{2}Cov\left(F_{1}^{i}\left(t\right),F_{1}^{j}\left(t\right)\right)+\epsilon^{3}Cov\left(F_{1}^{i}\left(t\right),F_{2}^{j}\left(t\right)\right)+\epsilon^{3}Cov\left(F_{2}^{i}\left(t\right),F_{1}^{j}\left(t\right)\right)\\
\\
 & +\epsilon^{4}Cov\left(F_{2}^{i}\left(t\right),F_{2}^{j}\left(t\right)\right)
\end{align*}
}}
\par\end{center}{\small \par}
\end{onehalfspace}

\noindent Due to the Cauchy-Schwarz inequality, we always have:

\begin{onehalfspace}
\begin{center}
{\small{
\[
\left[Cov\left(F_{i}\left(t\right),F_{j}\left(t\right)\right)\right]^{2}\leq Var\left(F_{i}\left(t\right)\right)Var\left(F_{j}\left(t\right)\right)
\]
}}
\par\end{center}{\small \par}
\end{onehalfspace}

\noindent namely $\left|Corr\left(F_{i}\left(t\right),F_{j}\left(t\right)\right)\right|\leq1$.
However, if we neglect the terms proportional to $\epsilon^{4}$,
as we did in Section \ref{sec:Correlation structure of the network},
this inequality is not guaranteed to hold anymore. Therefore even
if the approximations of the variance and covariance are good, the
correlation could be completely wrong. This is the origin of the problem
we have mentioned before. Moreover, it happens only when we deal with
the initial conditions and not with the other random variables, because
only for $\sigma_{2},\sigma_{4}\neq0$ and $\sigma_{1},\sigma_{3}=0$
do we have $4$th order terms and the variance and covariance converge
to zero for $t\rightarrow+\infty$, giving rise to an undefined correlation
of the form $\frac{0}{0}$.

\noindent The solution is to keep the $4$th order terms generated
by the initial conditions in the formula of the variance and covariance.
Now, for $\sigma_{2},\sigma_{4}\neq0$ and $\sigma_{1},\sigma_{3}=0$
we have:

\begin{onehalfspace}
\begin{center}
{\small{
\[
V_{i}\left(t\right)=Y_{0}^{i}\left(t\right)+\sigma_{2}Y_{2}^{i}\left(t\right)+\sigma_{4}Y_{4}^{i}\left(t\right)+\sigma_{2}\sigma_{4}Y_{2,4}^{i}\left(t\right)+\sigma_{4}^{2}Y_{4,4}^{i}\left(t\right)
\]
}}
\par\end{center}{\small \par}
\end{onehalfspace}

\noindent since it can be easily proved that $Y_{2,2}^{i}\left(t\right)=0$
$\forall t$. Therefore in this case the exact covariance function
is:

\begin{onehalfspace}
\begin{center}
{\small{
\begin{align}
 & Cov\left(V_{i}\left(t\right),V_{j}\left(t\right)\right)\nonumber \\
\nonumber \\
 & =\sigma_{2}^{2}Cov\left(Y_{2}^{i}\left(t\right),Y_{2}^{j}\left(t\right)\right)+\sigma_{4}^{2}Cov\left(Y_{4}^{i}\left(t\right),Y_{4}^{j}\left(t\right)\right)\nonumber \\
\nonumber \\
 & +\sigma_{4}\sigma_{2}^{2}\left[Cov\left(Y_{2}^{i}\left(t\right),Y_{2,4}^{j}\left(t\right)\right)+Cov\left(Y_{2,4}^{i}\left(t\right),Y_{2}^{j}\left(t\right)\right)\right]+\sigma_{4}^{3}\left[Cov\left(Y_{4}^{i}\left(t\right),Y_{4,4}^{j}\left(t\right)\right)+Cov\left(Y_{4,4}^{i}\left(t\right),Y_{4}^{j}\left(t\right)\right)\right]\nonumber \\
\nonumber \\
 & +\sigma_{2}^{2}\sigma_{4}^{2}Cov\left(Y_{2,4}^{i}\left(t\right),Y_{2,4}^{j}\left(t\right)\right)+\sigma_{4}^{4}Cov\left(Y_{4,4}^{i}\left(t\right),Y_{4,4}^{j}\left(t\right)\right)\label{eq:fixing-the-covariance}
\end{align}
}}
\par\end{center}{\small \par}
\end{onehalfspace}

\noindent The $4$th order term $\sigma_{2}\sigma_{4}^{3}Cov\left(Y_{2,4}^{i}\left(t\right),Y_{4,4}^{j}\left(t\right)\right)$
has not been taken into account because it is proportional to $Cov\left(Y_{2}^{k}\left(0\right)\frac{T_{ik}}{M_{i}},\frac{T_{jk}}{M_{j}}\frac{T_{kl}}{M_{k}}\right)$,
which is equal to zero, as proved below:

\begin{onehalfspace}
\begin{center}
{\small{
\begin{align*}
Cov\left(Y_{2}^{k}\left(0\right)\frac{T_{ik}}{M_{i}},\frac{T_{jk}}{M_{j}}\frac{T_{kl}}{M_{k}}\right)= & \mathbb{E}\left[Y_{2}^{k}\left(0\right)\frac{T_{ik}}{M_{i}}\frac{T_{jk}}{M_{j}}\frac{T_{kl}}{M_{k}}\right]-\mathbb{E}\left[Y_{2}^{k}\left(0\right)\frac{T_{ik}}{M_{i}}\right]\mathbb{E}\left[\frac{T_{jk}}{M_{j}}\frac{T_{kl}}{M_{k}}\right]\\
\\
= & \mathbb{E}\left[Y_{2}^{k}\left(0\right)\right]\left(\mathbb{E}\left[\frac{T_{ik}}{M_{i}}\frac{T_{jk}}{M_{j}}\frac{T_{kl}}{M_{k}}\right]-\mathbb{E}\left[\frac{T_{ik}}{M_{i}}\right]\mathbb{E}\left[\frac{T_{jk}}{M_{j}}\frac{T_{kl}}{M_{k}}\right]\right)\\
\\
= & 0
\end{align*}
}}
\par\end{center}{\small \par}
\end{onehalfspace}

\noindent We can simplify \ref{eq:fixing-the-covariance} further
by noticing that for $\sigma_{4}\neq0$ and $\sigma_{1},\sigma_{2},\sigma_{3}=0$
the problem of the correlation does not appear anymore if we calculate
it using the truncated covariance function \ref{eq:covariance}. Since
for these values of the perturbative parameters the covariance \ref{eq:fixing-the-covariance}
becomes simply:

\begin{onehalfspace}
\begin{center}
{\small{
\begin{align*}
Cov\left(V_{i}\left(t\right),V_{j}\left(t\right)\right)= & \sigma_{4}^{2}Cov\left(Y_{4}^{i}\left(t\right),Y_{4}^{j}\left(t\right)\right)\\
\\
 & +\sigma_{4}^{3}\left[Cov\left(Y_{4}^{i}\left(t\right),Y_{4,4}^{j}\left(t\right)\right)+Cov\left(Y_{4,4}^{i}\left(t\right),Y_{4}^{j}\left(t\right)\right)\right]+\sigma_{4}^{4}Cov\left(Y_{4,4}^{i}\left(t\right),Y_{4,4}^{j}\left(t\right)\right)
\end{align*}
}}
\par\end{center}{\small \par}
\end{onehalfspace}

\noindent which differs from formula \ref{eq:covariance} (calculated
for $\sigma_{4}\neq0$ and $\sigma_{1},\sigma_{2},\sigma_{3}=0$)
only in the $4$th order term $\sigma_{4}^{4}Cov\left(Y_{4,4}^{i}\left(t\right),Y_{4,4}^{j}\left(t\right)\right)$,
this means that there is no need to add this term in order to correct
the perturbative expansion. Therefore we see from \ref{eq:fixing-the-covariance}
that the only term which is required to alleviate the problem of the
correlation is $\sigma_{2}^{2}\sigma_{4}^{2}Cov\left(Y_{2,4}^{i}\left(t\right),Y_{2,4}^{j}\left(t\right)\right)$.
To conclude, the final formula for the covariance that we have to
use is:

\begin{onehalfspace}
\begin{center}
{\small{
\begin{align}
 & Cov\left(V_{i}\left(t\right),V_{j}\left(t\right)\right)\nonumber \\
\nonumber \\
 & =\sigma_{1}^{2}Cov\left(Y_{1}^{i}\left(t\right),Y_{1}^{j}\left(t\right)\right)+\sigma_{2}^{2}Cov\left(Y_{2}^{i}\left(t\right),Y_{2}^{j}\left(t\right)\right)\nonumber \\
\nonumber \\
 & +\sigma_{3}^{2}Cov\left(Y_{3}^{i}\left(t\right),Y_{3}^{j}\left(t\right)\right)+\sigma_{4}^{2}Cov\left(Y_{4}^{i}\left(t\right),Y_{4}^{j}\left(t\right)\right)\nonumber \\
\nonumber \\
 & +\sigma_{4}\left\{ \sigma_{1}^{2}\left[Cov\left(Y_{1}^{i}\left(t\right),Y_{1,4}^{j}\left(t\right)\right)+Cov\left(Y_{1,4}^{i}\left(t\right),Y_{1}^{j}\left(t\right)\right)\right]+\sigma_{2}^{2}\left[Cov\left(Y_{2}^{i}\left(t\right),Y_{2,4}^{j}\left(t\right)\right)+Cov\left(Y_{2,4}^{i}\left(t\right),Y_{2}^{j}\left(t\right)\right)\right]\right.\nonumber \\
\nonumber \\
 & \left.+\sigma_{3}^{2}\left[Cov\left(Y_{3}^{i}\left(t\right),Y_{3,4}^{j}\left(t\right)\right)+Cov\left(Y_{3,4}^{i}\left(t\right),Y_{3}^{j}\left(t\right)\right)\right]+\sigma_{4}^{2}\left[Cov\left(Y_{4}^{i}\left(t\right),Y_{4,4}^{j}\left(t\right)\right)+Cov\left(Y_{4,4}^{i}\left(t\right),Y_{4}^{j}\left(t\right)\right)\right]\right\} \nonumber \\
\nonumber \\
 & +\sigma_{2}^{2}\sigma_{4}^{2}Cov\left(Y_{2,4}^{i}\left(t\right),Y_{2,4}^{j}\left(t\right)\right)\label{eq:covariance-fixed}
\end{align}
}}
\par\end{center}{\small \par}
\end{onehalfspace}

\noindent where:

\begin{onehalfspace}
\begin{center}
{\small{
\begin{align}
 & Cov\left(Y_{2,4}^{i}\left(t\right),Y_{2,4}^{j}\left(t\right)\right)\nonumber \\
\nonumber \\
 & =e^{-\frac{2t}{\tau}}\left\{ \vphantom{{\displaystyle \sum\limits _{\substack{k,l=0\\
k\neq l
}
}^{N-1}}}\sum_{k=0}^{N-1}\mathbb{E}\left[\frac{T_{ik}T_{jk}}{M_{i}M_{j}}\right]\left[\int_{0}^{t}\widehat{\overline{J}}_{ik}\left(s\right)S'\left(Y_{0}^{k}\left(s\right)\right)ds\right]\left[\int_{0}^{t}\widehat{\overline{J}}_{jk}\left(s\right)S'\left(Y_{0}^{k}\left(s\right)\right)ds\right]\right.\nonumber \\
\nonumber \\
 & \left.+C_{2}{\displaystyle \sum\limits _{\substack{k,l=0\\
k\neq l
}
}^{N-1}}\mathbb{E}\left[\frac{T_{ik}T_{jl}}{M_{i}M_{j}}\right]\left[\int_{0}^{t}\widehat{\overline{J}}_{ik}\left(s\right)S'\left(Y_{0}^{k}\left(s\right)\right)ds\right]\left[\int_{0}^{t}\widehat{\overline{J}}_{jl}\left(s\right)S'\left(Y_{0}^{l}\left(s\right)\right)ds\right]\right\} \label{eq:covariance-correction-term}
\end{align}
}}
\par\end{center}{\small \par}
\end{onehalfspace}

\noindent We remind the reader that if he/she is interested only in
the calculation of the variance and covariance, the term $Cov\left(Y_{2,4}^{i}\left(t\right),Y_{2,4}^{j}\left(t\right)\right)$
is not important, but it must be used if he/she needs to evaluate
the correlation function. Indeed, using formula \ref{eq:covariance-fixed},
the problem of the correlation is corrected, as it can be seen from
Figure \ref{fig:Correlation-problem} (right-hand side).

\section{\label{sec:Fractal connectivity matrix}Fractal connectivity matrix}

\begin{figure}
\noindent \begin{centering}
\includegraphics[scale=1.4]{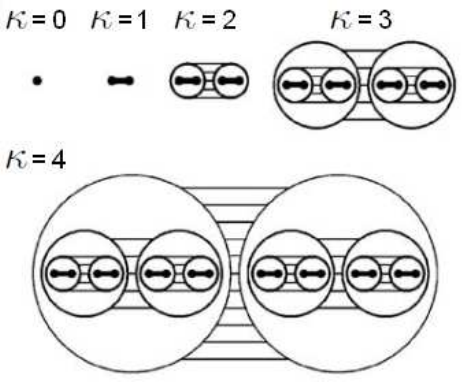}
\par\end{centering}

\caption[Sporns' algorithm for the fractal connectivity matrix.]{{\small{\label{fig:Sporns-algorithm}Sporns' algorithm for the fractal
connectivity matrix. At the level $\kappa=0$ a single dot represents
a group of $2^{\mu}$ fully connected neurons. At $\kappa=1$ we duplicate
this elementary block, obtaining two groups of $2^{\mu}$ neurons
which are linked together with a connection density $E^{-1}$. This
structure is generated again at the level $\kappa=2$, and connected
to the previous one with a connection density $E^{-2}$, and so on.
This figure has been taken and adapted from \protect{\cite{citeulike:1343837}.}}}}
\end{figure}

\begin{figure}
\begin{centering}
\includegraphics[scale=0.385]{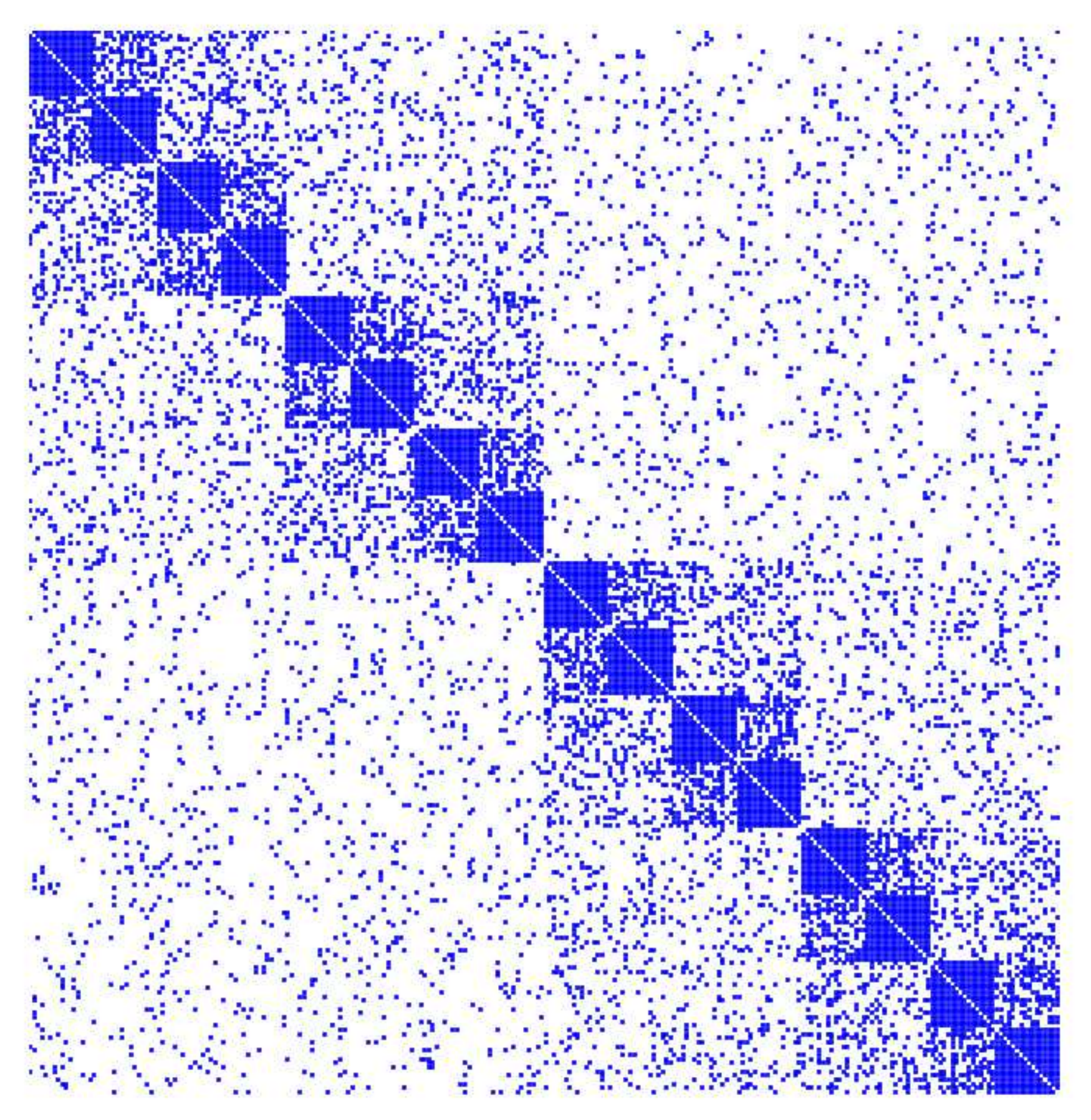}\includegraphics[scale=0.4]{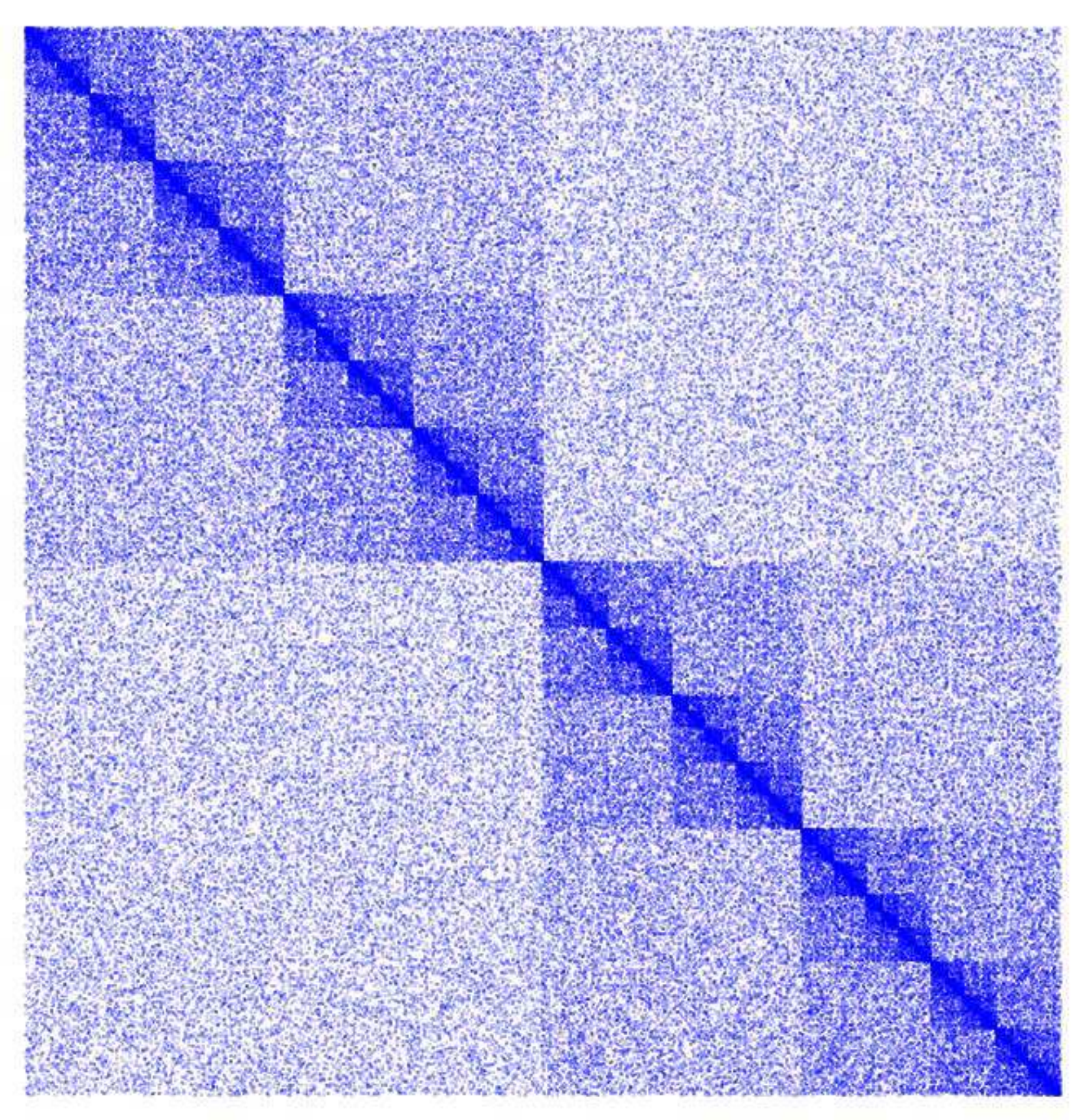}
\par\end{centering}

\caption[{\footnotesize{Fractal connectivity matrix}}]{{\small{\label{fig:Example-of-fractal-matrix}Two examples of fractal
matrix obtained with the Sporns' algorithm, for $\eta=8$, $\mu=4$
and $E=2.0$ (left-hand side) and for $\eta=11$, $\mu=2$ and $E=1.5$
(right-hand side). A blue dot corresponds to a $1$ in the topology
matrix, while the absence of the dot corresponds to a $0$. The figure
on the right-hand side has been resized in order to the have the same
spatial extension as the figure on the left-hand side. For this reason
it does not clearly show the diagonal white line corresponding to
$J_{ii}\left(t\right)=0$, namely to the absence of self-connections.}}}
\end{figure}

\noindent As we said in Section \ref{sec:Introduction}, the brain
is characterized by a small-world topology. A famous algorithm that
generates networks with this property has been introduced by Watts
and Strogatz \cite{watts1998cds}. Even if in principle it is possible
to calculate analytically the covariance structure of the neurons
over the random topology generated by this algorithm, in practice
it is not a simple task, because the exact evaluation of $\mathbb{E}\left[T_{ij}\right]$,
$\mathbb{E}\left[T_{ik}T_{jl}\right]$, $\mathbb{E}\left[T_{il}T_{jm}T_{kn}\right]$
etc, which is required for example by formula \ref{eq:Taylor-expansion-of-the-topology-1},
can be accomplished through a complicated combinatorial analysis.
Moreover this algorithm does not mimic the nested structure of the
connectivity matrix of the brain. In fact, Watts and Strogatz tried
to replicate only two features of the brain, namely its path length
(which represents the shortest distance between two vertices in terms
of the number of edges) and its clustering coefficient (which, for
a given vertex, quantifies the connectivity degree of its neighbourhood,
i.e. of the vertices directly connected to it), without taking into
account its nested structure. A more tractable algorithm, which reproduces
more biologically realistic connections, has been introduced by Sporns
in \cite{citeulike:1343837}. Since the connectome of the brain has
a nested structure, Sporns suggested to describe it using a fractal
connectivity matrix. One of the cases he studied is what he called
the \textit{fractal pattern }{\small{(frc).}} It is obtained by choosing
two integer numbers, $\mu$ and $\eta$ (Sporns called them $m$ and
$n$, but we prefer to use different symbols to avoid confusion with
the vector and matrix indices) with $\mu\leq\eta$, and a real non-negative
number $E$. The total number of neurons in the network is $N=2^{\eta}$,
and the different levels of the fractal structure are described by
a parameter $\kappa=0,1,...,\eta-\mu$ (Sporns called it $k$). As
shown in Figure \ref{fig:Sporns-algorithm}, we start with an elementary
block of $2^{\mu}$ neurons, which forms the level $0$ of the fractal
structure ($\kappa=0$). Within this block the neurons are fully connected
and without self-connections. Then we duplicate this block. The connection
density between the two elementary blocks is the number of actual
connections between them divided by the total number of possible connections.
So we connect them with a connection density $E^{-1}$ (here $\kappa=1$,
namely we are at the level $1$). This means that the number of connections
between the two blocks in one direction is the integer part of $4^{\mu}E^{-1}$.
We emphasize the fact that these connections are randomly chosen.
The resulting network is then ``duplicated'', namely we produce
another pair of groups with $2^{\mu}$ fully interconnected neurons
in each one, and interconnected between them with a connection density
$E^{-1}$ (the connections are chosen randomly again, so this is not
an identical copy). Then we connect the two \textquotedbl{}copies\textquotedbl{}
with a connection density $E^{-2}$ ($\kappa=2$), and so on and so
forth. The process is repeated iteratively until we reach the level
$\kappa=\eta-\mu$. It is also important to observe that these connections
are directed, therefore the connectivity matrix is generally not symmetric.
Two examples are shown in Figure \ref{fig:Example-of-fractal-matrix}.

\noindent According to \cite{citeulike:1343837}, the parameter $E$
determines the path length, the clustering coefficient and the complexity
of the network. The latter was first introduced in \cite{citeulike:6661358},
and quantifies the extent to which a system is both functionally segregated
and functionally integrated. This means that both the degree of independence
of the blocks and their level of cooperation are taken into account
by a single quantity, the complexity of the network, which for the
fractal topology is maximum when $E\approx2$.

\noindent Now we have to determine the quantities $\mathbb{E}\left[T_{ij}\right]$,
$\mathbb{E}\left[T_{ik}T_{jl}\right]$, $\mathbb{E}\left[T_{il}T_{jm}T_{kn}\right]$
etc. Therefore we need to analyze the algorithm that generates the
fractal connectivity matrix. If all the connections are at the level
$\kappa=0$, where the neurons are always fully connected, then we
trivially have:

\begin{onehalfspace}
\begin{center}
\begin{align*}
\mathbb{E}\left[T_{ij}\right]= & 1-\delta_{ij}\\
\\
\mathbb{E}\left[T_{ik}T_{jl}\right]= & \left(1-\delta_{ik}\right)\left(1-\delta_{jl}\right)\\
\\
\mathbb{E}\left[T_{il}T_{jm}T_{kn}\right]= & \left(1-\delta_{il}\right)\left(1-\delta_{jm}\right)\left(1-\delta_{kn}\right)\\
\vdots
\end{align*}

\par\end{center}
\end{onehalfspace}

\noindent because in this case the entries of the topology are deterministic.
Moreover, if we have an entry of the topology matrix, for example
$T_{ik}$, at the level $\kappa=0$, and another entry, for example
$T_{jl}$, at a different level, we obtain $\mathbb{E}\left[T_{ik}T_{jl}\right]=\left(1-\delta_{ik}\right)\mathbb{E}\left[T_{jl}\right]$,
and so on and so forth.

\noindent We next compute these statistical quantities when the connections
are not at the level $\kappa=0$. At a given level $\kappa>1$, the
total number of possible connections (in one direction) is $\alpha_{\kappa}=4^{\mu+\kappa-1}$,
among which the algorithm has to choose randomly $\beta_{\kappa}=\left\lfloor E^{-\kappa}\alpha_{\kappa}\right\rfloor $
connections.

\noindent At the level $\kappa$ the probability that $T_{ij}$ is
chosen at some time after $\beta_{\kappa}$ steps, regardless the
step at which it has been actually chosen, is:

\begin{onehalfspace}
\begin{center}
{\small{
\[
p\left(T_{ij}=1\right)=\frac{\beta_{\kappa}}{\alpha_{\kappa}}
\]
}}
\par\end{center}{\small \par}
\end{onehalfspace}

\noindent since we can draw uniformly among $\alpha_{\kappa}$ possible
connections, therefore:

\begin{onehalfspace}
\begin{center}
{\small{
\[
\mathbb{E}\left[T_{ij}\right]=0\times p\left(T_{ij}=0\right)+1\times p\left(T_{ij}=1\right)=\frac{\beta_{\kappa}}{\alpha_{\kappa}}
\]
}}
\par\end{center}{\small \par}
\end{onehalfspace}

\noindent Now we want to evaluate $\mathbb{E}\left[T_{ij}T_{kl}\right]$.
If, in the picture of the connectivity matrix, $T_{ij}$ and $T_{kl}$
are in two different squares, then clearly they are not correlated,
therefore in that case we have $\mathbb{E}\left[T_{ij}T_{kl}\right]=\mathbb{E}\left[T_{ij}\right]\mathbb{E}\left[T_{kl}\right]=\frac{\beta_{\kappa_{1}}}{\alpha_{\kappa_{1}}}\frac{\beta_{\kappa_{2}}}{\alpha_{\kappa_{2}}}$.
If instead they are in the same square, we have:

\begin{onehalfspace}
\begin{center}
{\small{
\[
\mathbb{E}\left[T_{ij}T_{kl}\right]=\frac{\beta_{\kappa}\left(\beta_{\kappa}-1\right)}{\alpha_{\kappa}\left(\alpha_{\kappa}-1\right)}
\]
}}
\par\end{center}{\small \par}
\end{onehalfspace}

\noindent since they are selected sequentially and independently from
each other. In general, for $n$ entries of the topology in the same
square, with $n\leq\beta_{\kappa}$, we obtain:

\begin{onehalfspace}
\begin{center}
{\small{
\[
\mathbb{E}\left[T_{i_{0}j_{0}}T_{i_{1}j_{1}}...T_{i_{n-1}j_{n-1}}\right]=\frac{\beta_{\kappa}\left(\beta_{\kappa}-1\right)...\left(\beta_{\kappa}-n+1\right)}{\alpha_{\kappa}\left(\alpha_{\kappa}-1\right)...\left(\alpha_{\kappa}-n+1\right)}=\frac{\beta_{\kappa}!\left(\alpha_{\kappa}-n\right)!}{\alpha_{\kappa}!\left(\beta_{\kappa}-n\right)!}
\]
}}
\par\end{center}{\small \par}
\end{onehalfspace}

\noindent \begin{flushleft}
thereby the problem of determining the correlation structure of the
neural network with the fractal connectivity matrix is solved.
\par\end{flushleft}

\section{\label{sec:Numerical experiments}Numerical experiments}

\begin{figure}
\begin{centering}
\includegraphics[scale=0.35]{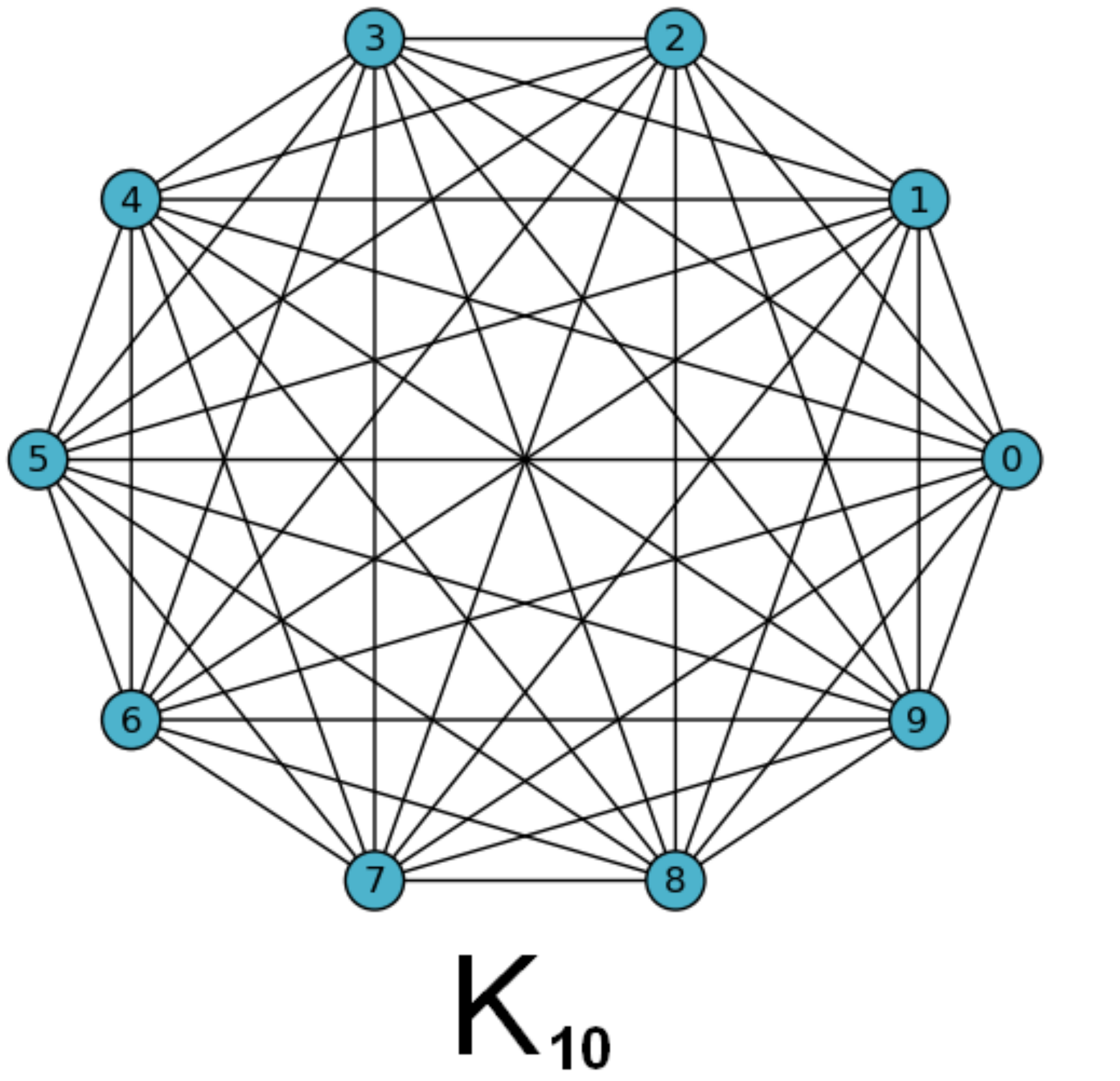}\includegraphics[scale=0.35]{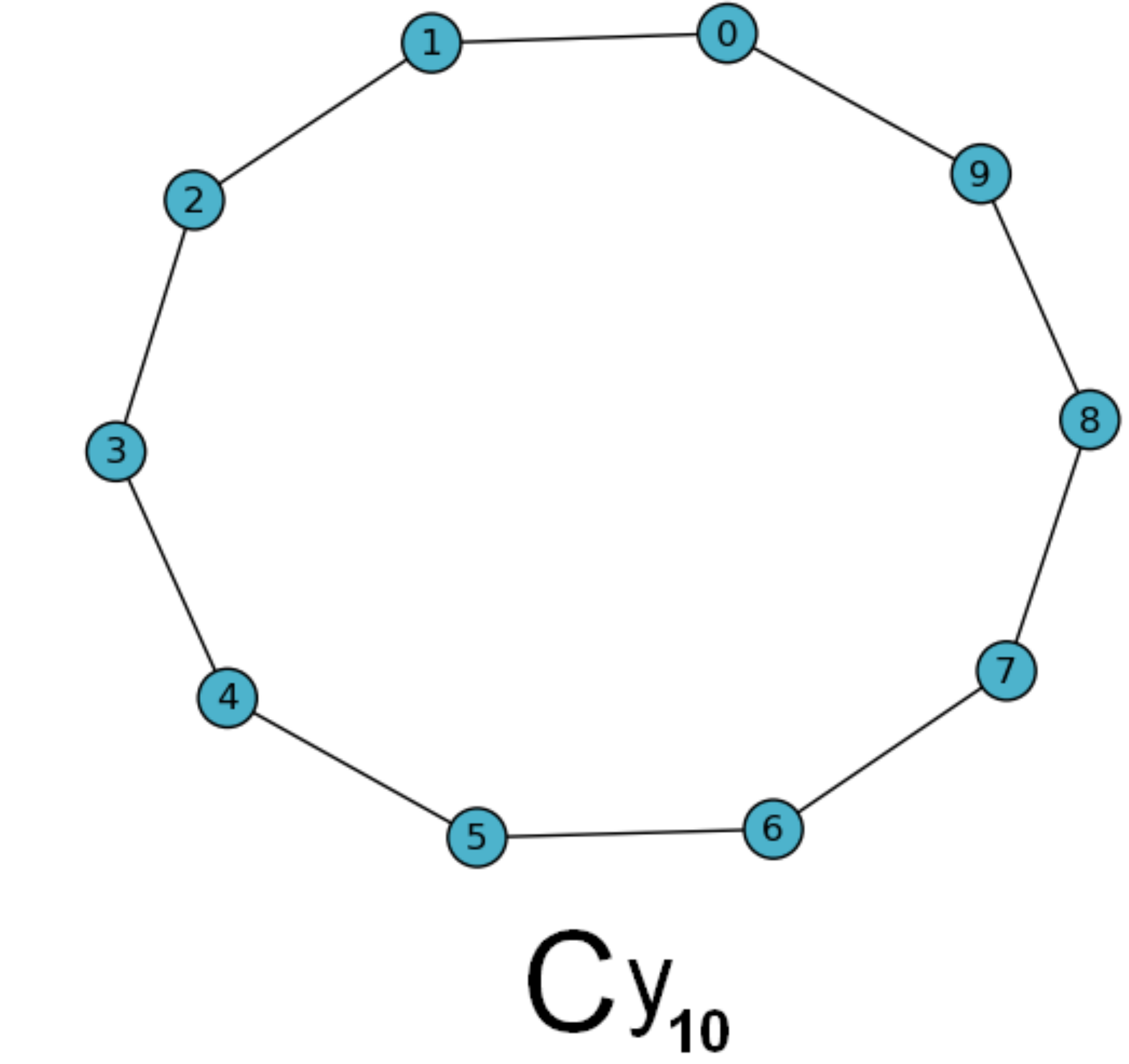}
\par\end{centering}

\caption[{\footnotesize{Some deterministic topologies}}]{{\small{\label{fig:Deterministic-topologies}Network topologies used
to generate Figures \ref{fig:Correlation-problem} and \ref{fig:Numerical-comparison-weak-1}
- \ref{fig:Numerical-comparison-weak-4}. In the context of Graph
Theory, $K_{N}$ is called }}\textit{\small{complete graph}}{\small{
and represents the topology of a fully connected network, while $Cy_{N}$
is called }}\textit{\small{cycle graph}}{\small{ and represents the
case when the neurons are connected to form a closed loop.}}}

\end{figure}

\begin{figure}
\noindent \begin{centering}
\includegraphics[scale=0.4]{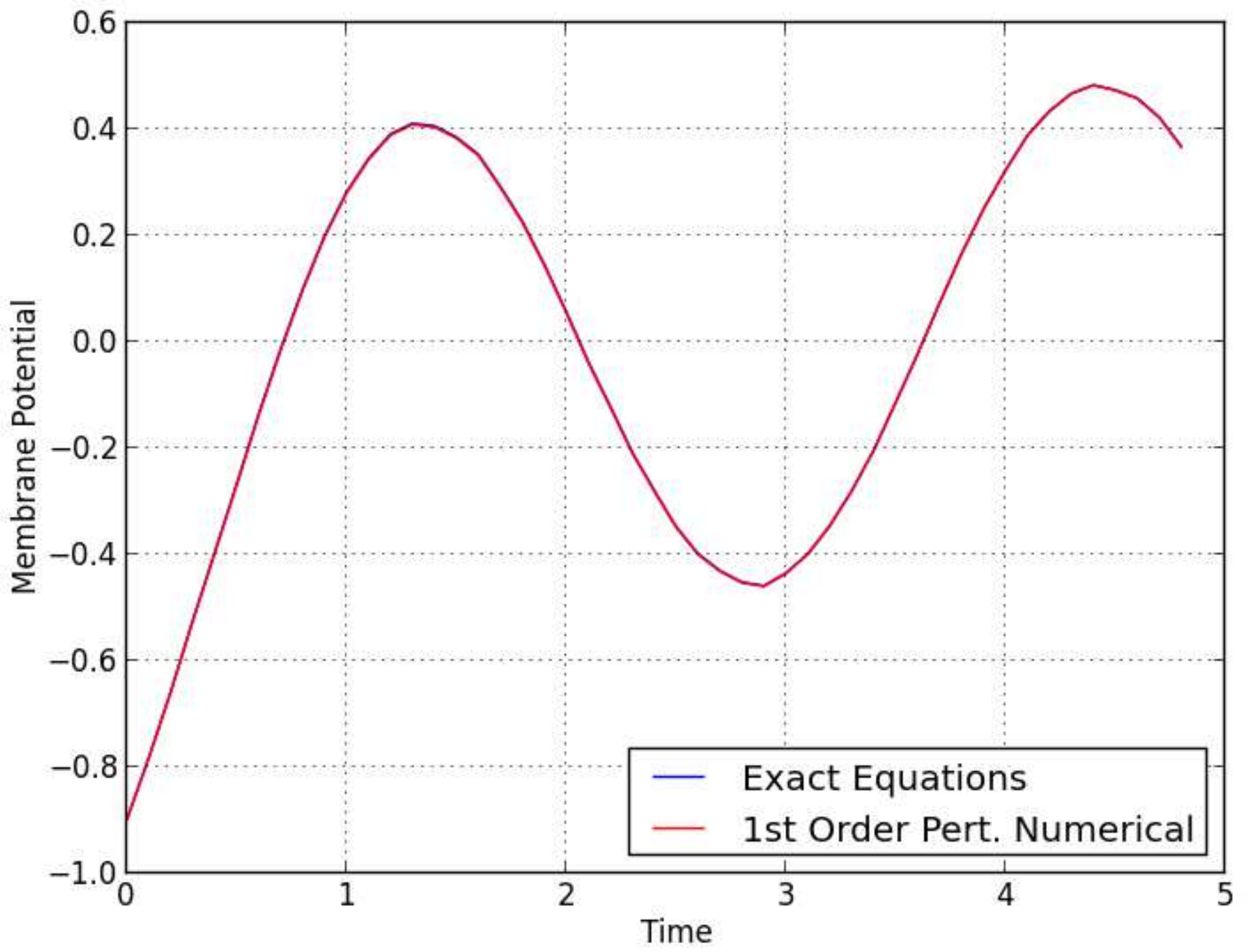}\includegraphics[scale=0.4]{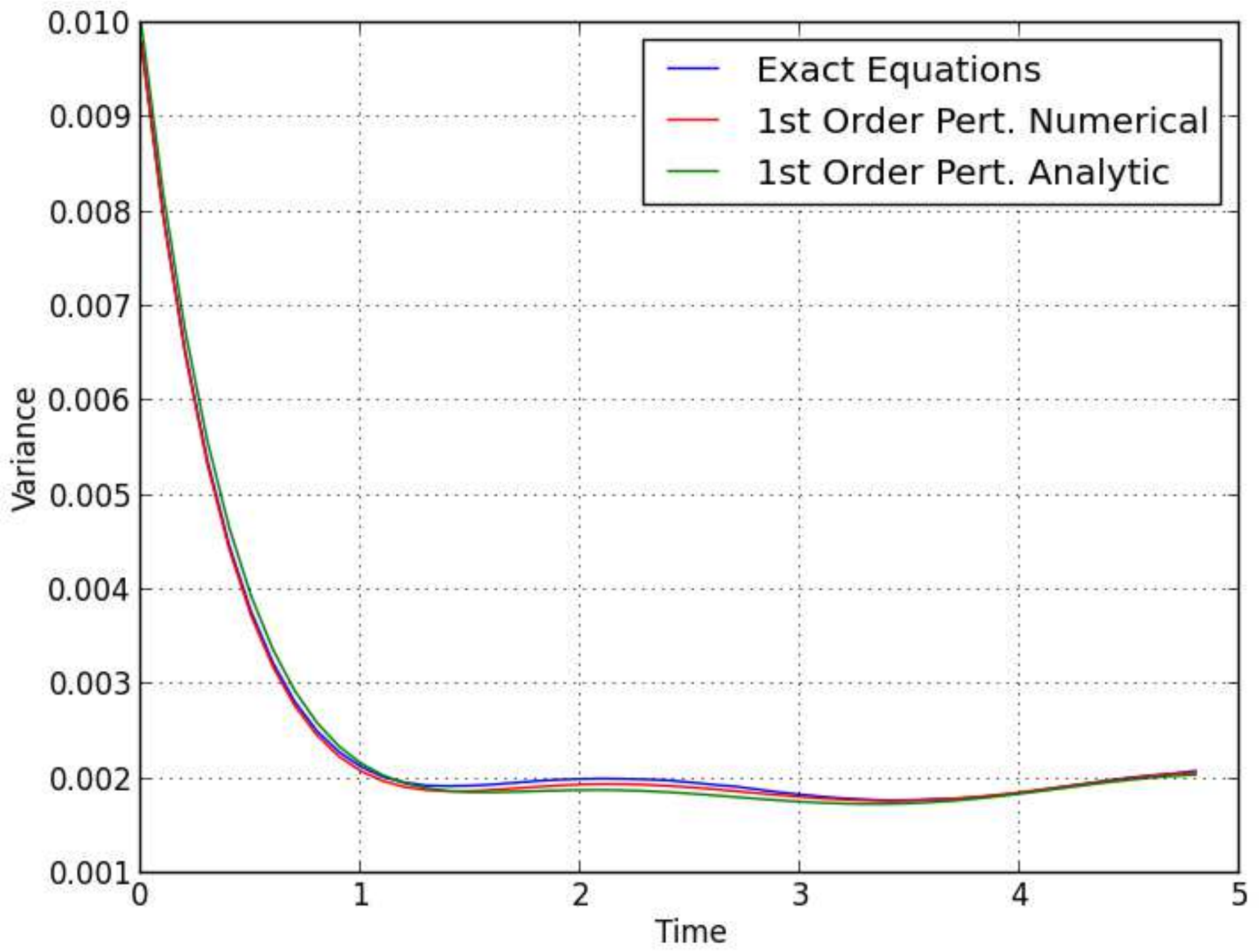}
\par\end{centering}

\noindent \begin{centering}
\includegraphics[scale=0.4]{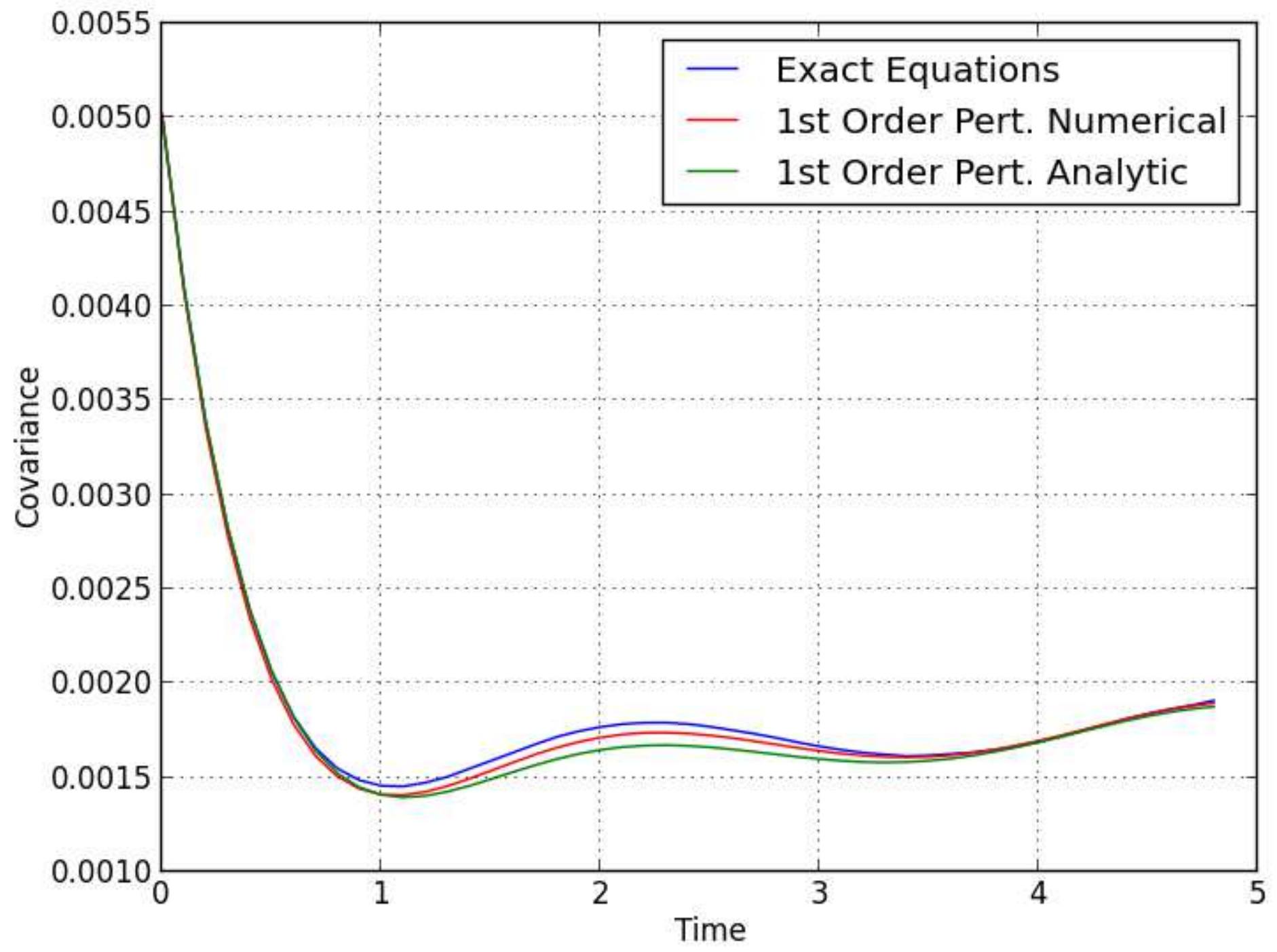}\includegraphics[scale=0.4]{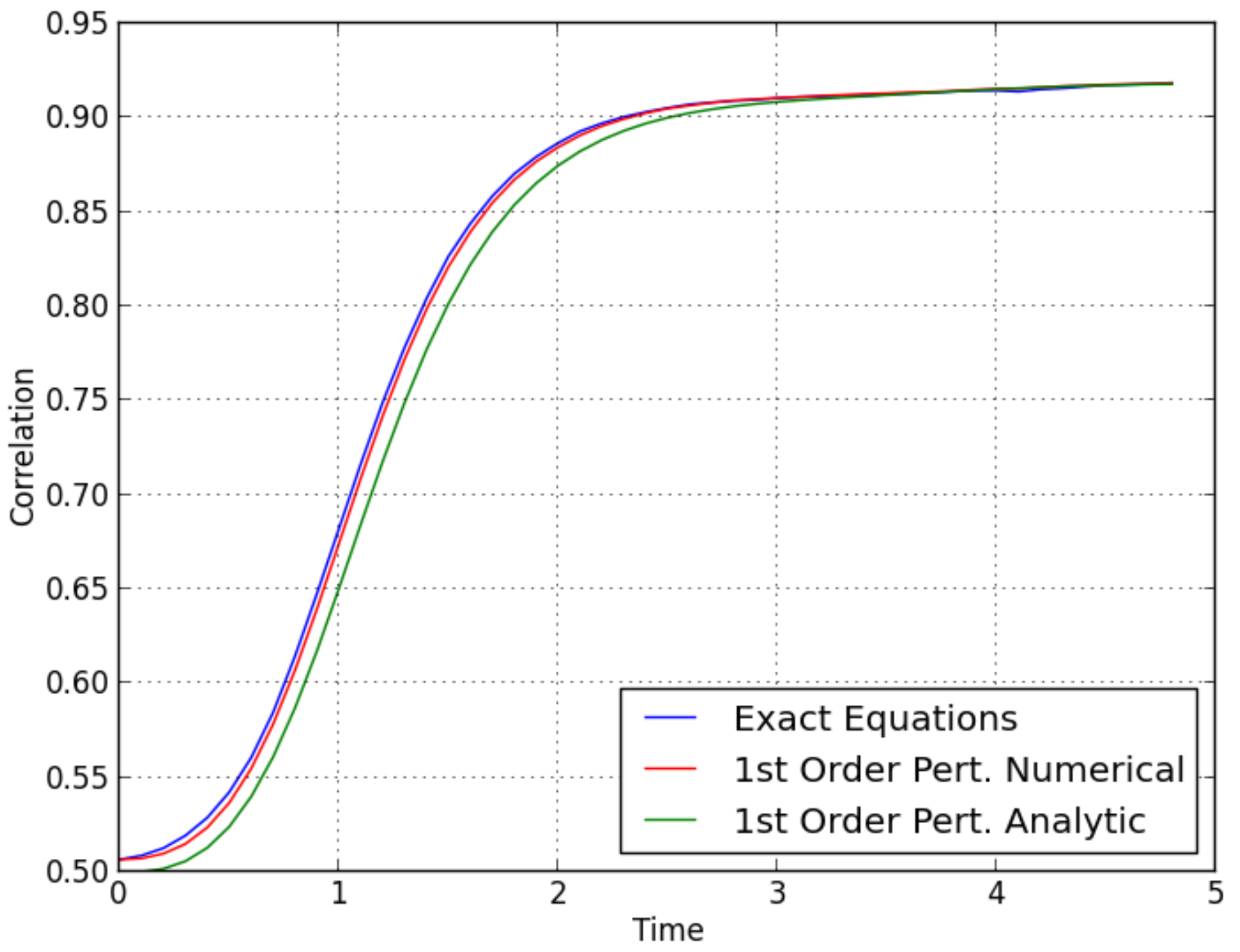}
\par\end{centering}

\caption[{\footnotesize{Numerical comparison of the perturbative expansion
with weak weights - 1}}]{{\small{\label{fig:Numerical-comparison-weak-1}Comparison of the
variance, covariance and correlation obtained from the simulation
of equations \ref{eq:exact-equation} (blue line), of equations \ref{eq:perturbative-equation-0}
- \ref{eq:perturbative-equation-4} (red line) and from formulae \ref{eq:covariance-part-1}
- \ref{eq:covariance-part-4}, \ref{eq:covariance-fixed} and \ref{eq:covariance-correction-term}
(green line). Therefore the perturbative expansion of the membrane
potential has been truncated at the first order, while those of the
variance and covariance at the second order. We have compared the
$0$th and the $1$st neuron, using the Euler-Maruyama scheme (blue
and red lines) and the trapezoidal rule (green line) with integration
time step $\Delta t=0.1$. The statistics have been evaluated with
$10,000$ Monte Carlo simulations, for the values of the parameters
reported in Table \ref{tab:simulation-parameters-weak}. The topology
is $K_{10}$ (see Figure \ref{fig:Deterministic-topologies}) and
therefore deterministic.}}}
\end{figure}

\begin{figure}
\noindent \begin{centering}
\includegraphics[scale=0.4]{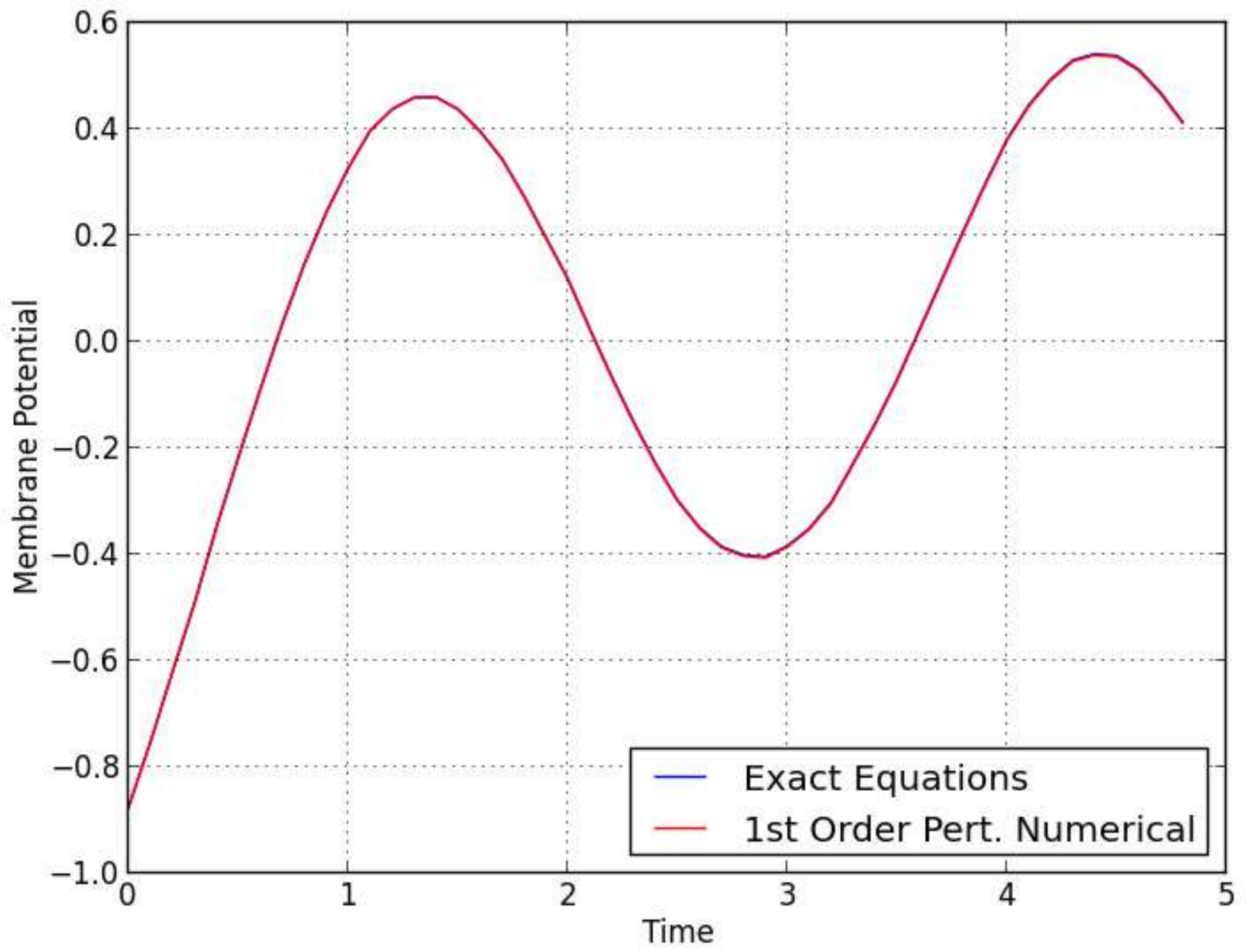}\includegraphics[scale=0.4]{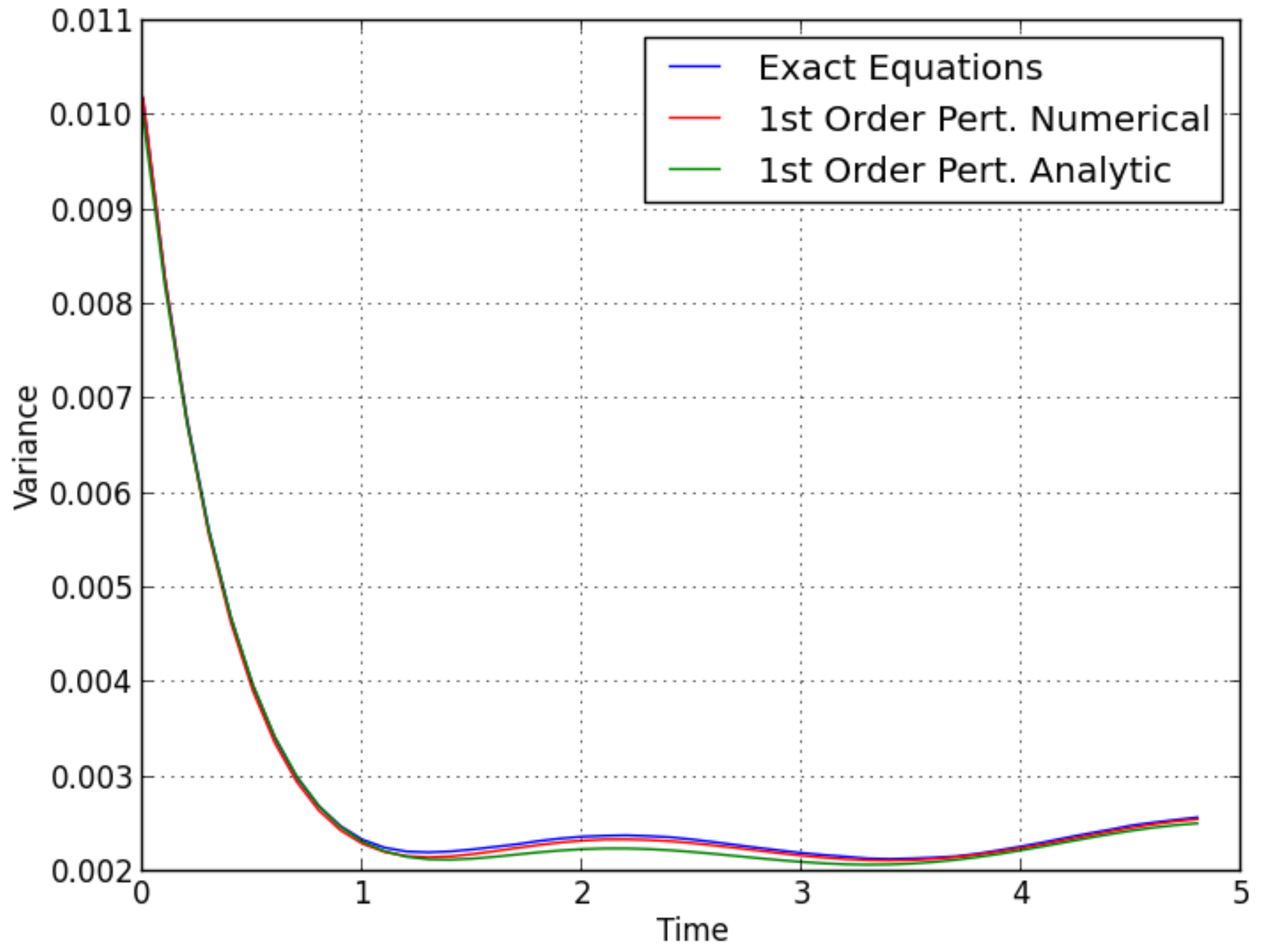}
\par\end{centering}

\noindent \begin{centering}
\includegraphics[scale=0.4]{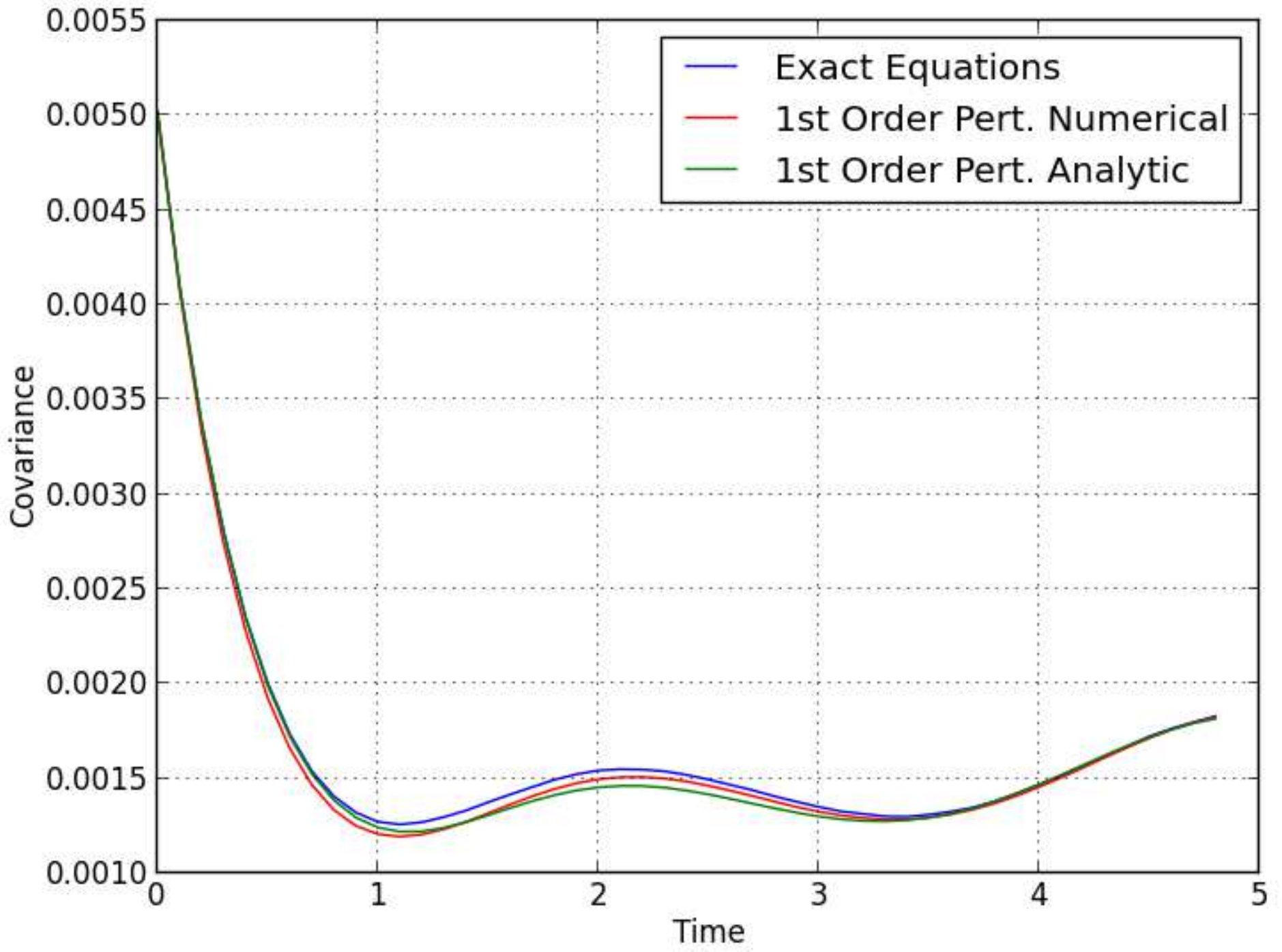}\includegraphics[scale=0.4]{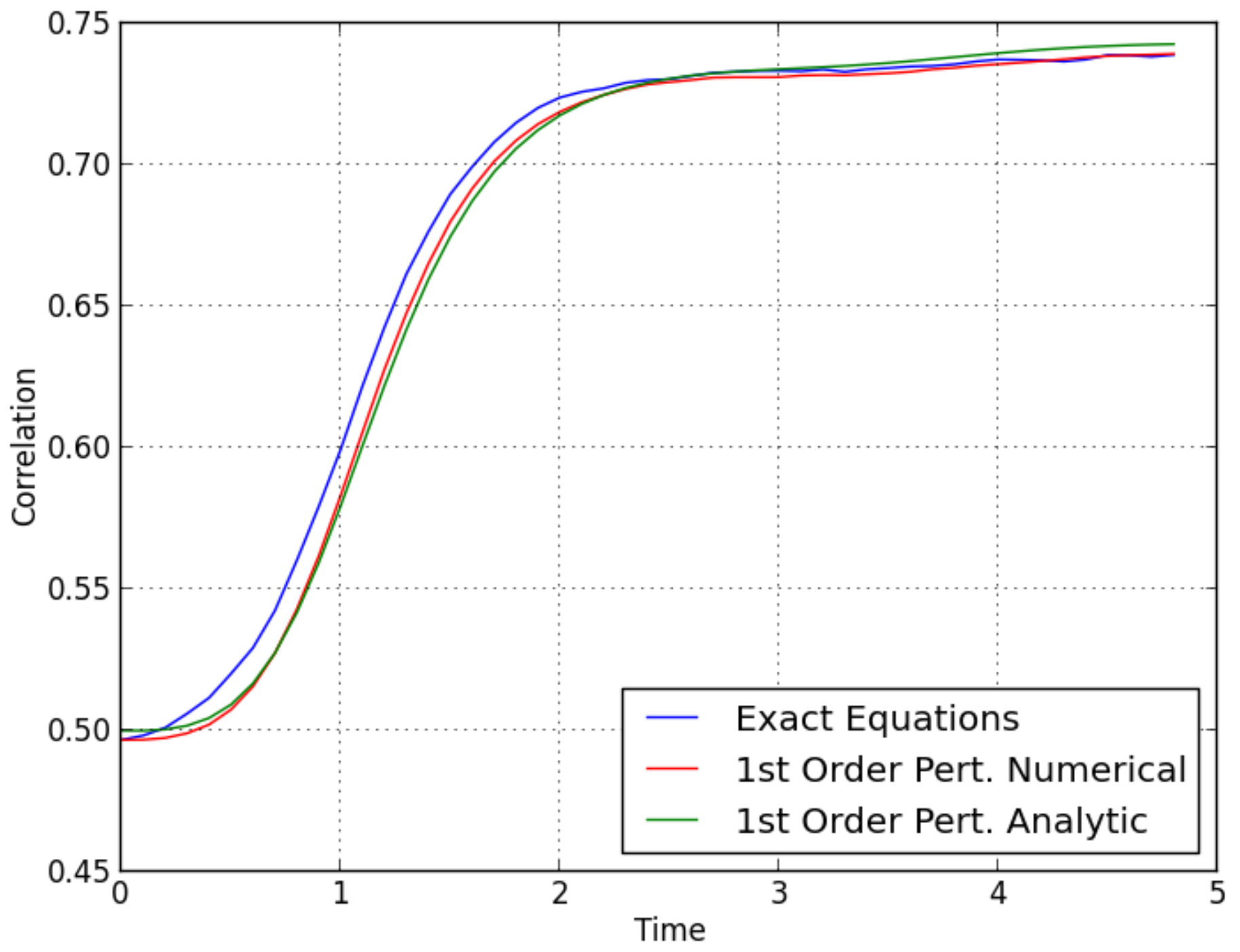}
\par\end{centering}

\caption[{\footnotesize{Numerical comparison of the perturbative expansion
with weak weights - 2}}]{{\small{\label{fig:Numerical-comparison-weak-2}Comparison of the
variance, covariance and correlation obtained for the deterministic
topology $Cy_{10}$ (see Figure \ref{fig:Deterministic-topologies}),
for the values of the parameters reported in Table \ref{tab:simulation-parameters-weak}.}}}
\end{figure}

\begin{figure}
\noindent \begin{centering}
\includegraphics[scale=0.4]{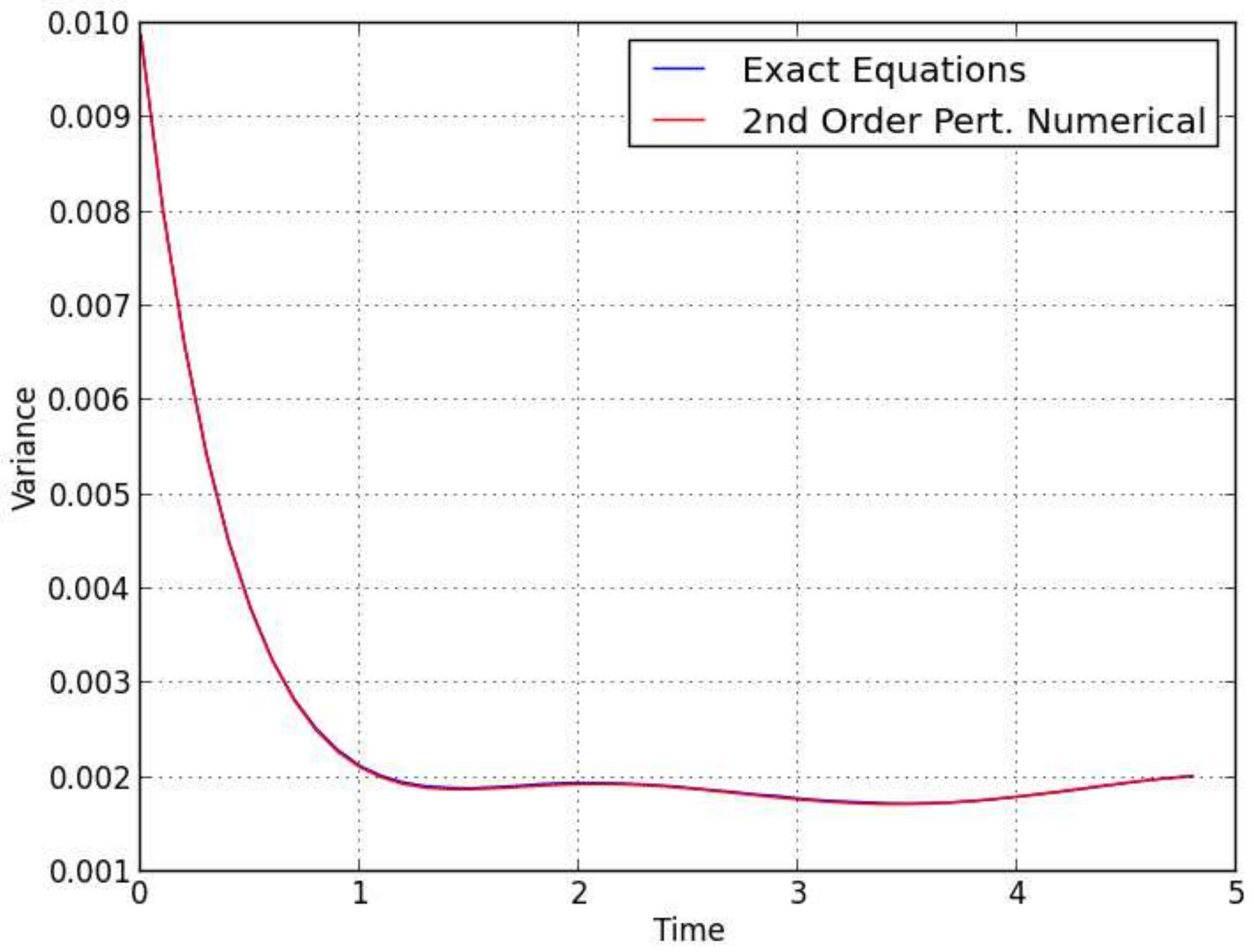}\includegraphics[scale=0.4]{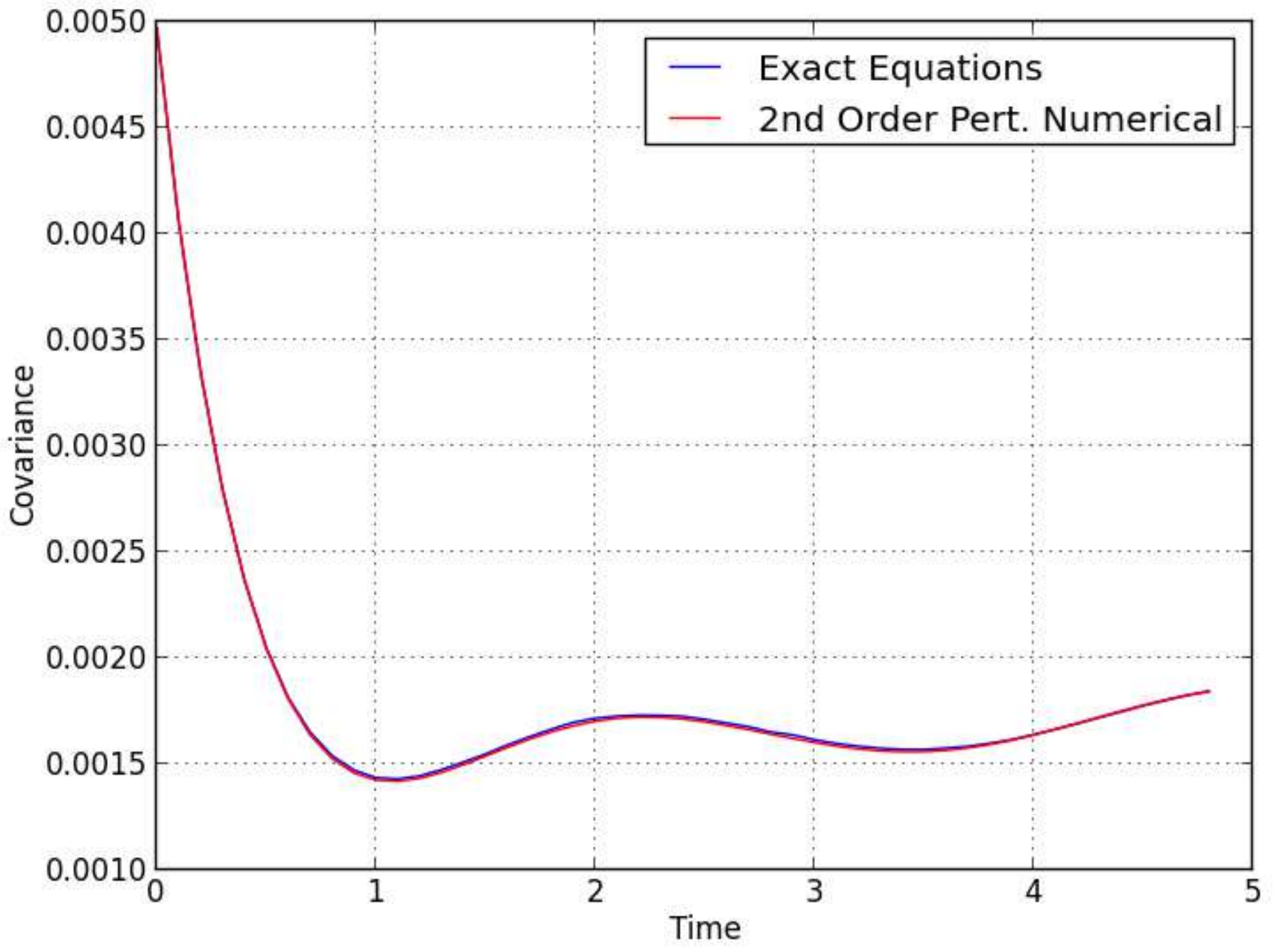}
\par\end{centering}

\noindent \begin{centering}
\includegraphics[scale=0.4]{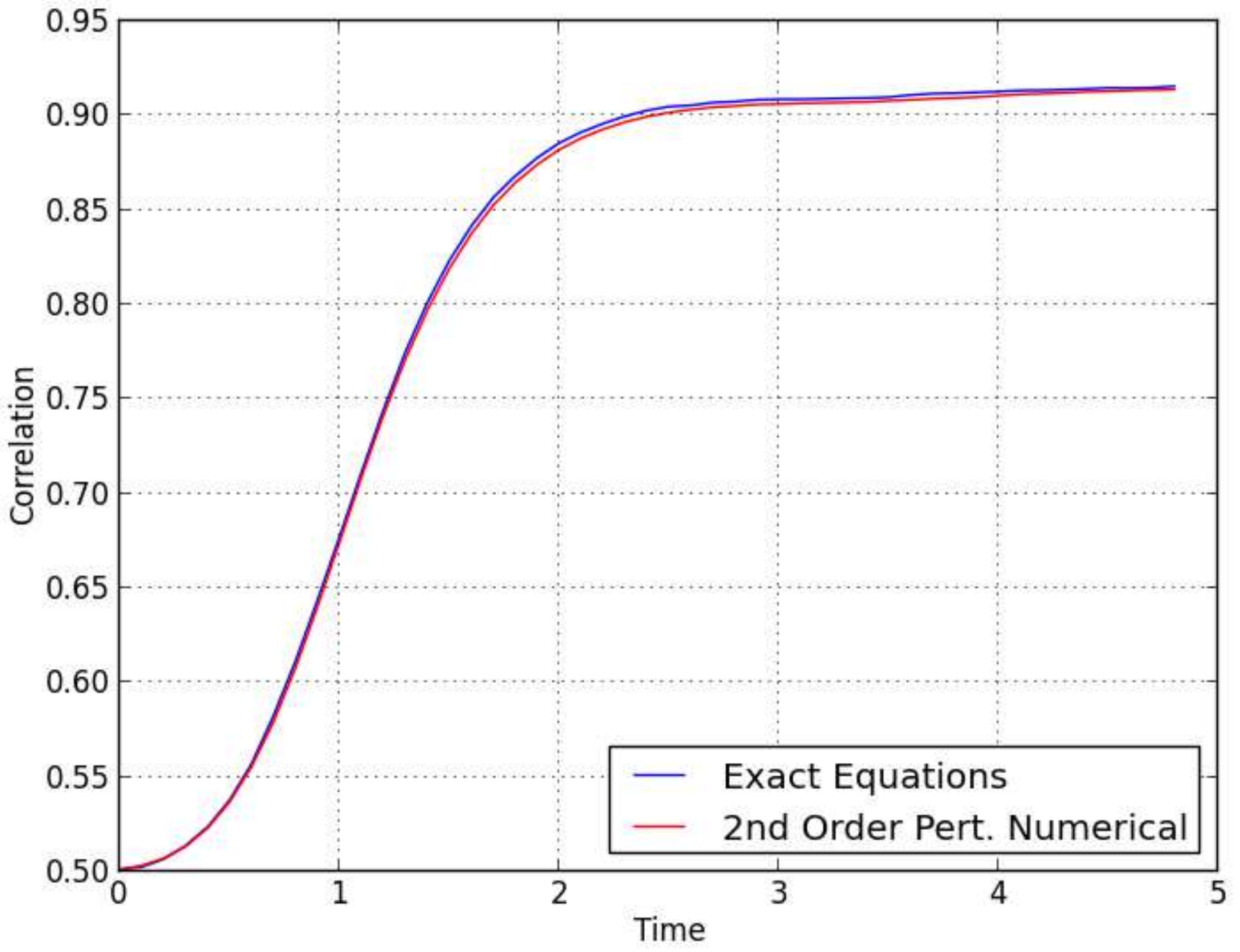}
\par\end{centering}

\caption[{\footnotesize{Numerical comparison of the perturbative expansion
with weak weights - 3}}]{{\small{\label{fig:Numerical-comparison-weak-3}Comparison of the
variance, covariance and correlation obtained for the deterministic
topology $K_{10}$, for the values of the parameters reported in Table
\ref{tab:simulation-parameters-weak}, but considering also the second
order corrections of the membrane potential. Clearly the match has
been improved by the addition of these terms, as the reader can easily
check from the comparison with Figure \ref{fig:Numerical-comparison-weak-1}.}}}
\end{figure}

\begin{figure}
\noindent \begin{centering}
\includegraphics[scale=0.4]{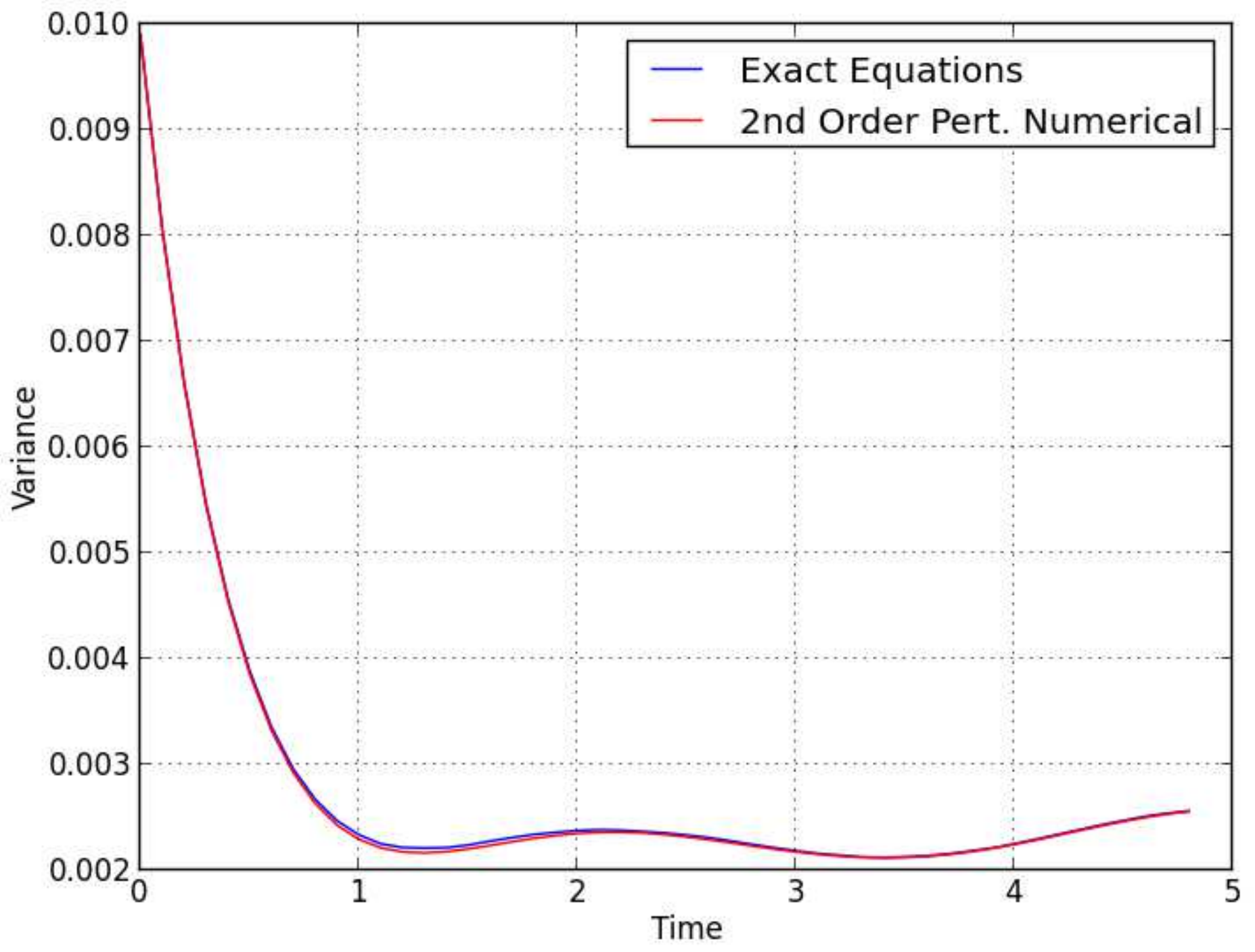}\includegraphics[scale=0.4]{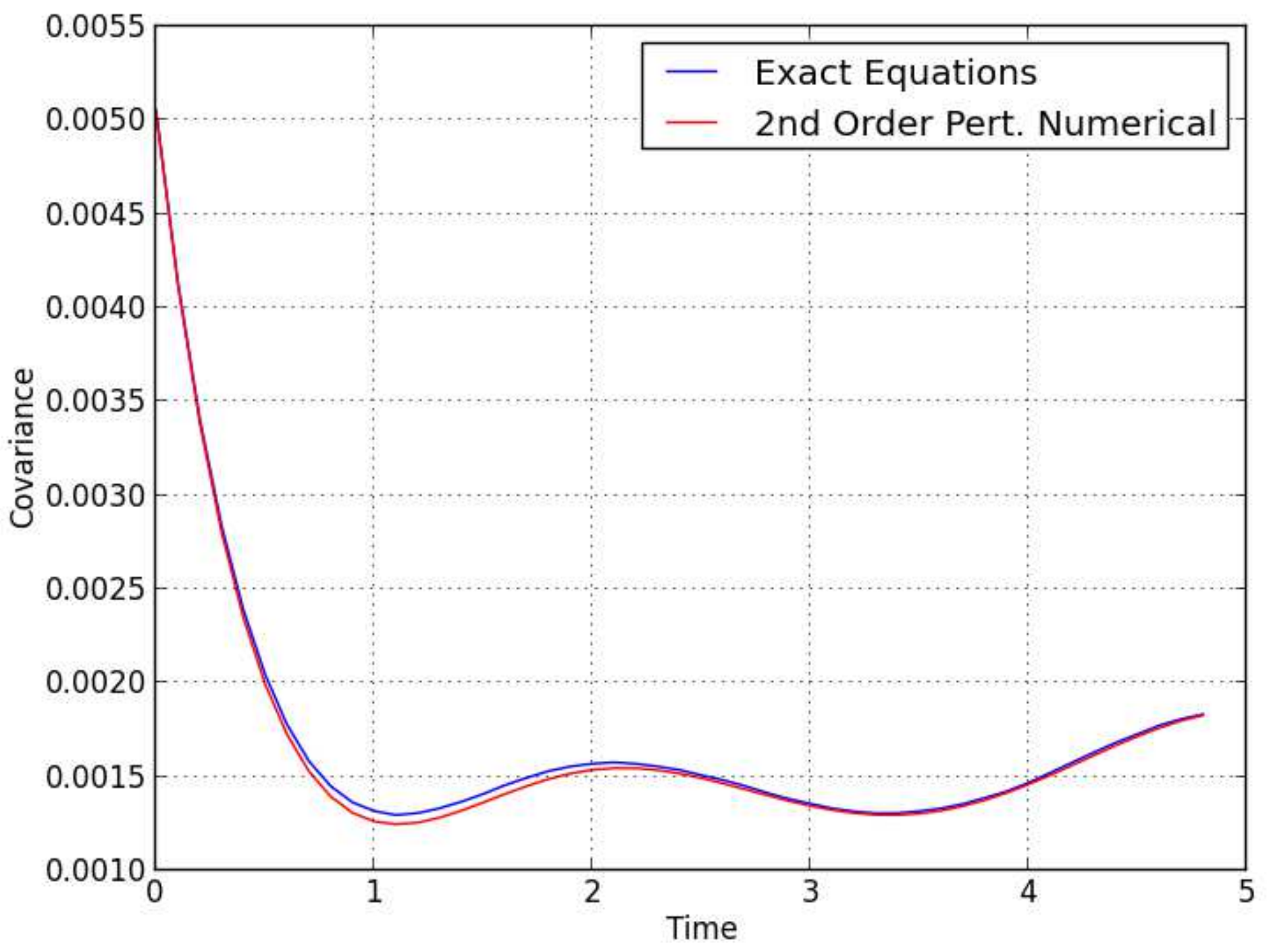}
\par\end{centering}

\noindent \begin{centering}
\includegraphics[scale=0.4]{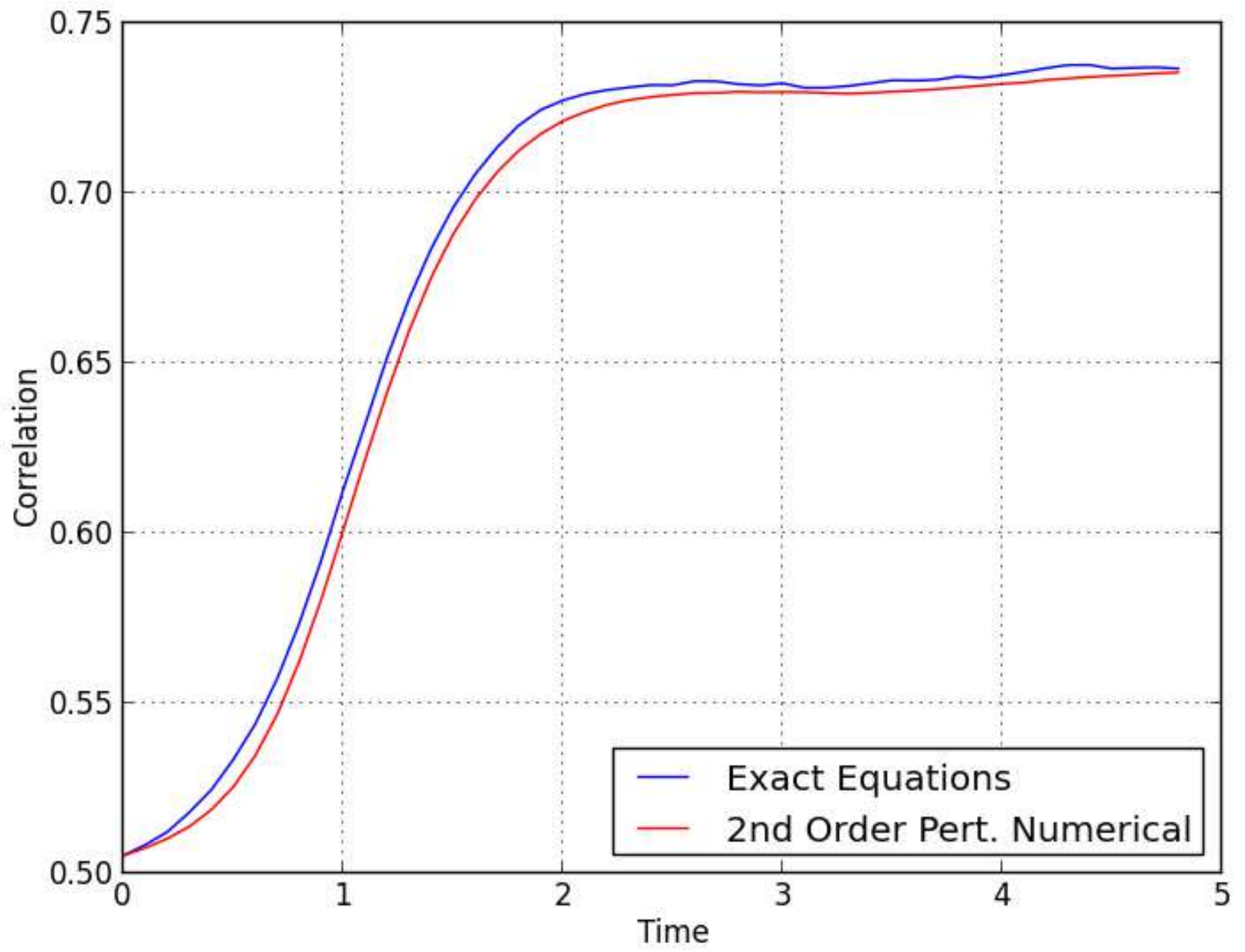}
\par\end{centering}

\caption[{\footnotesize{Numerical comparison of the perturbative expansion
with weak weights - 4}}]{{\small{\label{fig:Numerical-comparison-weak-4}Comparison of the
variance, covariance and correlation obtained for the deterministic
topology $Cy_{10}$, for the values of the parameters reported in
Table \ref{tab:simulation-parameters-weak}, but considering also
the second order corrections of the membrane potential. This time
the improvement of the match is not evident, if compared with Figure
\ref{fig:Numerical-comparison-weak-2}, which proves that the goodness
of the perturbative expansion depends also on the topology of the
network. It is important to observe that the second order corrections
are generally small, therefore their magnitude could be of the same
order of the numerical error introduced by the finite number of Monte
Carlo simulations.}}}
\end{figure}

\noindent We want to show that this perturbative expansion provides
a good match with the exact equations of the network. For this reason
in Figures \ref{fig:Numerical-comparison-weak-1} and \ref{fig:Numerical-comparison-weak-2}
we have shown the comparison between the membrane potential, variance,
covariance and correlation of pairs of neurons for two kinds of connectivity
matrices (fully connected and cycle graphs, see Figure \ref{fig:Deterministic-topologies}),
obtained from the simulation of equations \ref{eq:exact-equation}
(blue line), of equations \ref{eq:perturbative-equation-0} - \ref{eq:perturbative-equation-4}
(red line) and from formulae \ref{eq:covariance-part-1} - \ref{eq:covariance-part-4},
\ref{eq:covariance-fixed} and \ref{eq:covariance-correction-term}
(green line). Therefore we have obtained these figures without considering
the second order terms in the perturbative expansion of $V_{i}\left(t\right)$.
In other words, we have omitted the third order terms \ref{eq:covariance-part-5}
- \ref{eq:covariance-part-8} in the variance and covariance, due
to the difficulty of implementing them numerically. Instead in Figures
\ref{fig:Numerical-comparison-weak-3} and \ref{fig:Numerical-comparison-weak-4}
we have shown the comparison between equations \ref{eq:exact-equation}
(blue line) and equations \ref{eq:perturbative-equation-0} - \ref{eq:perturbative-equation-8}
(red line), therefore considering also the higher order terms, because
the numerical calculation of the variance and covariance through the
simulation of equations \ref{eq:perturbative-equation-5} - \ref{eq:perturbative-equation-8}
is much easier than the implementation of the terms \ref{eq:covariance-part-5}
- \ref{eq:covariance-part-8}.

\begin{figure}
\noindent \begin{centering}
\includegraphics[scale=0.4]{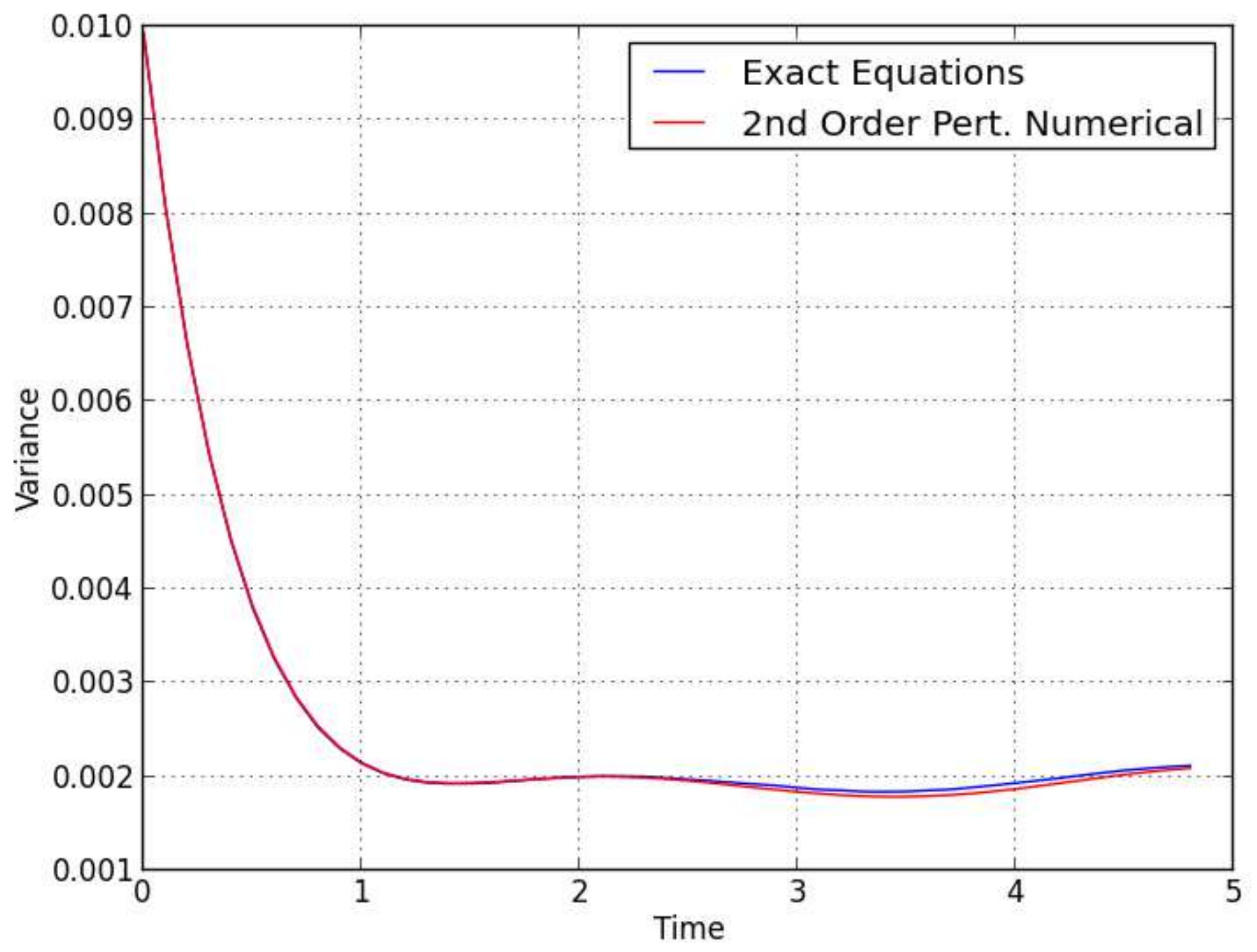}\includegraphics[scale=0.4]{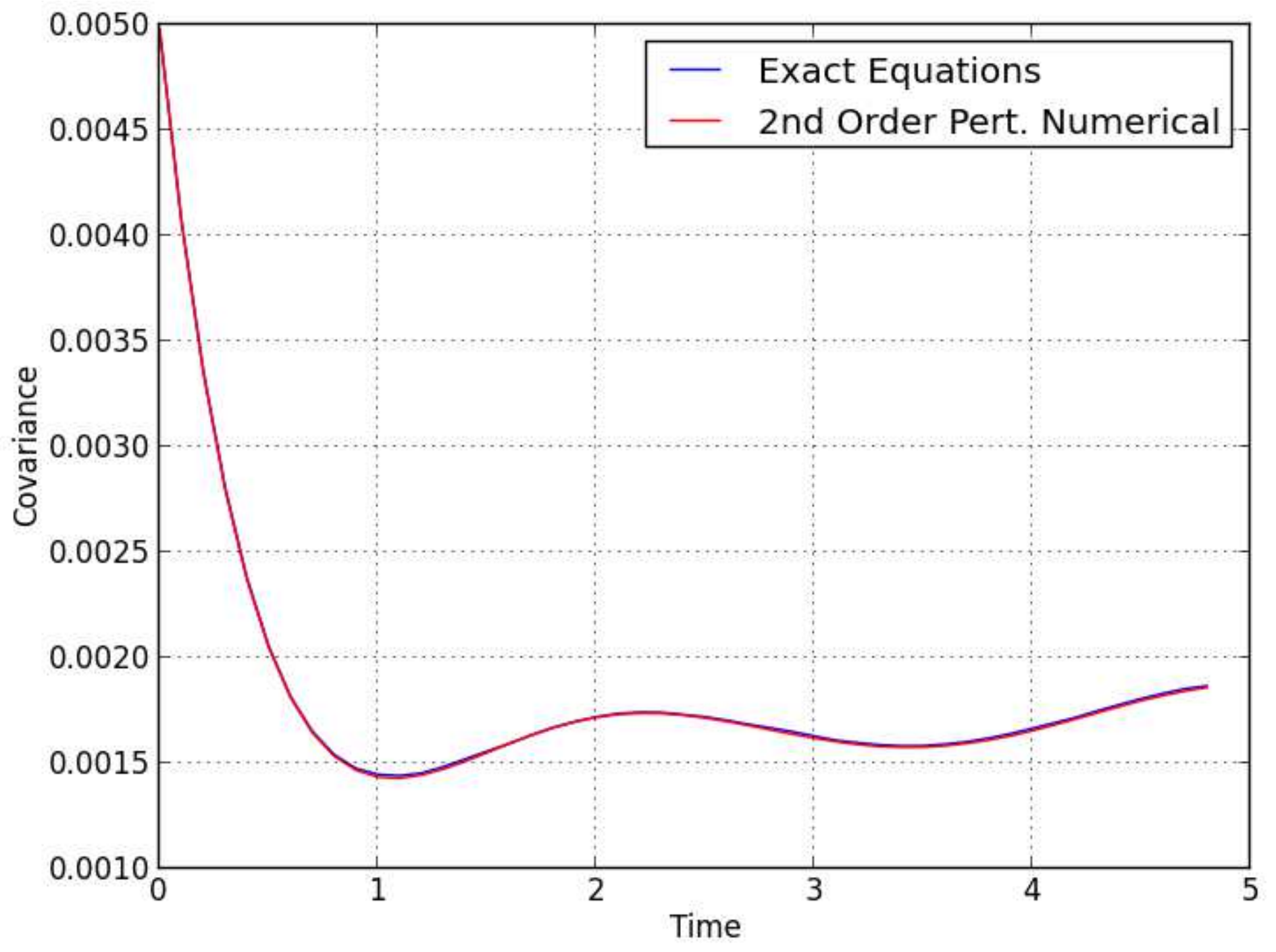}
\par\end{centering}

\noindent \begin{centering}
\includegraphics[scale=0.4]{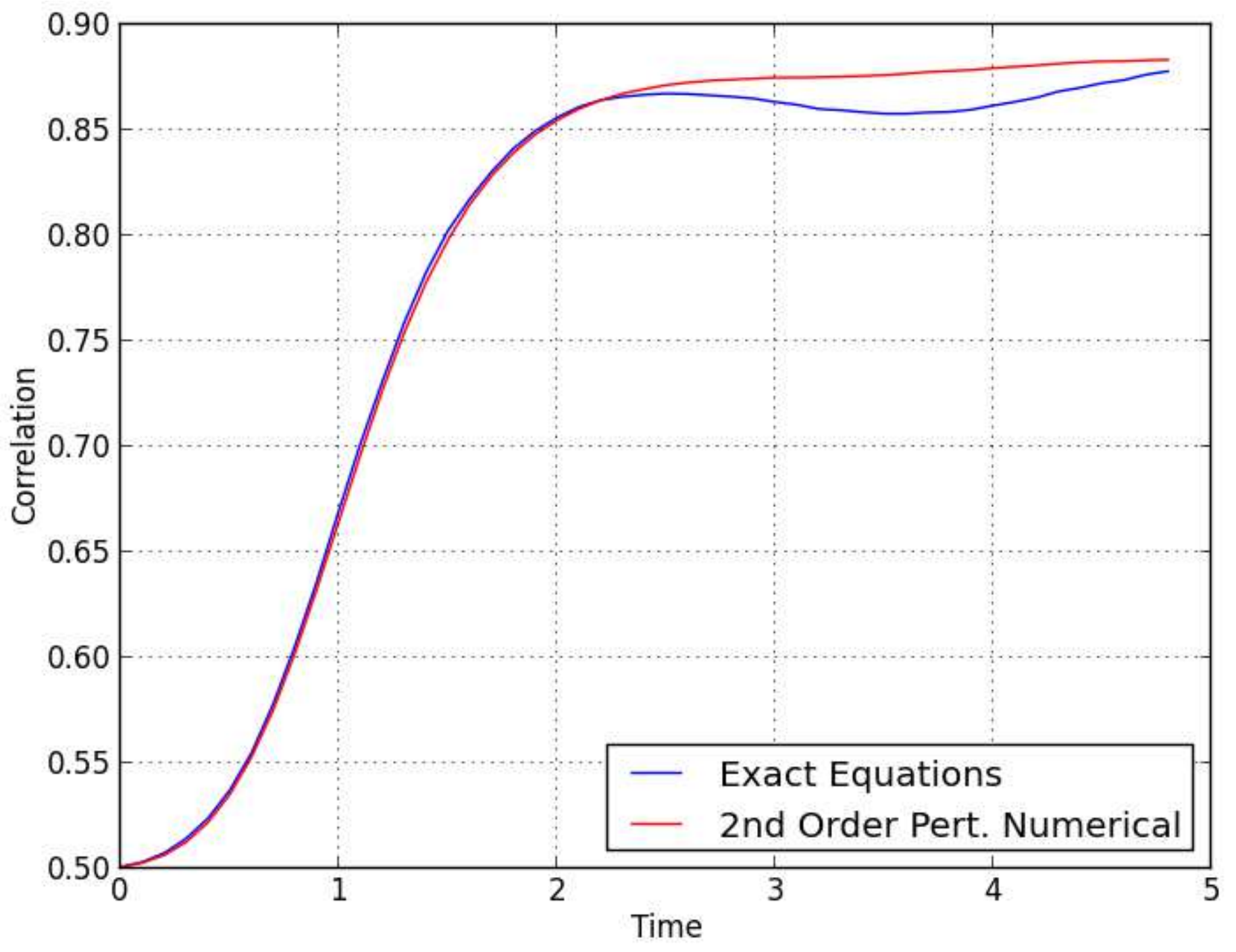}
\par\end{centering}

\caption[{\footnotesize{Numerical comparison of the perturbative expansion
with weak weights - 5}}]{{\small{\label{fig:Numerical-comparison-weak-5}Comparison of the
variance, covariance and correlation obtained for a random topology,
for the values of the parameters reported in Table \ref{tab:simulation-parameters-weak},
considering also the second order corrections of the membrane potential.
In detail, here we have assumed that each pair of neurons is connected
independently from the others and with probability $p=0.7$. Even
if the match of the variance and covariance is quantitatively very
good, the approximation of the correlation is not satisfying for $t>2$.
This is due to the fact that the ratio of small quantities (in this
case the variance and covariance) is very sensitive to small errors
in the numerator and denominator. Nevertheless the second order expansion
provides a satisfying result, because the variance and covariance
are in very good agreement with the exact neural equations. It is
important to observe that the discrepancy is also due to the finite
number of Monte Carlo simulations, which should be increased especially
for small values of the variance and covariance.}}}
\end{figure}

\begin{figure}
\noindent \begin{centering}
{\small{\includegraphics[scale=0.4]{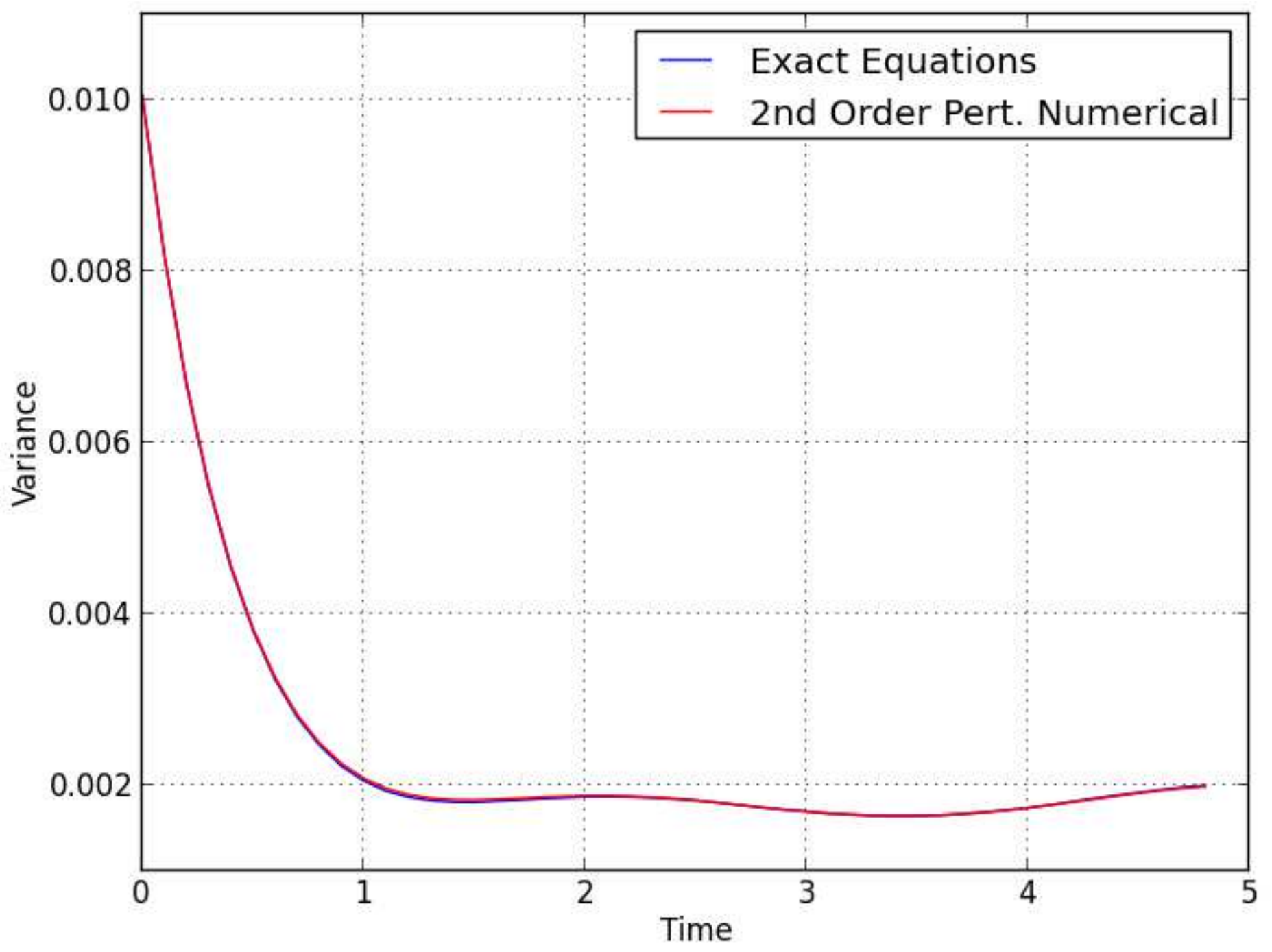}\includegraphics[scale=0.4]{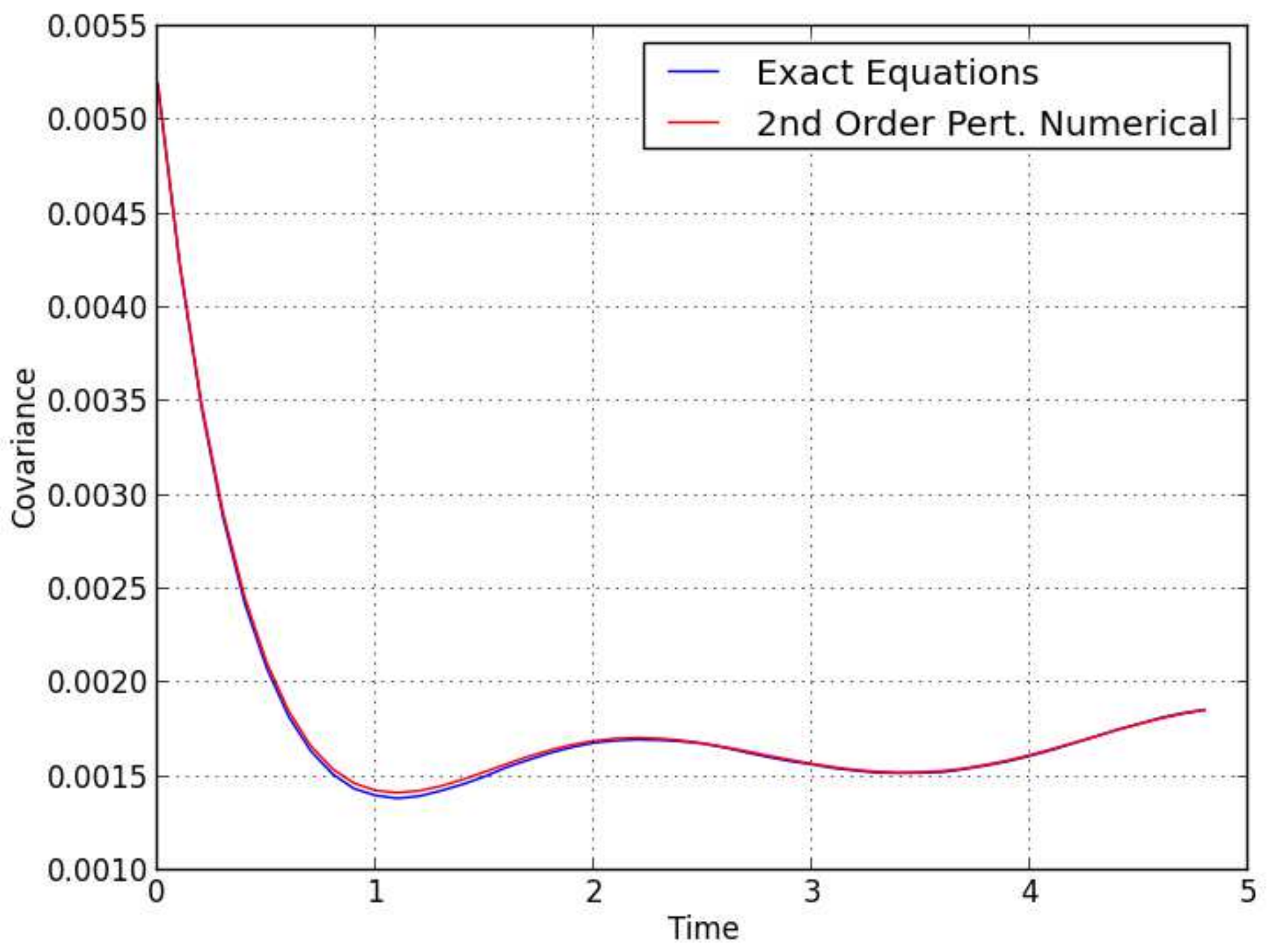}}}
\par\end{centering}{\small \par}

\noindent \centering{}\includegraphics[scale=0.4]{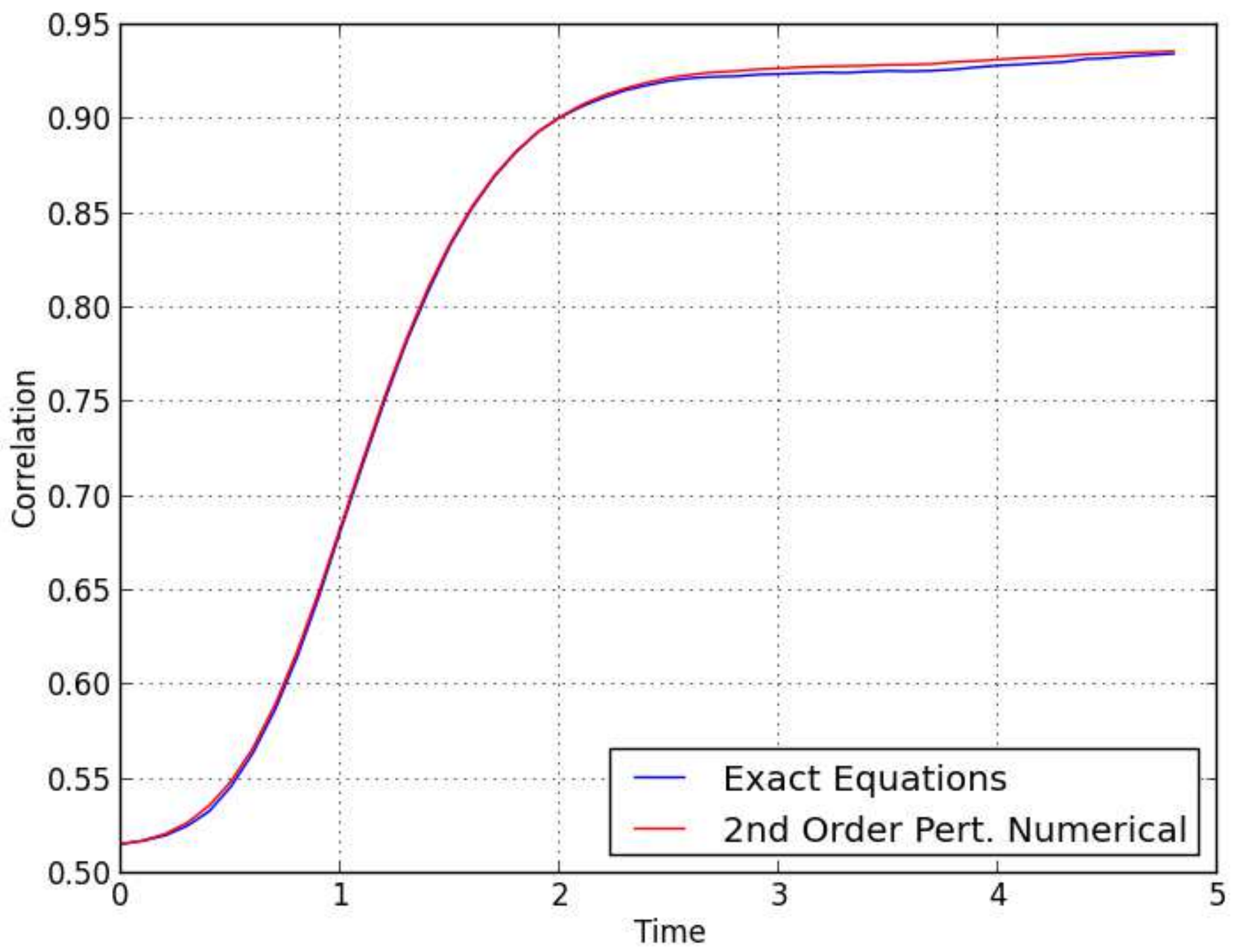}\caption[{\footnotesize{Numerical comparison of the perturbative expansion
with weak weights - 6}}]{{\small{\label{fig:Numerical-comparison-weak-6}Comparison of the
variance, covariance and correlation obtained for the Sporns' topology,
for the values of the parameters reported in Table \ref{tab:simulation-parameters-weak},
considering also the second order corrections of the membrane potential.
In this example we have set $\eta=4$ ($N=16$), $\mu=2$ and $E=1.1$,
therefore the network is almost fully connected. The two neurons are
in the same block, therefore they are connected at the level $\kappa=0$.}}}
\end{figure}

\begin{figure}
\noindent \begin{centering}
{\small{\includegraphics[scale=0.4]{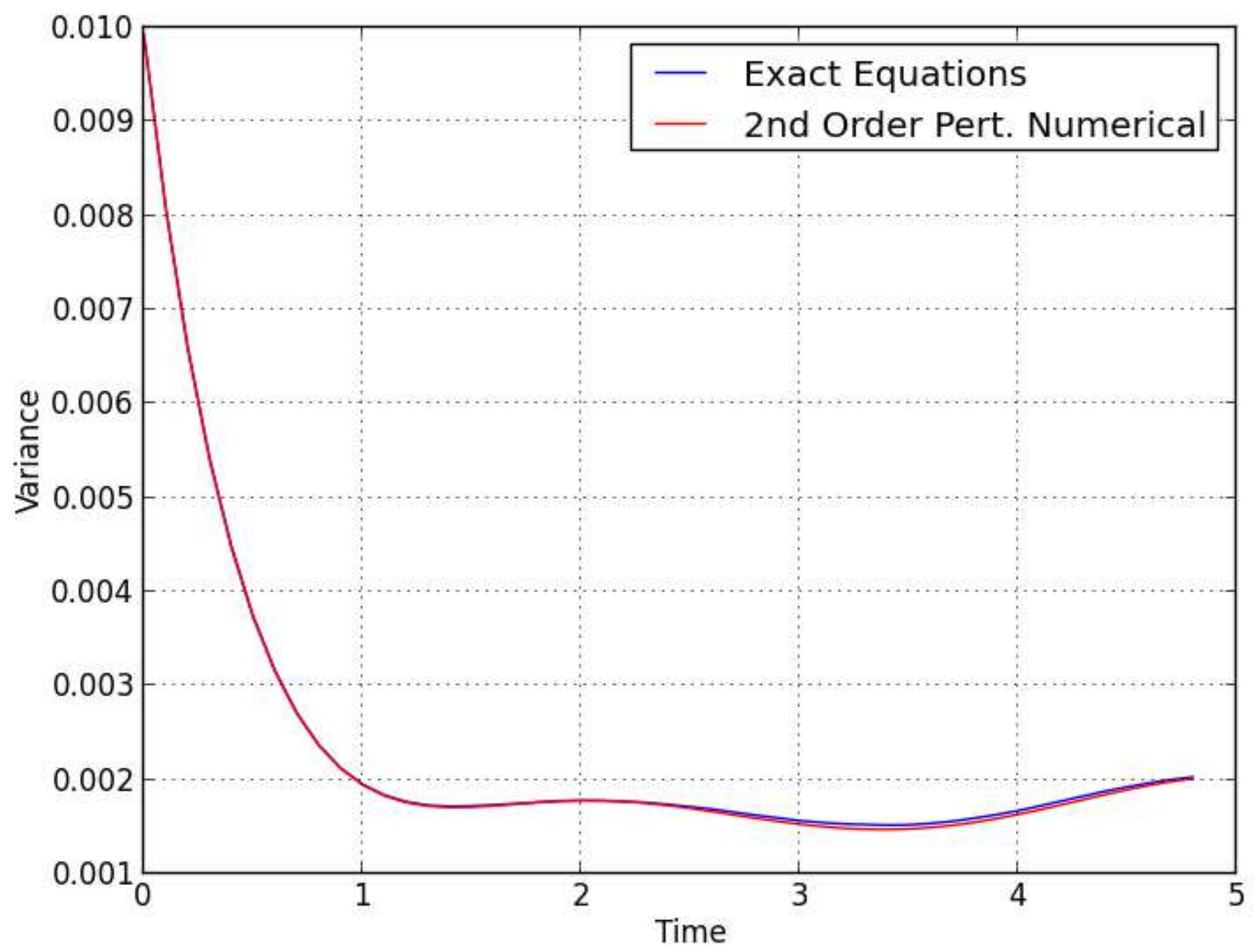}\includegraphics[scale=0.4]{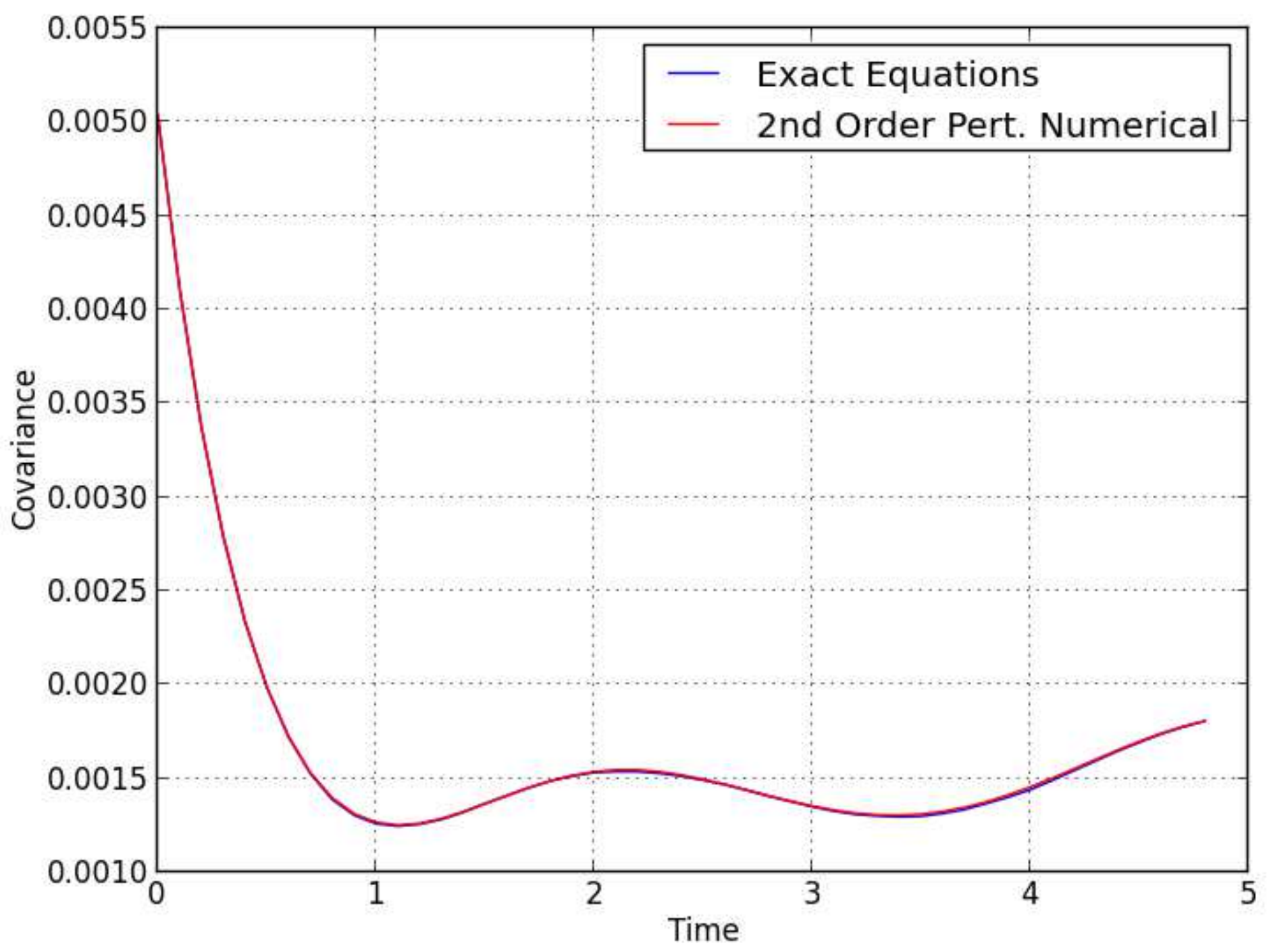}}}
\par\end{centering}{\small \par}

\noindent \centering{}\includegraphics[scale=0.4]{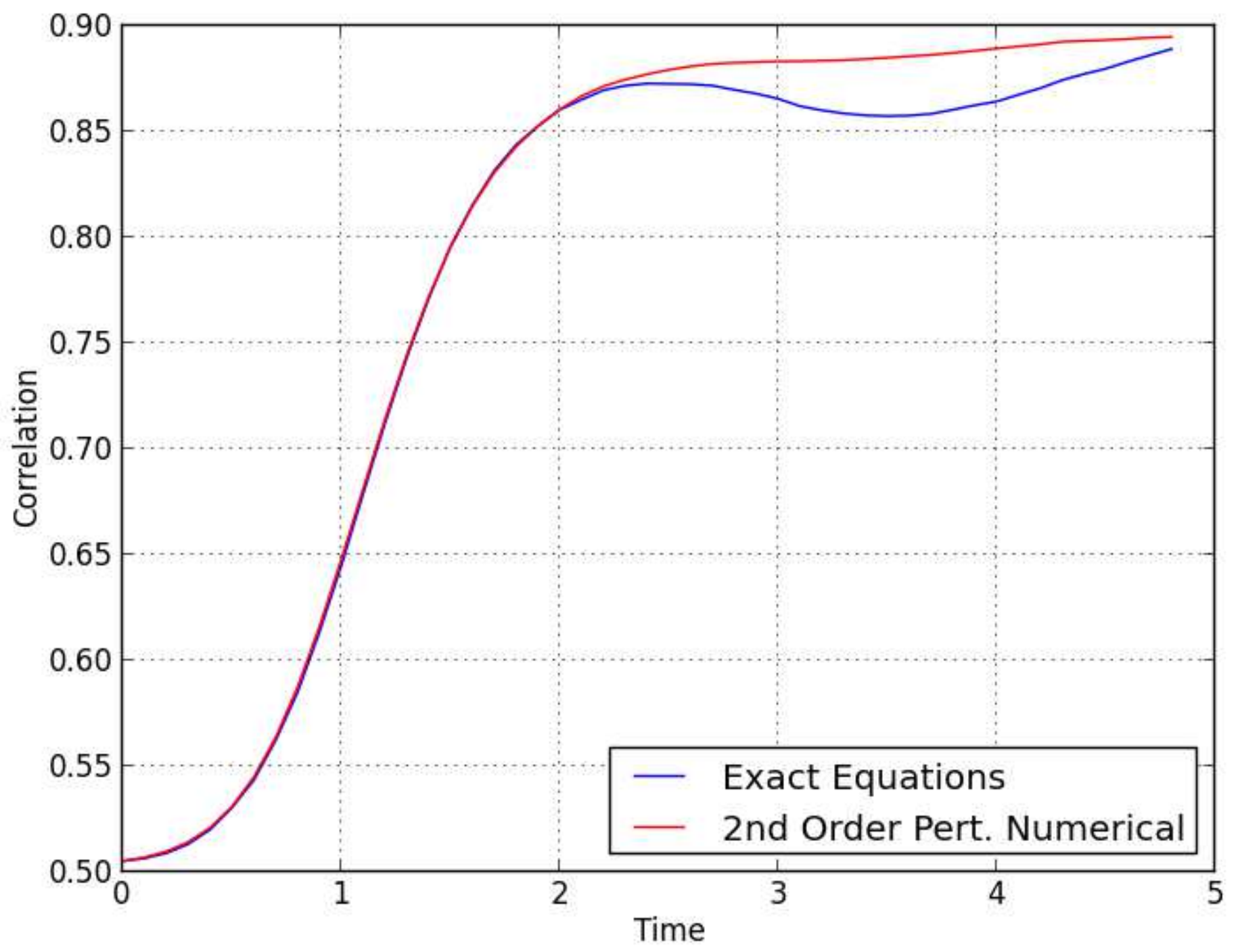}\caption[{\footnotesize{Numerical comparison of the perturbative expansion
with weak weights - 7}}]{{\small{\label{fig:Numerical-comparison-weak-7}As in the Figure
\ref{fig:Numerical-comparison-weak-6}, but with $E=2$. This, according
to \protect{\cite{citeulike:1343837}}, is approximately the point of maximum
complexity of the network, see text.}}}
\end{figure}

\begin{figure}
\noindent \begin{centering}
{\small{\includegraphics[scale=0.4]{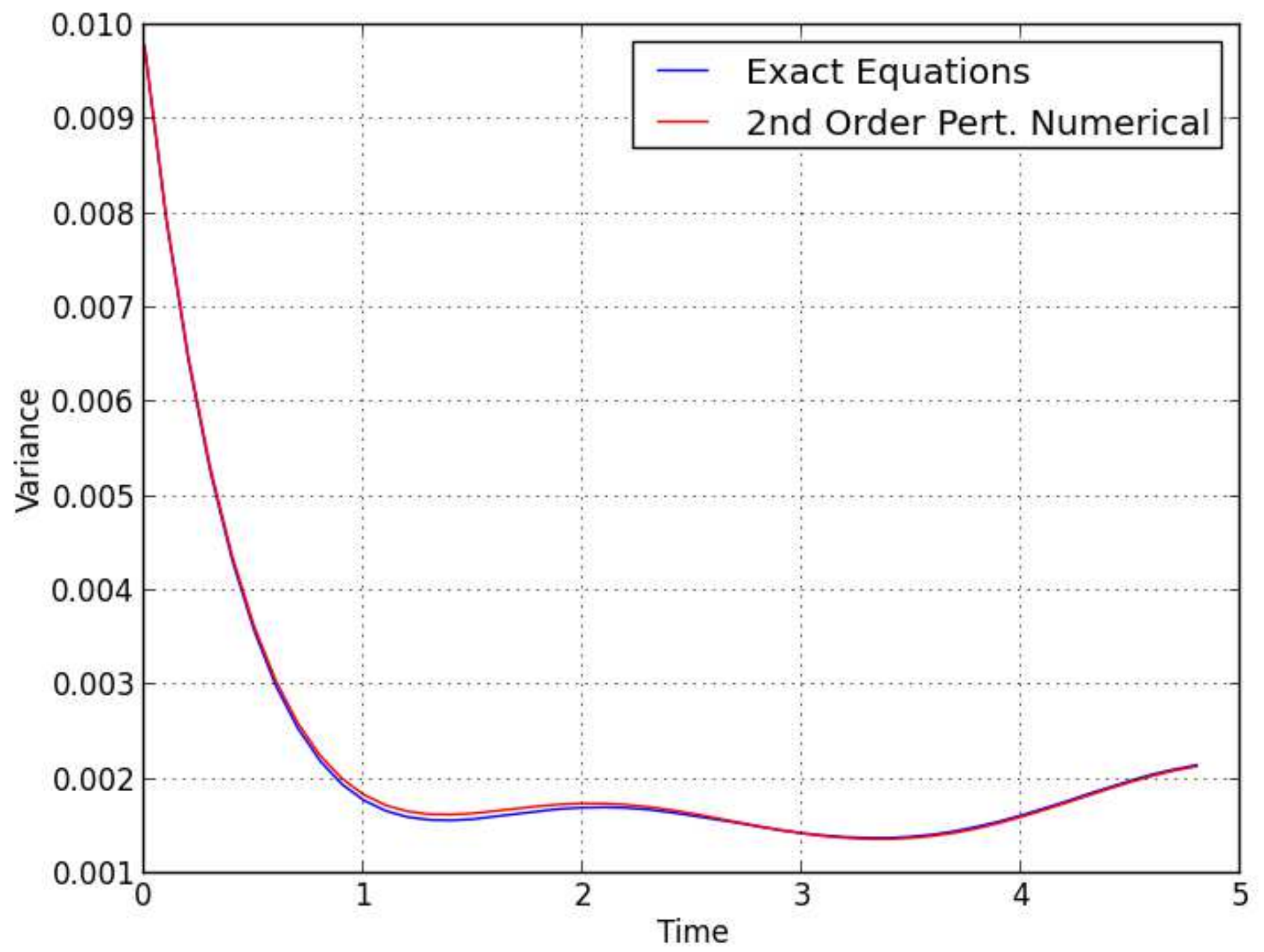}\includegraphics[scale=0.4]{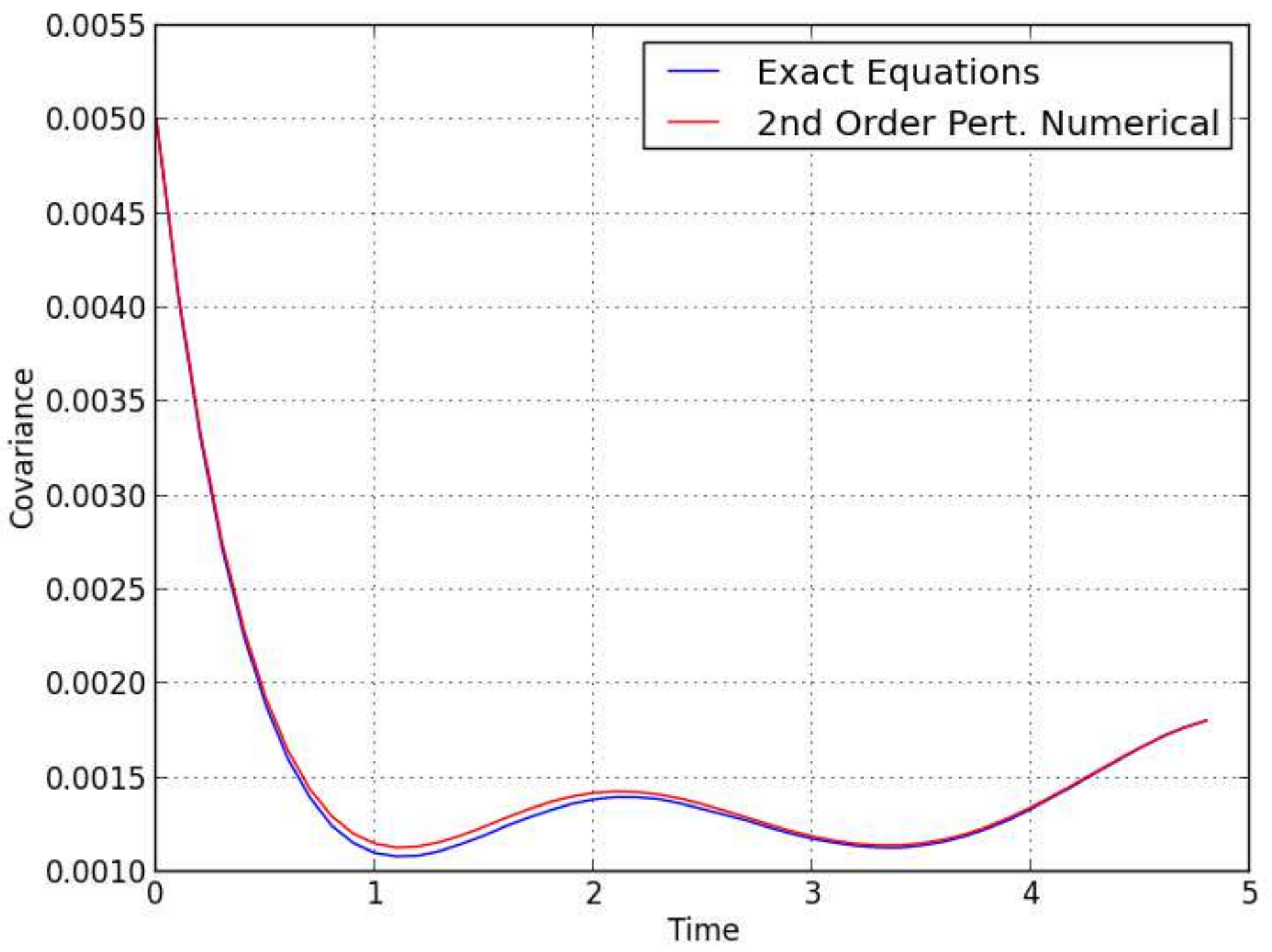}}}
\par\end{centering}{\small \par}

\noindent \centering{}\includegraphics[scale=0.4]{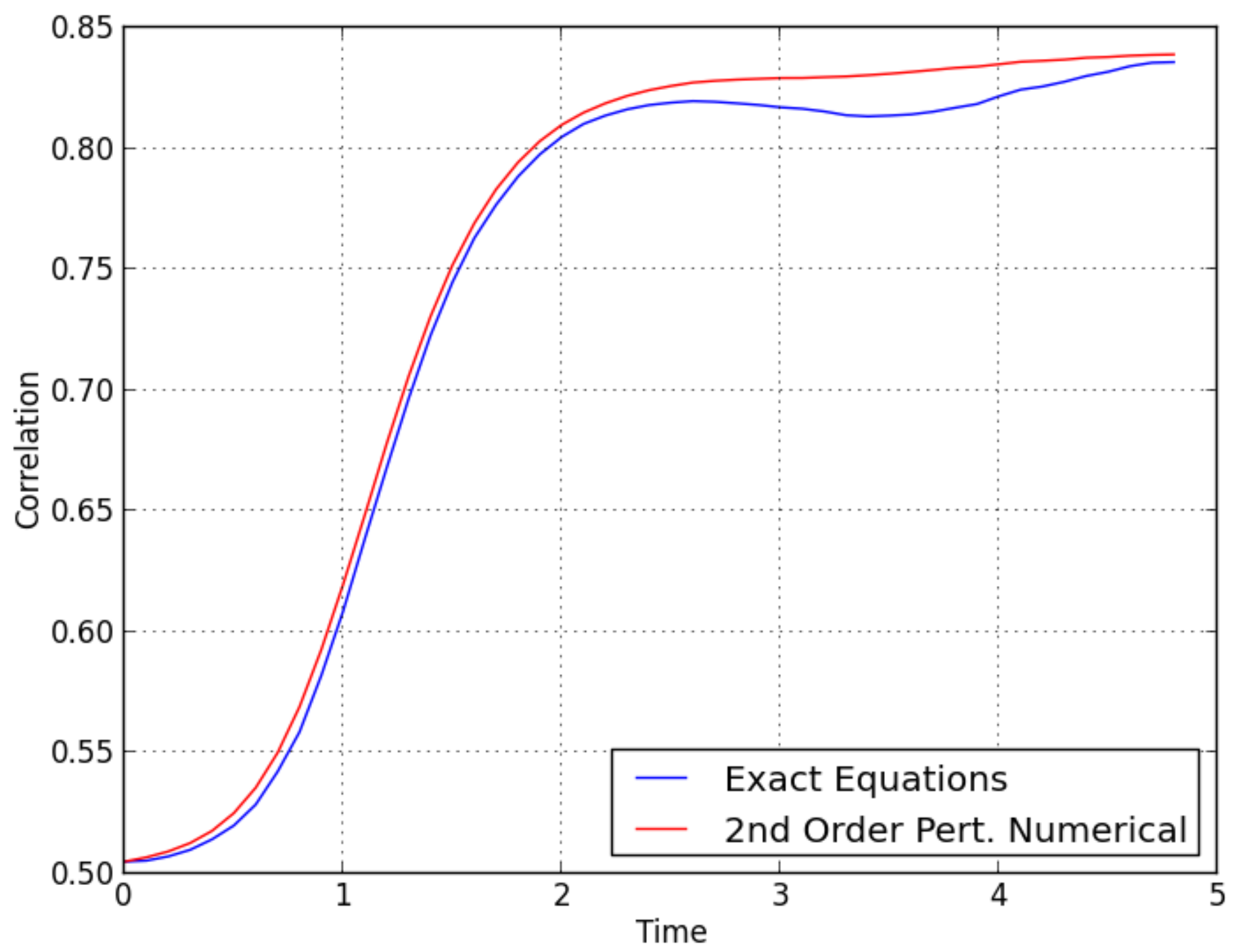}\caption[{\footnotesize{Numerical comparison of the perturbative expansion
with weak weights - 8}}]{{\small{\label{fig:Numerical-comparison-weak-8}As in the Figure
\ref{fig:Numerical-comparison-weak-6}, but with $E=5$. In this case
the blocks are almost completely disconnected. From the comparison
with Figures \ref{fig:Numerical-comparison-weak-6} and \ref{fig:Numerical-comparison-weak-7},
the reader can easily check that the increase of the parameter $E$
determines the reduction of the correlation at large $t$, as a consequence
of the diminution of the number of connections.}}}
\end{figure}

\begin{figure}
\noindent \begin{centering}
{\small{\includegraphics[scale=0.4]{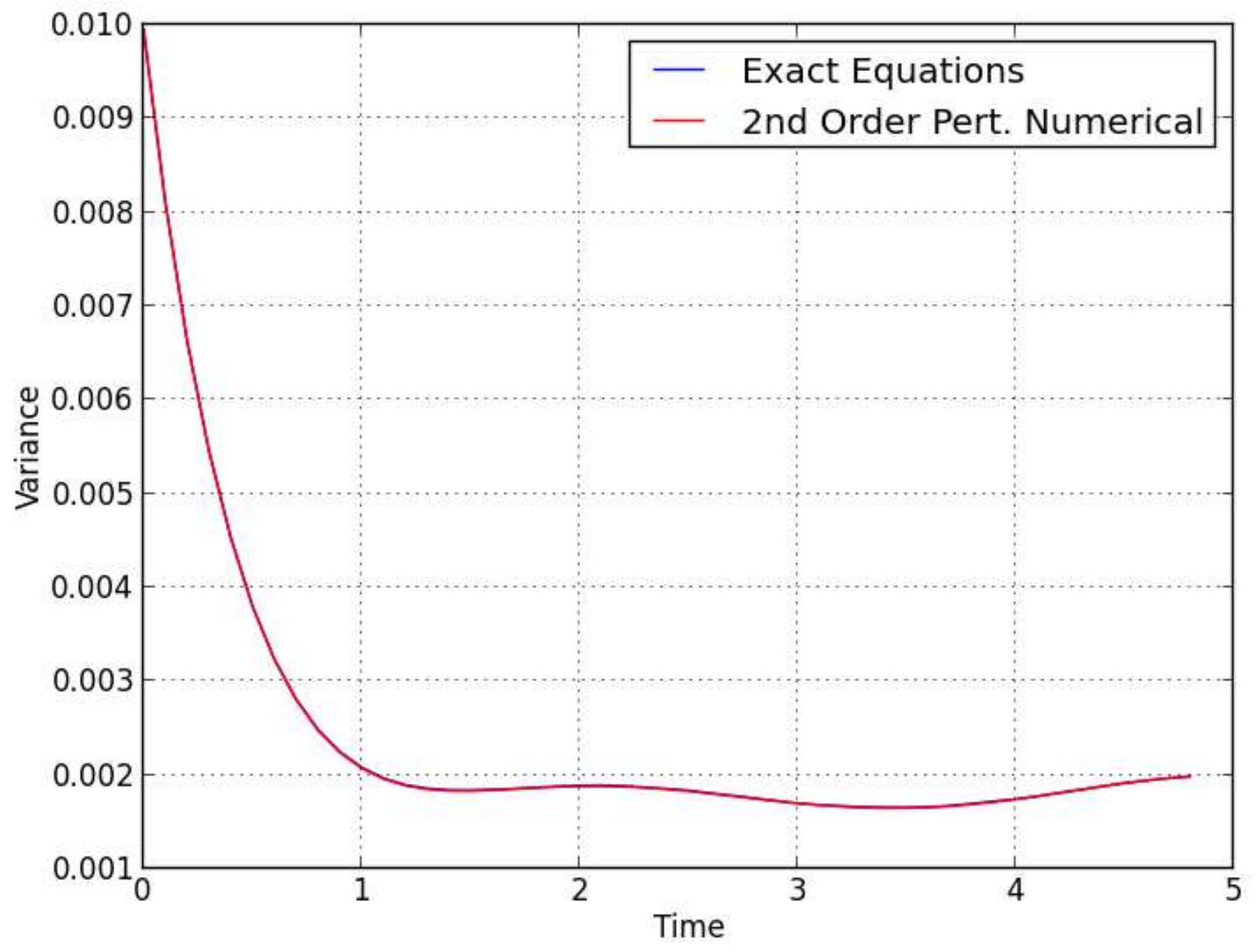}\includegraphics[scale=0.4]{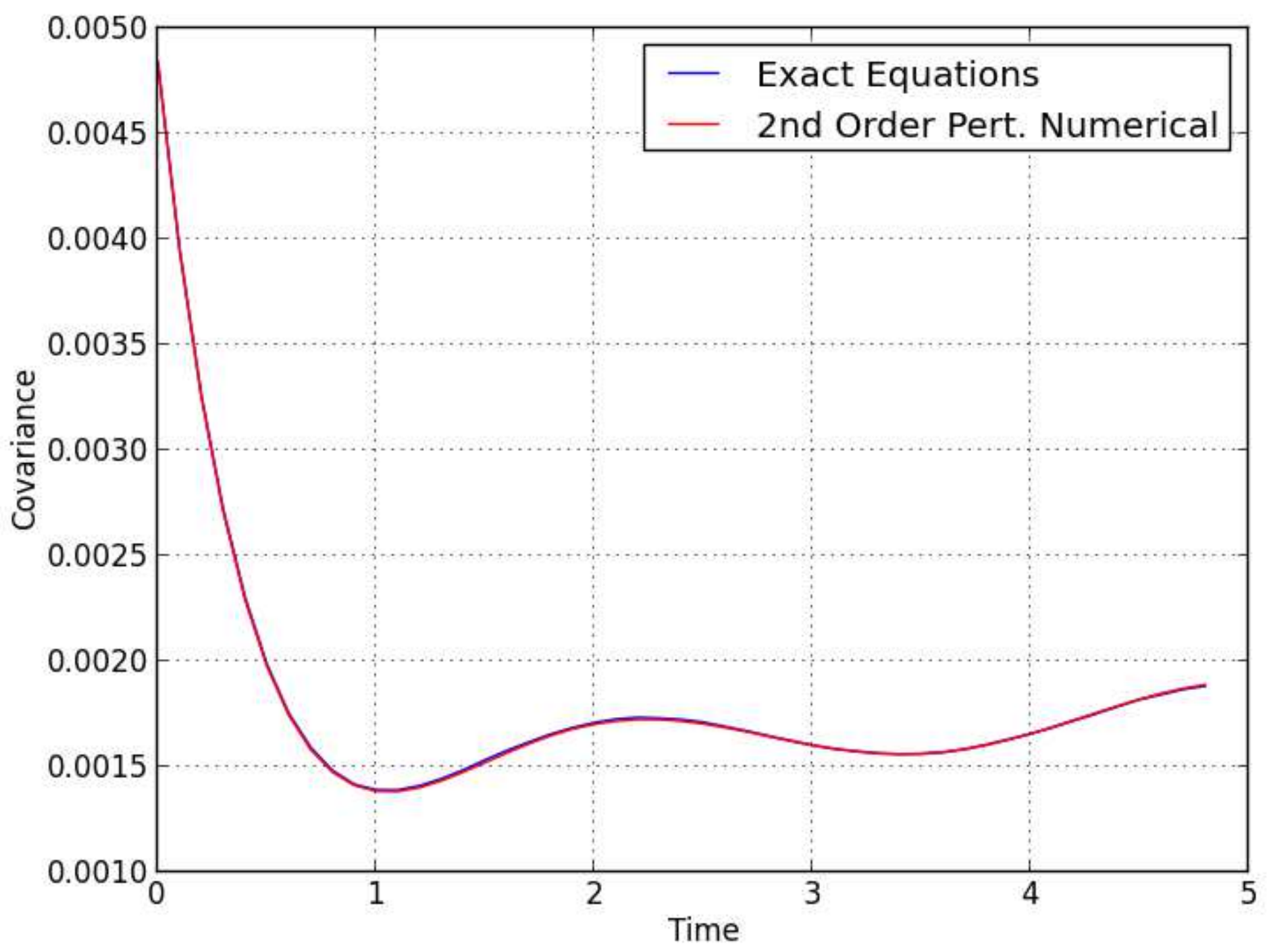}}}
\par\end{centering}{\small \par}

\noindent \centering{}\includegraphics[scale=0.4]{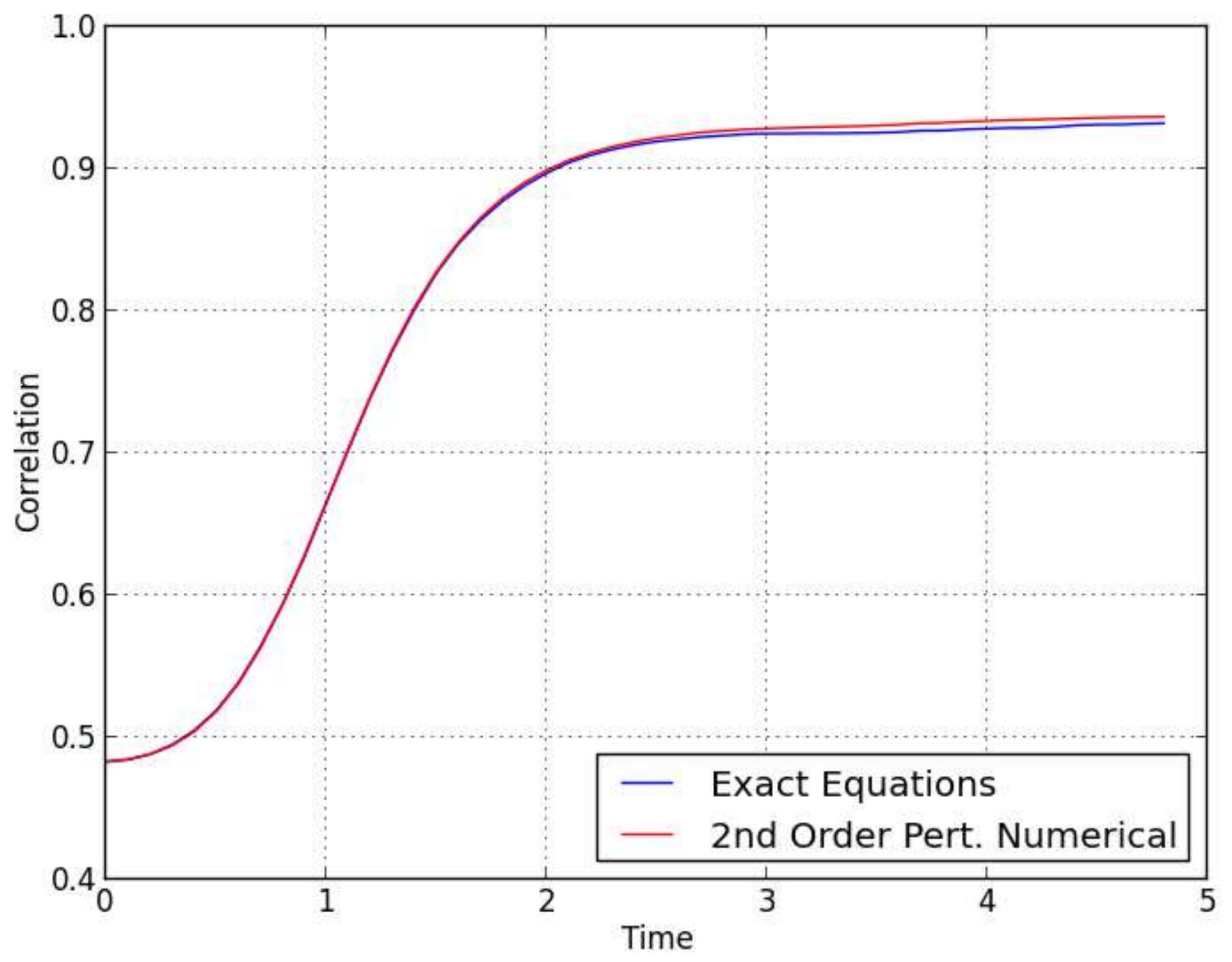}\caption[{\footnotesize{Numerical comparison of the perturbative expansion
with weak weights - 9}}]{{\small{\label{fig:Numerical-comparison-weak-9}Comparison of the
variance, covariance and correlation obtained for the Sporns' topology,
for the values of the parameters reported in Table \ref{tab:simulation-parameters-weak},
considering also the second order corrections of the membrane potential.
In this example we have set $\eta=4$ ($N=16$), $\mu=2$ and $E=1.1$,
as in Figure \ref{fig:Numerical-comparison-weak-6}, but now the neurons
are in two different blocks, and they are connected at the level $\kappa=2$.}}}
\end{figure}

\begin{figure}
\noindent \begin{centering}
{\small{\includegraphics[scale=0.4]{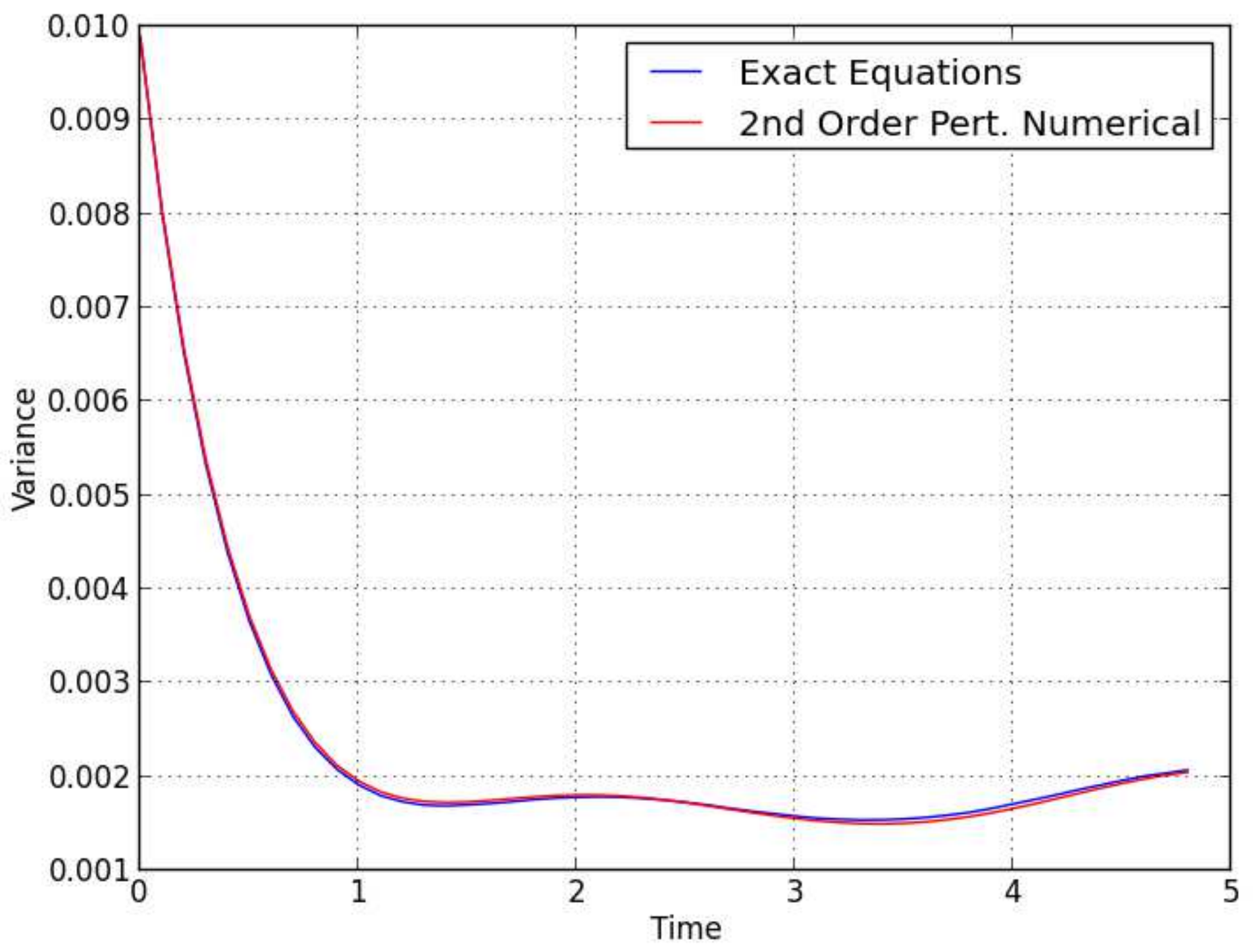}\includegraphics[scale=0.4]{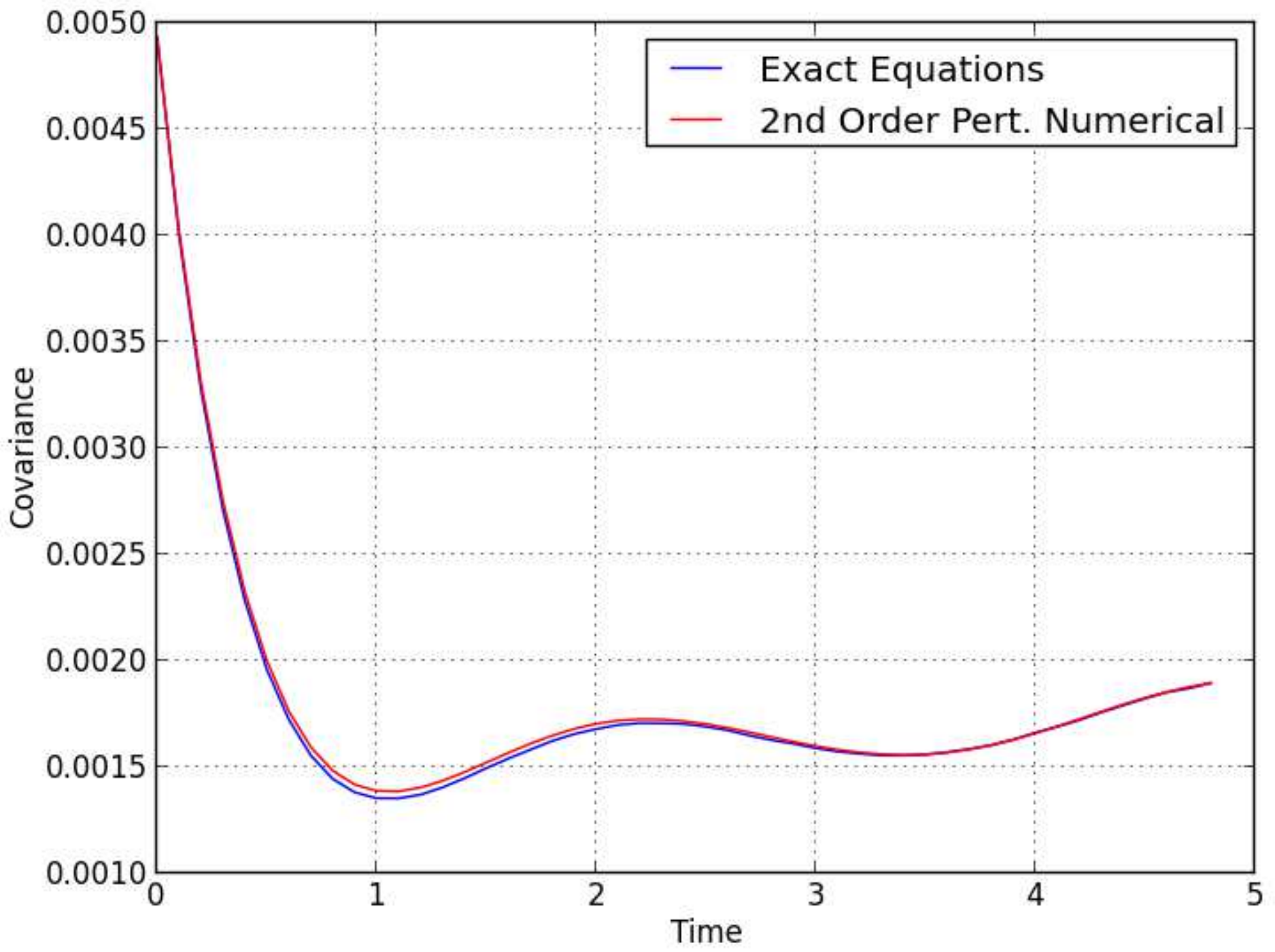}}}
\par\end{centering}{\small \par}

\noindent \centering{}\includegraphics[scale=0.4]{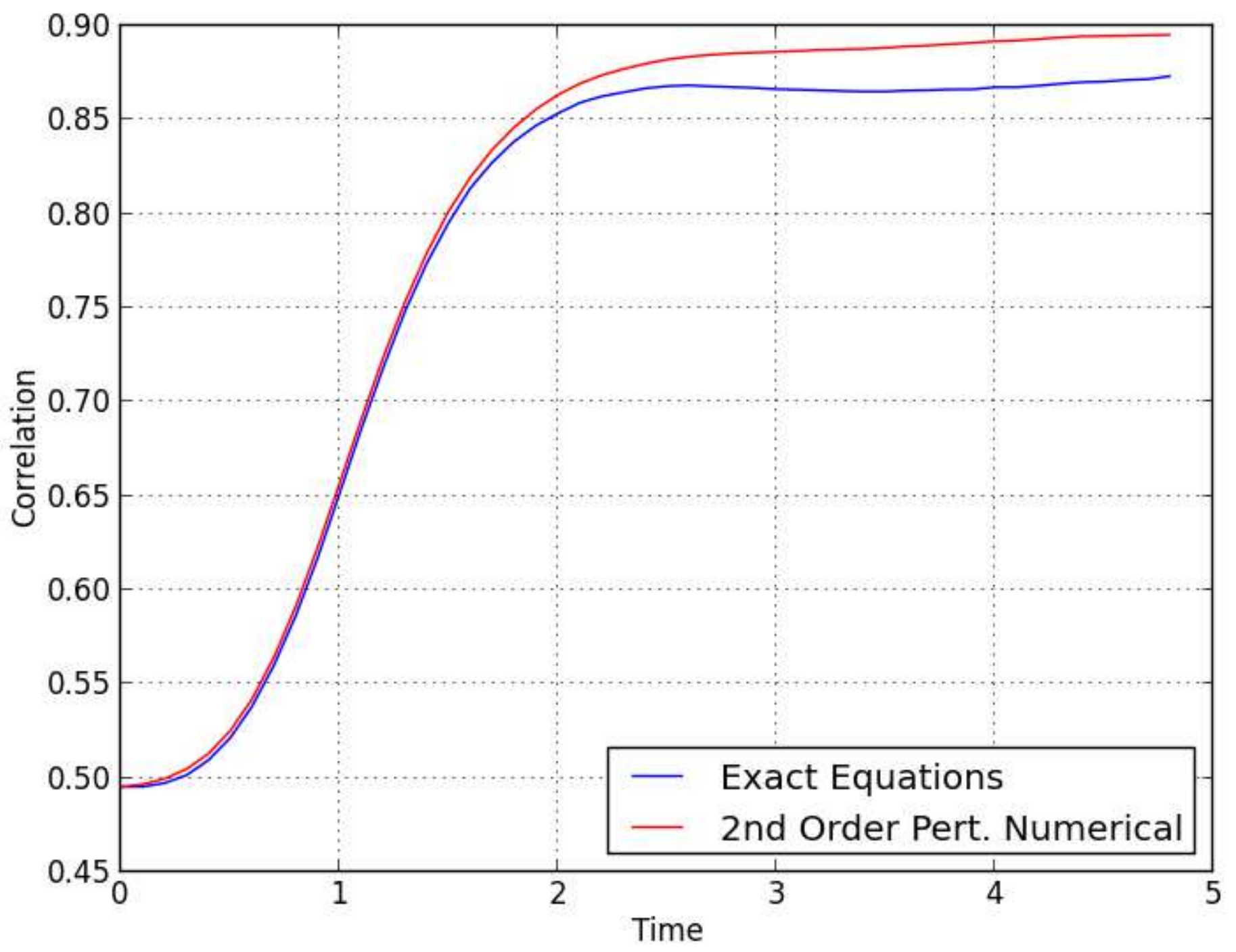}\caption[{\footnotesize{Numerical comparison of the perturbative expansion
with weak weights - 10}}]{{\small{\label{fig:Numerical-comparison-weak-10}As in the Figure
\ref{fig:Numerical-comparison-weak-9}, but with $E=2$.}}}
\end{figure}

\begin{figure}
\noindent \begin{centering}
{\small{\includegraphics[scale=0.4]{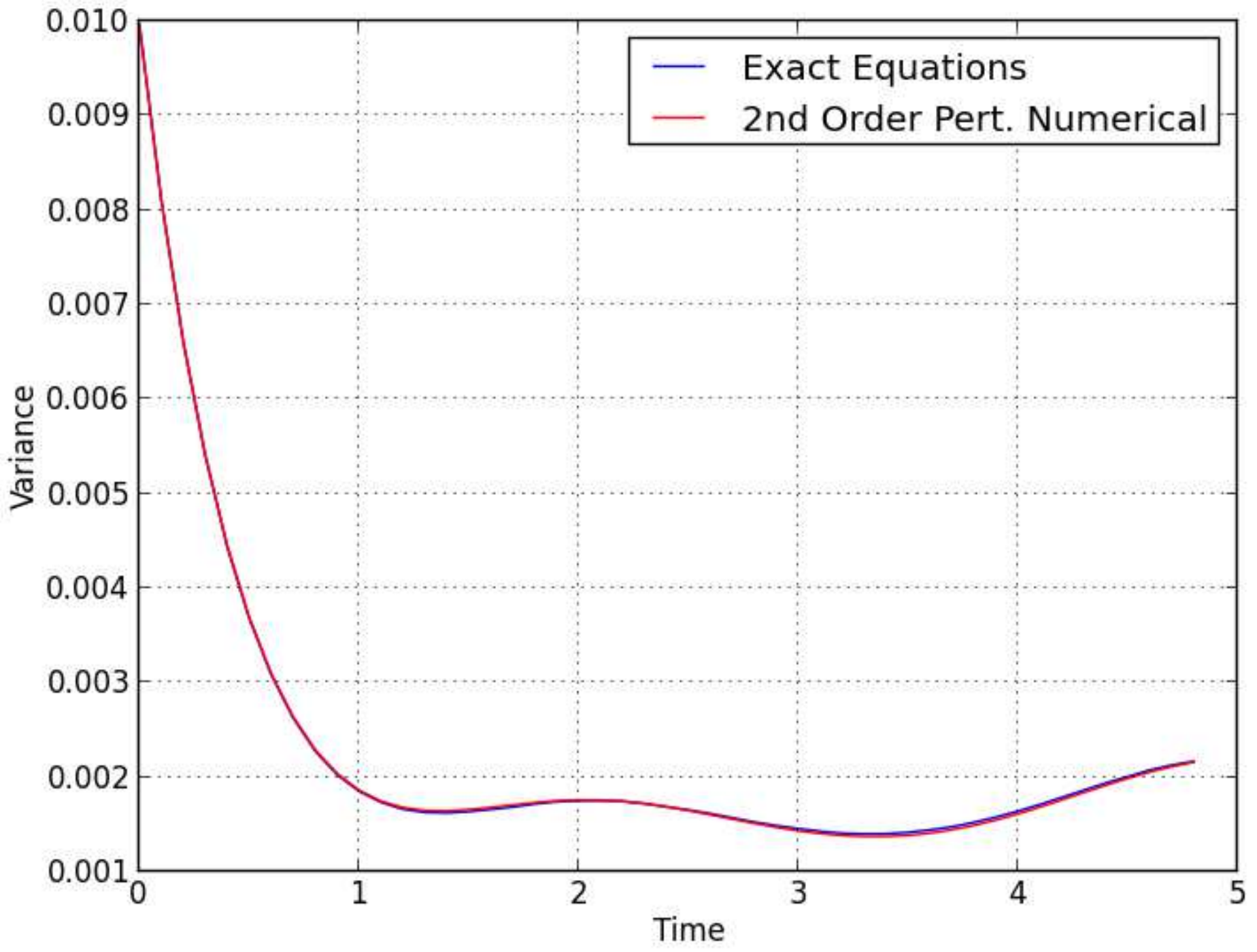}\includegraphics[scale=0.4]{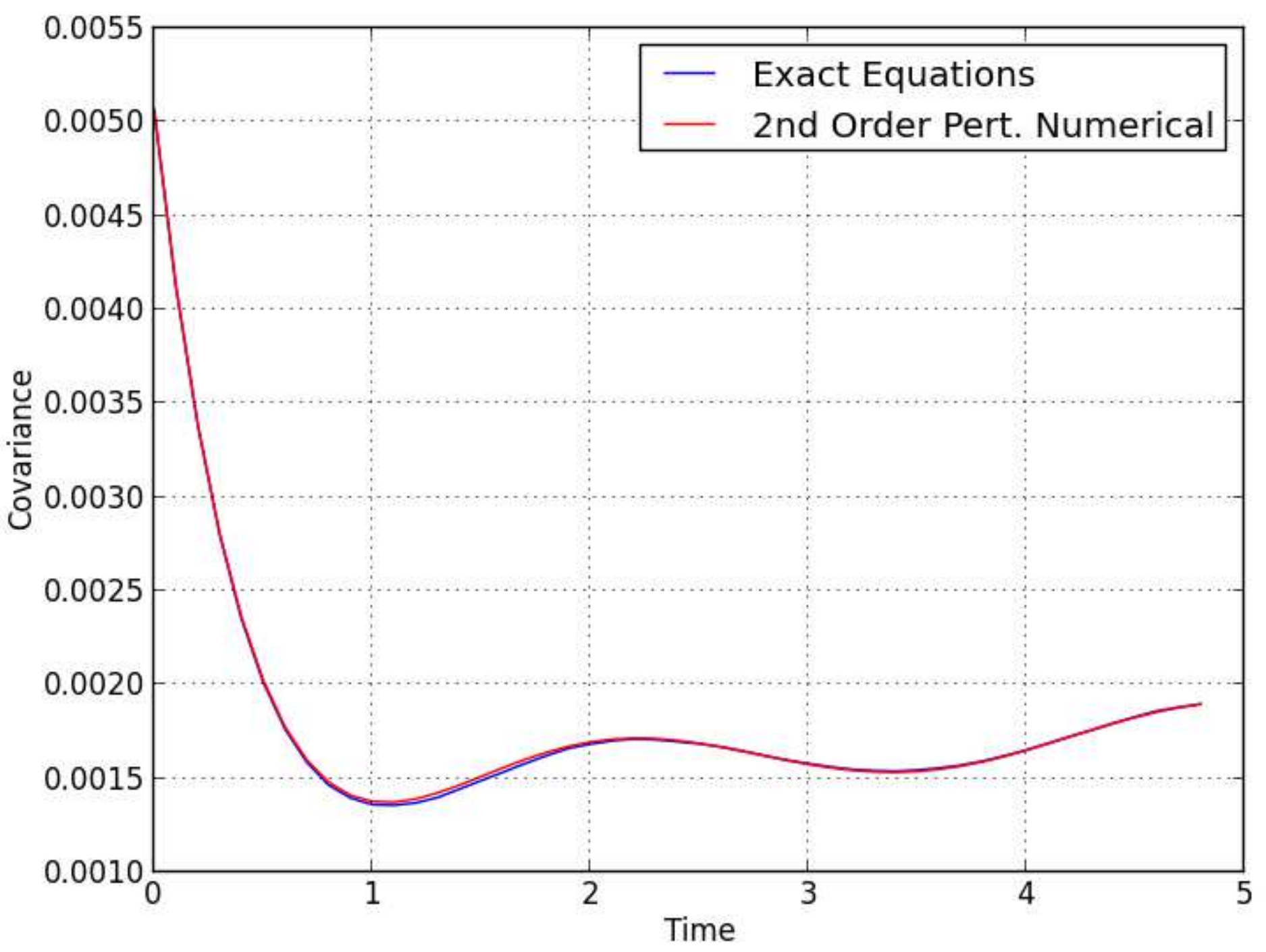}}}
\par\end{centering}{\small \par}

\noindent \centering{}\includegraphics[scale=0.4]{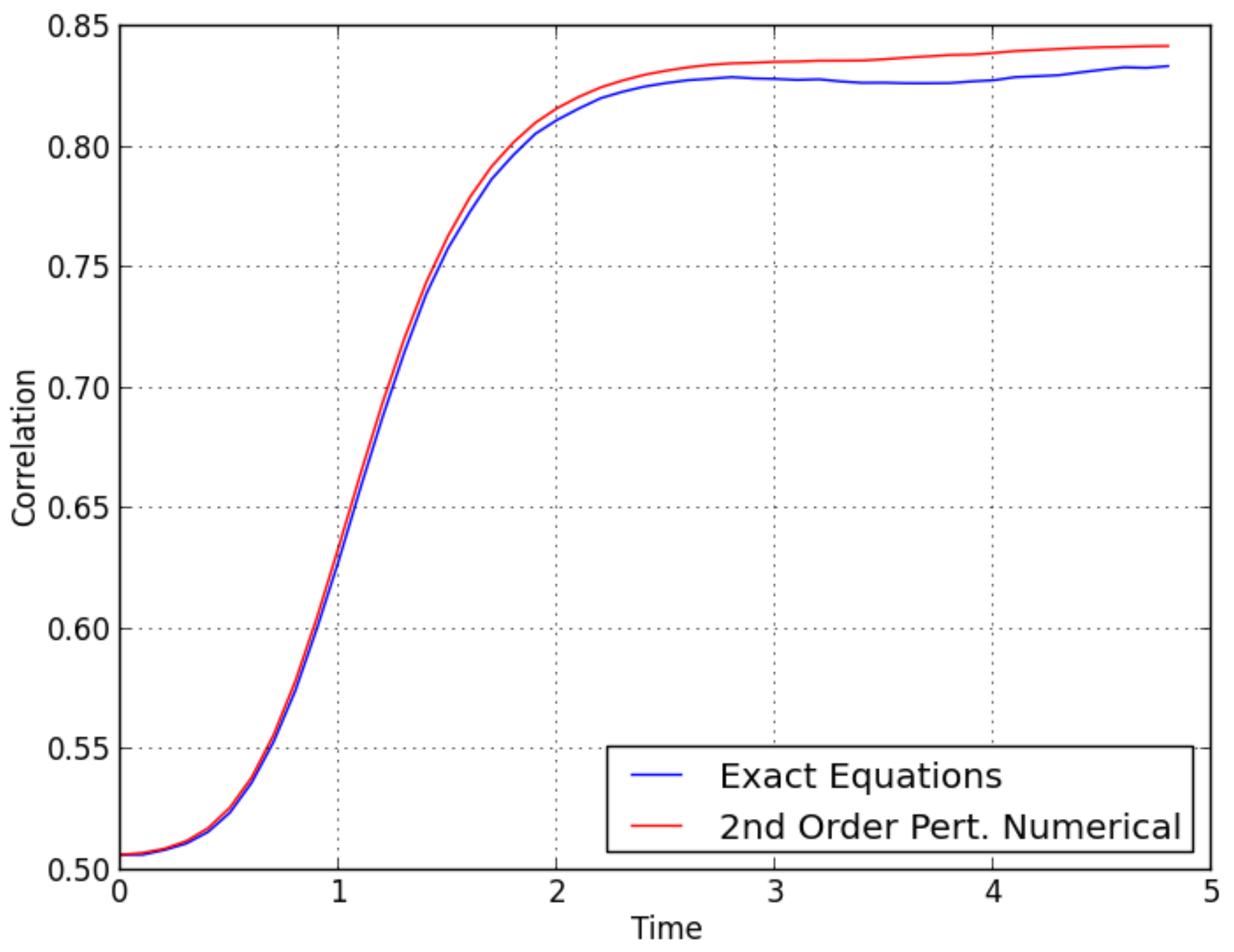}\caption[{\footnotesize{Numerical comparison of the perturbative expansion
with weak weights - 11}}]{{\small{\label{fig:Numerical-comparison-weak-11}As in the Figure
\ref{fig:Numerical-comparison-weak-9}, but with $E=5$. Again, the
increase of the parameter $E$ determines the reduction of the correlation
for large $t$. It is important to observe that the difference between
the two cases with the neurons in the same block or in two different
blocks is very small. This is due to the fact that the values of the
parameters $C_{1}$, $C_{2}$ and $C_{3}$ are relatively high (see
Table \ref{tab:simulation-parameters-weak}), therefore they strongly
determine the behavior of the correlation, for every topology. When
these parameters are set to zero, a richer behavior of the correlation
emerges. This analysis is not shown in the article, because the purpose
of this work is to develop mathematical tools that allow us to understand
a neural network, not the analysis of the consequences of the formulae.}}}
\end{figure}

\noindent For the networks with random topology, the analytic formulae
of the variance, covariance and correlation are rather complex to
implement. In fact usually the approximation of order $0$ of the
quantities $\mathbb{E}\left[\frac{T_{ij}}{M_{i}}\right]$, $\mathbb{E}\left[\frac{T_{ik}T_{jl}}{M_{i}M_{j}}\right]$
and $\mathbb{E}\left[\frac{T_{ik}T_{lm}T_{jl}}{M_{i}M_{j}M_{l}}\right]$
is not precise enough, forcing us to add the higher order corrections.
For example, for a network with independent random connections with
$p\left(T_{ij}=1\right)=p$ $\forall i,j:\, i\neq j$, the approximation
of order $0$ of $\mathbb{E}\left[\frac{T_{ij}}{M_{i}}\right]$ is:

\begin{onehalfspace}
\begin{center}
{\small{
\[
\mathbb{E}\left[\frac{T_{ij}}{M_{i}}\right]\approx\frac{\mathbb{E}\left[T_{ij}\right]}{{\displaystyle \sum_{k=0}^{N-1}}\mathbb{E}\left[T_{ik}\right]}=\frac{p}{\left(N-1\right)p}=\frac{1}{N-1}
\]
}}
\par\end{center}{\small \par}
\end{onehalfspace}

\noindent which does not depend on $p$ and therefore does not contain
information about the randomness of the topology. This means that
in general this approximation is a too poor description of the random
topology, and therefore the higher order corrections must be included.
Unfortunately, according to \ref{eq:Taylor-expansion-of-the-topology-2},
the approximations of order $1$ are always equal to zero, therefore
we have to extend the approximation up to the $2$nd order. In other
terms, we have to compute the second order derivatives in the multidimensional
Taylor expansions of $\mathbb{E}\left[\frac{T_{ij}}{M_{i}}\right]$,
$\mathbb{E}\left[\frac{T_{ik}T_{jl}}{M_{i}M_{j}}\right]$ and $\mathbb{E}\left[\frac{T_{ik}T_{lm}T_{jl}}{M_{i}M_{j}M_{l}}\right]$.
This is a feasible but complex task, and it is particularly hard for
the fractal connectivity matrix, since it depends on the blocks the
synaptic connections belong to. For this reason we have opted for
showing only the comparison between the numerical simulations of the
stochastic differential equations (red and blue lines), without using
the analytic formulae. Figures \ref{fig:Numerical-comparison-weak-5}
- \ref{fig:Numerical-comparison-weak-11} show these results for a
network with independent random connections and for the Sporns' fractal
matrix. The differential equations have been solved numerically using
the Euler-Maruyama scheme, while the integrals with respect to time
have been calculated using the trapezoidal rule, in both cases with
an integration time step $\Delta t=0.1$. All the statistics have
been evaluated with $10,000$ Monte Carlo simulations (where we have
independently generated repetitions of the four sources of randomness
of the system), while the remaining parameters are reported in Table
\ref{tab:simulation-parameters-weak}. The covariance and correlation
have always been calculated between the $0$th and the $1$st neuron.
The only exceptions are in Figures \ref{fig:Numerical-comparison-weak-9},
\ref{fig:Numerical-comparison-weak-10} and \ref{fig:Numerical-comparison-weak-11},
where the comparison is between the $0$th and the $8$th neuron.
Instead the membrane potentials and the variances have always been
reported only for the the $0$th neuron. In general we have obtained
a better agreement with the exact equations when we use also the second
order corrections of the membrane potential.

\noindent It is important to observe that a detailed analysis of the
error introduced by the perturbative expansion as a function of the
approximation order, the values of all the parameters of the system
and the infinitely many connectivity matrices is missing and is beyond
the purpose of this article.

\begin{table}
\begin{centering}
{\small{}}%
\begin{tabular}{|c|c|}
\hline 
{\small{Neuron}} & {\small{Input}}\tabularnewline
\hline 
\hline 
{\small{$\tau=1$}} & {\small{$I_{i}\left(t\right)=\begin{cases}
sin\left(2t\right), & \begin{array}{c}
i=0\div\frac{N}{2}-1\end{array}\\
0.2+e^{-t}, & \begin{array}{c}
i=\frac{N}{2}\div N-1\end{array}
\end{cases}$ }}\tabularnewline
\hline 
{\small{$C_{2}=0.5$}} & {\small{$C_{1}=0.4$}}\tabularnewline
\hline 
{\small{$\sigma_{2}=0.1$}} & {\small{$\sigma_{1}=0.01$}}\tabularnewline
\hline 
{\small{$\mu_{i}=\begin{cases}
-1, & \begin{array}{c}
i=0\div\frac{N}{2}-1\end{array}\\
0.5, & \begin{array}{c}
i=\frac{N}{2}\div N-1\end{array}
\end{cases}$}} & \tabularnewline
\hline 
\end{tabular}
\par\end{centering}{\small \par}

\selectlanguage{french}%
\noindent {\footnotesize{\smallskip{}
}}{\footnotesize \par}

\selectlanguage{english}%
\begin{centering}
{\small{}}%
\begin{tabular}{|c|c|}
\hline 
{\small{Synaptic Weights}} & {\small{Sigmoid Function}}\tabularnewline
\hline 
\hline 
{\small{$\widehat{\overline{J}}_{ij}\left(t\right)=\begin{cases}
\left(1+t^{2}\right)^{-1}, & \begin{array}{c}
i,j=0\div\frac{N}{2}-1\end{array}\\
\cos\left(t\right), & \begin{array}{ccc}
i=0\div\frac{N}{2}-1, &  & j=\frac{N}{2}\div N-1\end{array}\\
\sqrt{1+\frac{2}{\pi}\arctan\left(t\right)}, & \begin{array}{ccc}
i=\frac{N}{2}\div N-1, &  & j=0\div\frac{N}{2}-1\end{array}\\
e^{-t}\sin\left(t\right), & \begin{array}{c}
i,j=\frac{N}{2}\div N-1\end{array}
\end{cases}$}} & {\small{$T_{MAX}=1$}}\tabularnewline
\hline 
{\small{$C_{3}=0.6$}} & {\small{$\lambda=1$}}\tabularnewline
\hline 
{\small{$\sigma_{3}=0.1$}} & {\small{$V_{T}=0$}}\tabularnewline
\hline 
\end{tabular}
\par\end{centering}{\small \par}

\caption[{\footnotesize{Parameters for the simulation of the rate model}}]{{\small{\label{tab:simulation-parameters-weak}Parameters used to
generate Figures \ref{fig:Numerical-comparison-weak-1} - \ref{fig:Numerical-comparison-weak-11}}}}
\end{table}

\section{\label{sec:Conclusion}Conclusion}

\noindent We have shown how to study the correlation structure of
a stochastic neural network with a finite size and a generic connecivity
matrix.

\noindent This analysis has been performed using a second order perturbative
analysis in terms of the standard deviations of the sources of randomness
in the system and also in terms of the strength of the synaptic weights.

\noindent All the distributions are supposed to be normal, which has
allowed us to obtain analytic results using the Isserlis' theorem.

\noindent This calculation has been developed for both deterministic
and random topologies of the synaptic connections, and applied to
a biologically relevant case with fractal nested structure.

\noindent Moreover the numerical comparison with the exact neural
equations has shown a good agreement if the perturbative parameters
are small enough.

\noindent Therefore this technique can be used to study neurons with
complicated connections and to reveal the relation between the covariance
matrix of the membrane potentials, namely the \textit{functional connectivity}
of the system, and the matrix $J\left(t\right)$, also known as\textit{
structural} or \textit{anatomical connectivity.}

\noindent The perturbative analysis developed in this article can
also be extended easily to study the correlation structure of rate
neurons with synaptic plasticity or learning, namely when the intensities
of the synaptic weights are not chosen a priori, but are generated
by other differential equations.

\noindent This idea can also be extended to spiking neurons, like
those described by the FitzHugh-Nagumo \cite{citeulike:882569}\cite{4066548},
Morris-Lecar \cite{Morris1981} or Hodgkin-Huxley \cite{citeulike:851137}
equations.

\noindent The only problem with these models is that they are not
analytically solvable even when the neurons are disconnected, because
the differential equation of $Y_{0}^{i}\left(t\right)$ becomes non-linear.

\noindent Nevertheless the equations satisfied by the other functions
$Y_{m}^{i}\left(t\right)$ and $Y_{m,n}^{i}\left(t\right)$ are linear,
therefore in this case we can determine the correlation structure
of the system semi-analytically.

\noindent In other words, all the results will be expressed in terms
of analytic functions of $Y_{0}^{i}\left(t\right)$, which is not
analytically known, but must be solved numerically.

\noindent To conclude, we remind that the results of this article
can be applied only to the case of weak synaptic weights and when
the slope of the activation function is not too large, so the next
step will be the development of a theory which describes the behavior
of the network under more general hypotheses.

\section*{Acknowledgements}

This work was partially supported by the ERC grant \#227747 NerVi,
the FACETS-ITN Marie-Curie Initial Training Network \#237955 and the
IP project BrainScaleS \#269921.

\selectlanguage{french}%
\noindent {\footnotesize{\smallskip{}
}}{\footnotesize \par}

\selectlanguage{english}%
\bibliographystyle{unsrt}
\bibliography{Article}

\end{document}